\documentclass[useAMS,usenatbib]{mn2e}
\usepackage{graphicx}
\newcommand{\kms}{{\, \rm km~s}^{-1}}
\newcommand{\ho}{{\rm km~s}^{-1}~{\rm Mpc}^{-1}}
\newcommand{\reduceme}{\mbox{R\raisebox{-0.35ex}{E}D\hspace{-0.05em}
\raisebox{0.85ex}{uc}\hspace{-0.90em}\raisebox{-.35ex}{{m}}\hspace{0.05em}E}} 
\title[On-axis spectroscopy of quasar hosts]
 {On-axis spectroscopy of the host galaxies of 20 optically luminous quasars at
$z\sim0.3$
\thanks{Based on observations made with the ESO Very Large Telescope Unit 1
(ANTU/UT1) at ESO Paranal Observatory, Chile, under program IDs 65.P-0361, and
66.B-0139.}} 
\author[G. Letawe et al.]
{G. Letawe,$^1$\thanks{E-mail: gletawe@ulg.ac.be}
P. Magain,$^{1}$ F. Courbin,$^{2}$ P. Jablonka,$^{2,3}$ 
\newauthor
K. Jahnke,$^{4}$, G. Meylan,$^{2}$ and  L. Wisotzki$^{5}$\\
$^{1}$ Institut d'Astrophysique et  G\'eophysique, Universit\' e de
  Li\`ege, All\'ee du 6 Ao\^ut, 17, Sart Tilman (Bat. B5C), B-4000 Li\`ege,
Belgium\\
$^{2}$ Laboratoire d'Astrophysique, Ecole Polytechnique F\'ed\'erale de
Lausanne (EPFL),
   Observatoire,   CH-1290 Sauverny,   Switzerland \\
$^{3}$ Observatoire, Universit\'e de  Gen\`eve, CH-1290 Sauverny,   Switzerland
\\
$^{4}$  Max-Planck-Institut f\"{u}r Astronomie, K\"{o}nigstuhl 17, D- 69117
Heidelberg, Germany \\
$^{5}$ Astrophysikalisches Institut Potsdam, An der Sternwarte 16, 
D-14482 Potsdam, Germany}

\pagerange{\pageref{firstpage}--\pageref{lastpage}} \pubyear{2007}
\begin{document}
\maketitle
\label{firstpage}
\begin{abstract} We present the analysis of a 
 sample  of  20 bright low-redshift  quasars ($M_B<-23$
and $z  < 0.35$) observed  spectroscopically with  the VLT.  The FORS1
spectra, obtained in Multi Object Spectroscopy (MOS) mode, allow  to
observe simultaneously the quasars and several reference stars used to
spatially deconvolve the data.  Applying the MCS deconvolution method,
we are able to separate the individual spectra of the
quasar and of  the underlying host galaxy.  Contrary to some previous claims, we  find that luminous quasars are not exclusively hosted by massive ellipticals. Most quasar host
galaxies harbour large amounts of gas, irrespective of their morphological
type. Moreover, the stellar
content  of half of the hosts is a young Sc-like population, associated with 
a rather low metallicity interstellar medium. 
A significant fraction of the galaxies contain gas ionized at
large  distances by the quasar  radiation. This large distance ionization is
always associated with signs of gravitational interactions 
(as detected from images or disturbed rotation curves).
The spectra of the quasars themselves provide evidence that gravitational 
interactions bring dust and gas in the immediate surrounding of the super
massive 
black hole, allowing to feed it.  The quasar activity might thus be triggered
(1) in young gas-rich spiral galaxies by local events and (2) in more evolved
galaxies by gravitational interactions or collisions. The latter mechanism
gives rises to the most powerful quasars.
Finally, we derive mass models for the isolated spiral host galaxies and we
show that the most reliable estimators of the systemic redshift in the quasar
spectrum are the tips of the  H${\alpha}$  and H${\beta}$ lines.

\end{abstract}


\section{Introduction}

Close to half a century has  passed since  the first detailed spectroscopic
observations  of  a      quasi-stellar-object  (QSO)     (\citealt{oke};
\citealt{green}),  and two decades since the first clear spectroscopic 
evidence  that quasars  are located  in  the centres  of  much larger
stellar envelopes or {\it  host galaxies} (\citealt{boroson84}).  After this
pioneering spectroscopic  work, most studies have concentrated on
imaging,  attempting to measure the shape and  the luminosity of quasar
host galaxies   (e.g.,   \citealt{bahc97}; \citealt{dunlop03}).   With
better image processing techniques,  it has become possible  to derive
information about the stellar content of  quasar hosts using  multicolour
imaging (\citealt{ronnback96}; \citealt{jahnke04}; \citealt{sanchez}),
but very  little effort has  been spent  on genuine deep spectroscopy,
mainly  because of   the difficulty  to  accuratly  decontaminate  the galactic
spectrum  from the light of the central quasar.

The quasar-host separation problem has been  circumvented, at least in
part, in  three  ways.  One is to   restrict  oneself to  samples  of
obscured quasars  (\citealt{kauff}), with  the  obvious  drawback that
samples built in that way are biased, and that there is no easy way to
accurately recover   the  unobscured    quasar  luminosity.    Another
technique is to carry out off-axis spectroscopy: the spectra are
obtained with  the slit of  the spectrograph  located a few arcseconds
away from the quasar (e.g., \citealt{hutch90}; \citealt{nolan};
\citealt{miller}). However, residual contamination by the quasar remains
difficult to  remove completely, the velocity  information on the host
is lost, and only the external parts of  the host can be studied.
Finally, the 1D template-spectra of non-active galaxies can be used as
a basis to  perform eigenvector  decomposition (\citealt{vdb05}).  The
latter method is efficient  in  drawing general conclusions using   very large
numbers of objects,  but relies on a  restricted number of eigenspectra.
This prevents the discovery of unexpected spectral features, e.g., due
to the specific spatial distribution of the gas in each galaxy, to the
starburst  activity, or to  winds and  jets escaping  from the central
AGN.

 In the present work, we take  advantage of a  new approach, where
the quasar  spectra are  taken  on-axis,  i.e.   with the slit   lying
directly accross the nuclear  point source.  Our quasar-host separation
technique relies on a spatial deconvolution  of the spectra, using the
spectra of neighbouring PSF stars.   This method has already been exploited
in  three    previous  papers  (\citealt{courb02};    \citealt{let04};
\citealt{mag05})   and allows to  go beyond  previous  studies: 1- the
spectral properties of the hosts can be determined down to the central
kpc; 2- the stellar  and gaseous contributions to the total spectrum can be
accurately separated, which is not the case in multicolour imaging;  3- no
prior assumption on the host, neither spectral  nor morphological, is necessary
during the deconvolution process; 4- the velocity curve of the host around its
central quasar can be measured.

\begin{figure}
\centering
\includegraphics[width=8.5cm]{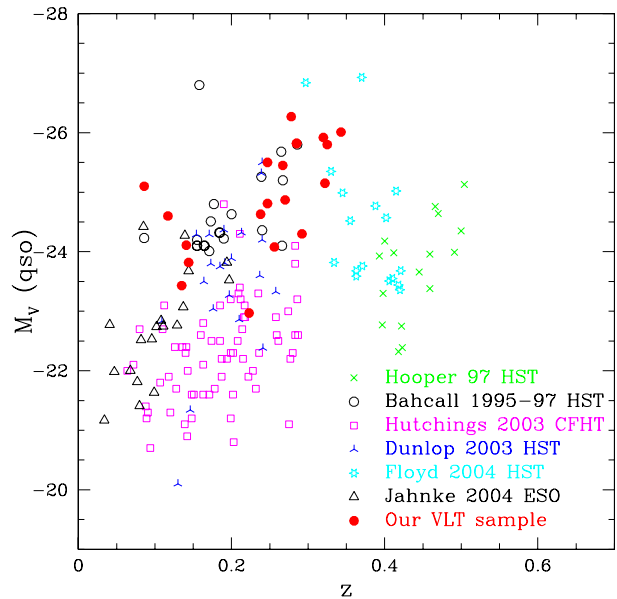}
\caption{Location of our quasar sample in the absolute magnitude vs. 
redsdhift  plane, compared   with other  samples   observed in imaging
either with the {\it HST} or from the ground (CFHT, ESO).}
\label{ech}
\end{figure}

\begin{figure}
\centering
\includegraphics[width=8.5cm]{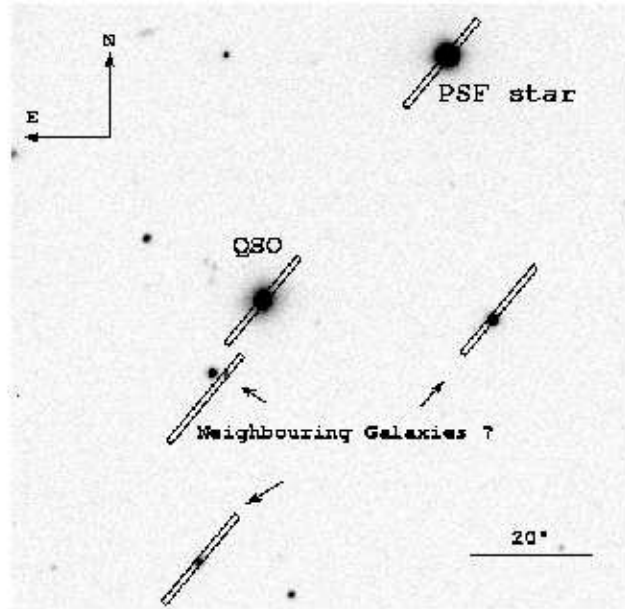}
\caption{Example of a MOS mask used to carry out the FORS1 observations,
for   the    quasar  HE\,1503+0208.   One  slitlet  (19 $\times$
1 arcsec) is positioned on  the  quasar, and  several others on   PSF
stars.  The   remaining slitlets are  used   to obtain the  spectra of
neighbouring galaxies.}
\label{slits}
\end{figure}

Recent  studies  have argued that  luminous  quasars reside in massive
elliptical   galaxies  \citep{dunlop03}   with   globally old  stellar
populations, but with   slightly  bluer colours than   their quiescent
counterparts, indicative  of an additional younger  stellar component
(\citealt{kauff}; \citealt{jahnke04}; \citealt{vdb05}). We show that these
general  conclusions  should  be taken with caution,  as important information
is lacking from their data, such as separation of the gas and stars, velocity
fields, gas ionization,\ldots\,  
In particular, we investigate how  the central AGN affects  the evolution
of  the whole  galaxy by carrying  out  a comparative  study between 
samples of galaxies with  and without  quasars.   We compare their  stellar
populations      and    the      ionization        state    of   their
Interstellar Medium (ISM). Whenever possible,  we  infer the dynamical mass  of
the host galaxies and we trace distorsions of their velocity field due
to past or present interactions with nearby companions.

Our work is  a first step towards  a  larger spectroscopic study  of quasar
hosts aimed  at understanding why some galaxies  are active while others
are not.  Many studies have characterized the morphology of the hosts,
but the morphology alone does not reflect the actual stellar and gaseous
content of the hosts. The aforementioned blue colours of quasar hosts
confirm that their evolution  does not follow a classical
path.  It is thus important to understand how their stage of evolution is
linked to the nuclear activity.  Because of the complexity   and because of
the variety of physical  phenomena at  work in  quasars and their  host
galaxies, the observational solutions of this puzzle  can only come from
combined spectroscopic and imaging studies.

In  Section~\ref{sample}, we  present  our  sample  of 20 low-redshift
optically  luminous quasars.   In Sections~\ref{obs} and \ref{obsspec}
we describe    the    observations  and data processing.   Imaging is treated
in Section~\ref{anima}, while we   give in
Section~\ref{overv} an overview of the spectral analysis, further detailed in
Section~\ref{anispec}.  Section~\ref{dyn} is devoted
to the dynamics  of quasar hosts and to  the detection of gravitational
interactions with  companions,  while  Section~\ref{quasar} studies the
links    between   the     nuclear  and   host   properties.  Finally,
Section~\ref{redsh}  establishes  the  relation between  the
redshift of the quasars and that of  their host galaxy. Throughout the
paper, we adopt the  following cosmological parameters: $H_0=65~\ho $,
$\Omega_m=0.3$ and $\Omega_{\Lambda}=0.7$.

\begin{table*}
\caption[]{Our QSO sample sorted by increasing right ascension: (1-) Object
name,  
(2-) redshift  determined from  the  H${\alpha}$ emission  line of the
quasar (see Section~\ref{redsh}), (3-) integrated absolute $B$ magnitude
from  the HES catalogue, (4-5) absolute  $V$-magnitude  for the quasar
and host  as measured from  our  VLT observations,  (6-) observing run
(A = April 2000; B = November 2000), (7-) Galactic
extinction  from \citet{schlegel98}, (8-9)  seeing  value and exposure time for
each  grism
(G600B,    G600R,   G600I),  (10-)   host   morphology as
determined from other studies (see  Section~\ref{anima}) and interaction signs.
D stands for
disc dominated, E for    elliptical, U for undefined, the number coding stands
for: 0 = no  interaction signs, 1 = close
companion outside the   galaxy, 2 = tails,   3 = merger, (11-)  quasar
properties:  objects with   radio emission  are indicated  by  R (from
NVSS), but only HE1302-1017 is radio-loud. Narrow line quasars
are labelled NL (see Section~\ref{quasar}).}
\begin{tabular}{lcccccccclc}
\hline
Object name & z(H${\alpha}$)  & M$_B $(tot) & M$_V$(QSO)& M$_V$(host) & Run &
E$(B-V)$ & Seeing          & $t_{exp}$        & Gal& QSO   \\ 
            &  quasar      &   HES       &    VLT    &     VLT     &     &       
& arcsec: $B,R,I$ & seconds: $B,R,I$&    &       \\ 
\hline
HE\,0132$-$4313  & 0.237 & -24.37 & -24.63 & -22.63 & B &
0.017&0.65-0.71-0.67&3000-1800-1800&U, 0&NL\\
HE\,0203$-$4221  & 0.319 & -24.88 & -25.92 & -23.10 & B &
0.014&0.71-0.75-0.60&3000-1800-1800&U, 2&\\
HE\,0208$-$5318  & 0.344 & -24.76 & -26.01 & -23.00 & B &
0.026&0.61-0.50-0.50&3000-1800-1800&U, 0&\\
HE\,0306$-$3301  & 0.247 & -24.32 & -24.81 & -22.41 & B
&0.014&0.68-0.66-0.86&3000-1800-1800&D, 2,3&\\
HE\,0354$-$5500  & 0.267 & -24.70 & -25.45 & -23.47 & B &
0.016&0.52-0.67-0.48&3600-1800-1800&D, 3&\\
HE\,0441$-$2826  & 0.155 & -24.62 &  -     &    -   & B &
0.035&0.74-0.47-0.53&3000-1800-2400&E, 0&R\\
HE\,0450$-$2958  & 0.286 & -25.24 & -25.82 & -  & B &
0.015&0.54-0.67-0.57&3600-1800-1800&U, 1&NL,R\\
HE\,0530$-$3755  & 0.293 & -24.07 & -24.30 & -20.70 & B
&0.027&0.70-0.68-0.65&3000-1800-1000&U, 1,2&\\
HE\,0914$-$0031  & 0.322 & -25.01 & -25.15 & -22.02 & A &
0.035&2.06-2.11-2.24&1200-1200-1200&D, 0&\\
HE\,0956$-$0720  & 0.326 & -24.79 & -25.80 & -23.04 & A &
0.055&0.53-0.65-0.59&1200-1200-1200&U, 0&\\
HE\,1009$-$0702  & 0.271 & -24.28 & -24.87 & -22.94 & A
&0.033&0.69-0.64-0.70&1800-1800-1800&D, 0&NL\\
HE\,1015$-$1618  & 0.247 & -25.13 & -25.50 & -22.55 & A &
0.078&1.20-0.92-1.01&1200-1200-1200&D, 0&\\
HE\,1029$-$1401  & 0.086 & -25.07 & -25.10 & -22.53 & A &
0.067&1.61-1.54-1.94&1200-1200-1200&E, 3&R\\
HE\,1228+0131  & 0.118 & -24.33 & -24.60 & -21.44 & A &
0.019&0.91-0.76-0.72&1200-1200-1200&E, 0&NL\\ 
HE\,1302$-$1017  & 0.278 & -26.15&  -26.27&  -23.40 & A &
0.043&0.60-0.60-0.50&2400-1800-1800&U, 3&R\\
HE\,1434$-$1600  & 0.144 & -24.34 & -23.82 & -22.87 & A &
0.126&0.58-0.60-0.49&1200-1200-1200&E, 1&R\\
HE\,1442$-$1139  & 0.257 & -23.81 & -24.08 & -23.23 & A
&0.079&0.98-1.20-1.10&1200-1200-1200&D, 0&\\
HE\,1503+0228  & 0.135 & -23.02 & -23.43 & -22.78 & A
&0.05&0.60-0.68-0.72&1200-1200-1200&D, 0&\\ 
HE\,2258$-$5524  & 0.141 & -23.71 & -24.11 & -23.11 & B
&0.013&0.88-0.70-0.76&3000-1800-1800&U, 3&NL\\
HE\,2345$-$2906  & 0.223 & -23.09 & -22.97 & -22.84 & B
&0.019&0.86-0.83-0.67&3000-1500-1500&D, 0&\\
\hline
\end{tabular}
\label{sampletab}
\end{table*}


\section{The sample}
\label{sample}
Our quasar sample is selected from the flux-limited Hamburg-ESO Survey
(HES) \citep{wiso00}. The HES is designed to detect optically luminous
quasars and  Seyferts  1  in the  Southern   hemisphere, with  optical
magnitudes $13  <  B  < 17.5$   and   redshifts $ z   <  3.2$.   The
classification of the candidates as  quasars is made  spectroscopically
from  the   broad-band   features.  Out of this sample, we  selected   the
instrinsically luminous quasars,  with absolute magnitude $M_B < -23$  and   
redshift $ z< 0.35$ (this redshift limit is set so that the H${\alpha}$ line
stays in our useful spectral range).  From the 32 QSOs matching  these
criteria, we could observe 20.  No  spectral or morphological selection of 
the object was applied prior to the observations.  The only  relevant factor
in our choice of
an object  was  its  visibility at  the  time  of the  observing runs.
Table~\ref{sampletab} gives the  main  characteristics of our  sample.
Figure~\ref{ech} shows the location of all our objects in the absolute
magnitude vs. redshift  plane and compares them  with previous studies  that
all use  imaging   rather   than   spectroscopy  (e.g.  \citealt{floyd04};
\citealt{dunlop03}; \citealt{bahc97}). Our sample  concentrates on the 
brightest part of the quasar luminosity function and at rather low redshift.
Its characteristics compare well with those of \citet{floyd04} and
\citet{bahc97}.


\section{Observations}
\label{obs}
\subsection{Spectroscopy}

Our spectroscopic  observations  were obtained with the  ESO  Very Large
Telescope (VLT) and the FORS1  instrument in Multi Object Spectroscopy
(MOS) mode. The data were acquired during two observing runs, in April
and November 2000.  All objects were observed with 3 medium resolution
grisms  (G600B,  G600R  and G600I),  thus  covering   the full optical
spectral range 3500-9000\AA.  The seeing  varied between 0.5 and
2.2 arcsec  (see    Table~\ref{sampletab}), and the   pixel   scale  was
0.2 arcsec,  using   the standard  resolution  collimator.   The slitlet
(19  $\times$  1 arcsec) of the   MOS mask  closest to  the
optical axis  was centred   on  the quasar  itself
(on-axis).  Several  slitlets were used to  obtain the spectra of PSF
stars with the same spectral range as the quasar, which is crucial for
subsequent deconvolution of  the data using the spectroscopic  version
of the MCS algorithm (\citealt{mcs}; \citealt{courb00}, see Section
\ref{obsspec}). The remaining slitlets were  placed on galaxies in the
field, in order to look  for possible   companions,  groups  or clusters  (see
Section~\ref{interac}).  We show in   Fig.~\ref{slits} a typical MOS
mask configuration for  the 5 slitlets  closest to  the quasar.   As the
slits are parallel, the orientation of the mask relative to the major axis
of  the  host  galaxy (when visible  prior  to deconvolution) is
imposed by  the necessity to  observe simultaneously  at least one PSF
star   (in fact two  for the   majority of the  fields  observed) with
the same spectral coverage as for the quasar.

The  reduction of  the spectra was  carried out  with the standard IRAF
packages,  leading      to bias-subtracted,    flat-fielded, and
sky-subtracted spectra.   The wavelength  calibration was  done using
Ne-Ar-He  spectra  obtained  in  the same  slit  configuration  as the
observations. It has been performed on the two  dimensional spectra in order
to   correct  for  slit curvature   and   to  ease the  sky
subtraction. All  spectra were rebinned to a  pixel scale of 1 \AA\, in
the spectral  direction and 0.2 arcsec in the   spatial direction.  When
several exposures were available  in a given  grism, they were  combined
into  one deeper spectrum  using a cosmic-ray  rejection algorithm. 
Spectroscopic standard stars were observed during the two runs in order to
allow for an accurate flux calibration.

\subsection{Imaging}
\label{analima}
The imaging  part   of the sample  was   limited to acquisition  images
obtained during  the spectroscopic  runs, with 30s exposure time  through the
$R$  or $V$ filters.  The seeing
spanned  a broad  range, 0.5  to  1.5 arcsec.    The pixel scale   is
0.2 arcsec, as in  the   spectra.  In one  case  the  QSO   is saturated
(HE\,0441-2826).  The reduction of all  images was performed with the IRAF
package.
It consisted in  bias   subtraction, flat-fielding and   sky
subtraction.    Standard stars  were  also observed   in each run for
magnitude determination.

\begin{figure}
\centering
\includegraphics[width=8.5cm]{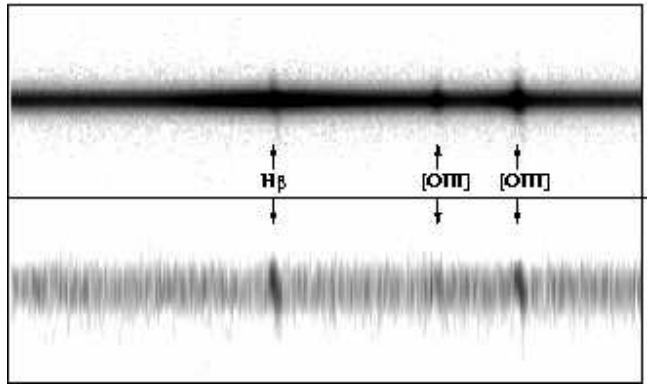}
\caption{Example of spectra decomposition using the MCS algorithm, for the 
$z=0.223$ quasar HE\,2345-2906. The    spectrum shown here  spans  the
wavelength interval  5800-6250 \AA. The top panel
displays the original data with  the VLT spectrum  of the quasar  plus
host galaxy. The  bottom panel  is the  result  of the  decomposition,
where only the  2D-spectrum of the host  is shown. The narrow emission
lines of the host can be traced down to the centre  of the galaxy. The
height of each panel is 19 arcsec.}
\label{exdec}
\end{figure}


\section{Deconvolution of the data}
\label{obsspec}

The  whole sample of 2D spectra  and images has been deconvolved using
the     MCS   algorithm \citep{mcs} and    its    extension to spectra
\citep{courb00}. In the MCS deconvolution algorithm, 
the final resolution is chosen  by the user, but  is constrained to be
compatible with the sampling of the  deconvolved data.  In this way, the
resolution  of the processed data
is improved  but not  infinitly high,  and the deconvolved data do  not show
any of
the undesirable  ``deconvolution artefacts'' often  found using  other
algorithms.   In addition, the  MCS method is particularly well suited
to the  quasar host studies because  of its ability  to separate point
sources (here the  quasars)  from diffuse  ones (the  host  galaxies).
Taking advantage of the availability of several stars to construct the
PSF, we could as well subsample  our deconvolved spectra and images, leading to
a
final sampling of 0.1 arcsec per pixel.

The deconvolution of     spectra with  MCS,  as  every   deconvolution
procedure, is highly    sensitive to the accuracy of    the PSF.  Poor
deconvolution results are always  associated with problems in the  PSF
construction.   Several   weaknesses  of  the    process  account  for
uncertainties sometimes found in the     deconvolution results  and
difficulties    in
separating both components.  (1) The stars are often  not bright enough in
the bluer part of  the G600B grism, in  comparison to the quasar. This
introduces extra noise at the  shortest wavelengths in the deconvolved
spectra. (2)  The orientation and angular separation   of the slits being
imposed by the geometry of the mask (see Fig.~\ref{slits}), it may happen
that parts of the PSF wings are lost on one side because the star could
not be centred on   the slit. (3) The spectral  types  of the stars  can
influence  the quality  of  the PSF,  smoother stellar  spectra  being
preferred to spectra with many  strong absorption lines. (4) Finally, the
slight variation of  the  PSF profile  accross  the field   is
minimized  by  taking  the  stars closest to  the quasar  for  PSF
construction or  by the  use of several  stars on   both sides  of the
quasar  to average the  profile variation. For  85 \% of the objects in
our sample,  the simultaneous observation of  several  stars allows to
construct reliable PSFs.  The main characteristics of the PSF stars are
summarized in Table~\ref{psfchar}.

Figure~\ref{exdec}  shows an  example of  deconvolution.   The reduced
spectrum   is shown   on the  top   panel  of  the  figure,  while the
2D-spectrum of the  host galaxy alone is shown   in the bottom  panel.
The  MCS method is  used  here essentially  in   order to efficiently separate
the nuclear and diffuse components of the data  rather than for sharpening
the data in the spatial  direction. Depending on the signal-to-noise ratio 
(S/N), this may result  in some smoothing of
the  extended  source  or, in  any  case,  in  no  significant gain  in
resolution. Figure~\ref{exdec}   nevertheless clearly shows that
no significant residual flux from  the quasar is left  in the spectrum of the
host galaxy.  For example, the spectrum of the host does not display any residual
from the broad quasar emission lines.

Once the data are deconvolved, we extract 1D spectra by integrating
fluxes over the spatial direction,  in the quasar spectrum
and in the host  spectrum. The final atlas of  quasar and host spectra
is shown in Fig.~\ref{spectro1}.

\subsection{Estimate of the uncertainties on quasar/host separation}
\label{error}
While the outcome of the deconvolution process shows no traces of contamination
from the quasar lignt in the host spectrum (as illustrated by Fig.
\ref{exdec}), it nevertheless appears useful to estimate the maximum effects an
imperfect separation between host galaxy and nuclear spectra may have on the
various results presented hereafter.

For this purpose, we adopt a very conservative approach, deciding not to trust
the quasar/host separation given by the deconvolution process, and arbitrarily
changing the magnitude of the quasar spectrum to be subtracted from the total
(quasar+host) spectrum, until the shape of the host spectrum shows clear signs
of (either positive or negative) contamination.  We thus determine the limits
for over- or undersubtraction of the nuclear spectrum.

In practice, we concentrate on the spectral region containing the H$\beta$ and
[{O}{III}] lines, as these spectral features are of prime importance for the
derivation of the galactic properties.

The limit of oversubtraction of the quasar spectrum is set, for each object separately, when one of the
following conditions is met: (1) the 2D host  spectrum presents a depression
in the centre, (2) a broad H$\beta$ component appears in absorption on the
integrated host spectrum, (3) the continuum flux of the host reaches zero. The
limit of undersubtraction of the nuclear component is set when: (1) the
spatial profile of the host spectrum clearly shows a central point source, or
(2) a broad H$\beta$ or H$\gamma$ emission appears in the integrated host
spectrum. 

Such error bars, set when visual inspection clearly shows a problem, should be
considered as roughly amounting to 2 $\sigma$ (which means that we estimate to
be able to detect such a wrong subtraction in 95$\%$ of the cases). The
percentage of nuclear flux that has to be added to reach the limit of
oversubtraction or removed to detect an undersubtraction is given, for each
object, in Table~\ref{psfchar}.

Each galactic property derived hereafter has been evaluated from the 
deconvolved spectra {\it and} from the associated upper and lower limit
spectra derived as explained above, thus allowing a quantitative estimate of
the very conservative uncertainties related to the deconvolution (or, rather, to the quasar/host
separation). These upper and lower limit spectra are represented as 
dotted curves on the atlas of Fig.~\ref{spectro1}

\begin{table}
\caption[]{Quantitative estimates of uncertainties in the deconvolution
process and characteristics of the PSF stars used for deconvolution. We give
here the percentage of the quasar flux that can be added to (+\%) or subtracted
from (-\%) the galactic spectra before a clear nuclear signature appears (see
text and Fig. \ref{spectro1}), the angular separation $\theta$ between the quasar and the PSF star(s) and indicative PSF star(s) magnitude(s) in the B band, to be compared to the
quasar apparent magnitude.}
\begin{tabular}{lccccc}
\hline
Object name & +\%& -\% & $\theta$(min:sec)    & B(PSF) &  B(QSO)      \\ 
\hline
HE\,0132$-$4313  & 1 & 1 & 1:09/2:16& 17.8/16.2 &16.4\\
HE\,0203$-$4221  & 0.5 & 0.2 & 1:02/2:05 & 14.6/13.8&16.0 \\
HE\,0208$-$5318  & 1 & 1 & 1:11/1:35 & 16./14.8&16.1\\
HE\,0306$-$3301  & 1 & 1 & 1:34/2:02 &16.5/15.4&15.3\\
HE\,0354$-$5500  & 2 & 0.5 & 2:22/1:41& 16.1/17.&15.6\\
HE\,0441$-$2826  & 1 & 0.6 &  1:53/2:14 & 18./14. &15.2\\
HE\,0450$-$2958  & 0.5 & 0.2 & 1:59 & 15.3 &14.8\\
HE\,0530$-$3755  & 0.8 & 0.5 & 1:49/2:03 & 14.8/16.6&16.1\\
HE\,0914$-$0031  & 2 & 1 & 1:04 & 16.&16.9\\
HE\,0956$-$0720  & 1 & 1 & 1:12 1:43 &15.2 16.9 &15.2\\
HE\,1009$-$0702  & 1.5 & 1 & 1:29/1:22& 15.9/16.3&15.9\\
HE\,1015$-$1618  & 1 & 1 & 1:46/0:55 &18.5/18.7 &15.2\\
HE\,1029$-$1401  & 1 & 0.5 & 1:23 &15.6 &13.5\\
HE\,1228+0131  & 0.8 &1  & 2:06 & 14.7&14.4\\
HE\,1302$-$1017  & 0.5 & 0.2& 1:01/1:08&16./15.7 & 14.5\\
HE\,1434$-$1600  & 3 & 2 & 1:39 & 17.8&14.8 \\
HE\,1442$-$1139  & 5 & 2 & 0:36/1:46 & 16.5/16.9 &15.5\\
HE\,1503+0228  & 3 & 2 & 0:51 & 16.1 &15.8\\
HE\,2258$-$5524  & 2 & 1 & 2:57 &13.6  &14.7\\
HE\,2345$-$2906  & 4 & 2 &0:39/1:08 & 17.2/16.7 &16.1\\
\hline
\end{tabular}
\label{psfchar}
\end{table}

\begin{figure}
\centering
\includegraphics[width=0.45\textwidth]{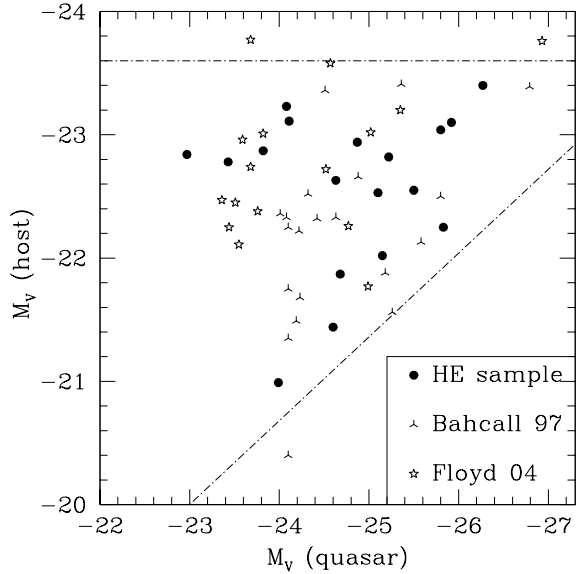}
\caption{Absolute M$_V$ magnitude of the host galaxies versus absolute 
M$_V$  magnitude of the quasar. For  comparison, we also  plot the two
samples of  quasars covering the  same luminosity
range,   from \citet{floyd04}   and  \citet{bahc97}, converted  to our
cosmology. The dotted lines show the  envelope discussed in the
text.}
\label{mag}
\end{figure}

\begin{figure*}
\centering
\includegraphics[width=17.5cm]{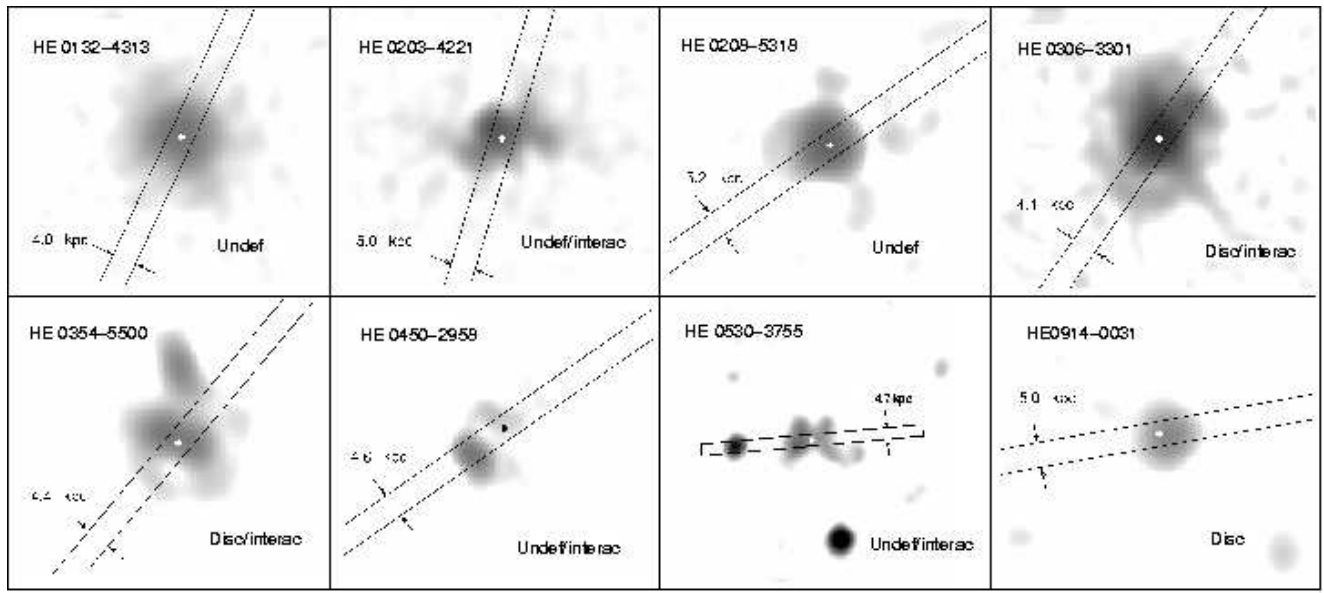}
\caption{Deconvolved VLT images of the host galaxies, in the $R$-band (H
\,0132-4313, HE\,0203-4221 
and HE\,0914-0031)  or in the $V$-band  (HE\,0208-5318, HE\,0306-3301,
HE\,0354-5500, HE\,0450-2958 and HE\,0530-3755).  The quasar light has
been removed from  these  images, which  are displayed in a logarithmic intensity
scale.
Crosses indicate the location of the  quasar.  The 1 arcsec-wide slits
used in spectroscopy are overplotted, and  their physical scale on the
sky is indicated.  Global   morphologies  are   not estimated from these
images but are derived following a   2D
isophotal fitting of  near-IR and/or HST/ACS data  from other
studies (see  Table~\ref{morph} in Section~\ref{anima}).  In each case
we
indicate possible signs of interactions,  as  can be  spotted from images and
from the  radial velocity curves  described in  Section~\ref{rc}.  All
images are 12 arcsec-wide, except HE\,0530-3755 which is 24 arcsec on
a side. The intensity cuts are chosen to optimize the contrast and are
not identical in all the images. North is up and East to the left.}
\label{img1}
\end{figure*}
\begin{figure*}
\centering
\includegraphics[width=17.5cm]{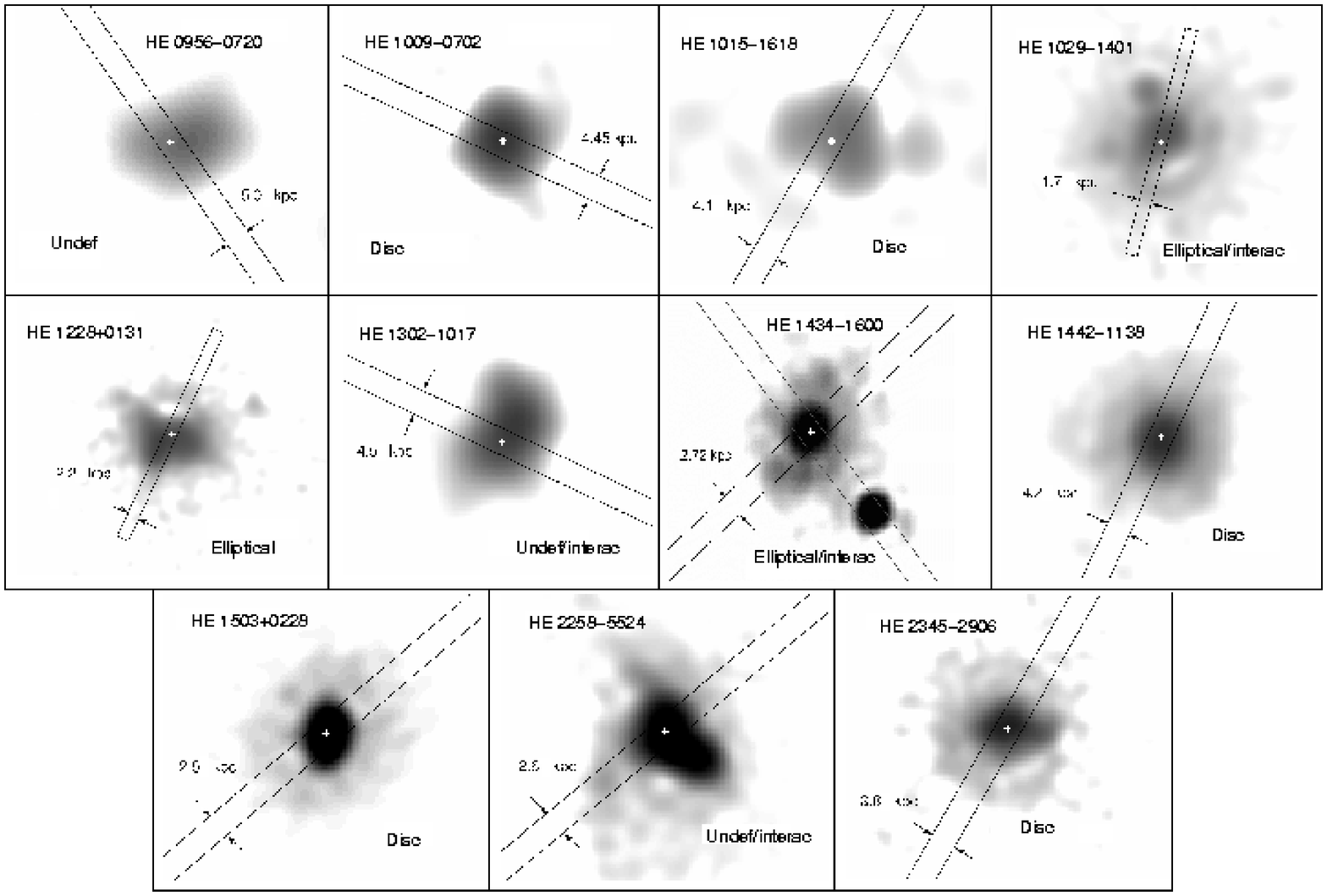}
\contcaption{All images are in the $R$ filter. 
All images  are 12 arcsec-wide, except HE\,1029-1401 and HE\,1228+0131
that    are   24 arcsec-wide.   North   is   up    and   East  to  the
left. HE\,1434-1600 was re-observed  with a different slit orientation
in order to get the spectrum of a companion galaxy at the same redshift
\citep{let04}.}
\end{figure*}

\section{Image analysis}

\label{anima}

A short 30-seconds $R$ or $V$ acquisition image was taken before a long
spectroscopic   exposure.   Even  though  these  images    are of  low
$S/N$ and  were not meant  to be  used for scientific purposes, they
are generally good  enough to determine  some basic properties of
the quasars and of their hosts. We deconvolved all the available images
and used  them in order to determine  the magnitudes  of the quasar and of the 
host, and to unveil any  obvious sign of interaction
such as  tails or  companions.  Recent {\it HST}/ACS   imaging of some of  our
objects actually confirms  that the shapes  given by the deconvolution
of  these    images   are   accurate    enough   for     that  purpose.
Two dimensional fitting
on these low $S/N$ data remains unfortunately hazardeous.

Figure~\ref{img1} shows the   deconvolved images of the  quasar  hosts
after removal of  the  quasar itself. HE\,0441-2826 is missing because the
quasar light reached the saturation of the detector.  The host images  are
used to measure  the
apparent magnitude of the host  galaxies from aperture photometry.  We
then compute absolute magnitudes (Eq.
\ref{amag})  taking into account the  K-correction between the $R$ (or
$V$)      filter  and   the      restframe  $V$-band  ($RV=R_{{\rm
obs}}-V_{{\rm rf}})$.  This is  done without any assumption on the
nature of the hosts since we compute their integrated fluxes over the
appropriate wavelength bands, directly from our  VLT FORS1 spectra. The
computations can be summarized as follows:

\begin{eqnarray}
M_V&=&m_{R}-5\log{d}+5-A-2.5\log{(1+z)}-RV\nonumber\\
m_{R}&=&m_{R_0} - 2.5 \log{(F_R)}-a\tau
\label{amag}
\end{eqnarray}

where $d$ is the distance  to the object in  parsecs, according to the adopted
cosmology,  $A$ the Galactic  extinction (from  \citealt{schlegel98}),
$z$ is  the redshift, $F_R$ is the  observed $R$-band flux, $a$ is the
airmass, $\tau$ the atmospheric optical depth, and    $m_{R_0}$   is the zero 
point   determined   from the
observation of    standard    stars.  

The quality  of the deconvolutions  can be  checked by inspecting  the
``residual maps'', as  defined  in  \citet{mcs}.  These maps  are  the
absolute differences between  the  data and the deconvolved  images, reconvolved
with the PSF, and in units of the photon noise per pixel. A successful
deconvolution must  show a flat  residual map,  equal to 1 on average.
Based on these residual maps and on numerical simulations, we estimate
that the quasar-host decomposition  of  our VLT acquisition images  is
accurate to 0.01  mag.  However, given that the angular scale of the PSF is 
 similar to that of the bulge component, part of the host flux might be left in
the quasar light. This may introduce a systematic underestimate of the flux in
the host, especially in cases where it has a compact bulge, which may be
difficult to separate from the point source.
 The magnitudes given
in Table~\ref{sampletab} are not corrected  for such systematic errors. For
this  reason, we  avoid  any interpretation  relying on host magnitudes
alone.  However, Fig.~\ref{mag} shows that our absolute magnitudes for
hosts and nuclei   are   consistent with other  studies   in  the same
luminosity  range (\citealt{bahc97}; \citealt{floyd04}) and are useful
as a guide   to  compare our sample  with  others.   While no  obvious
correlation  is apparent between    the   host and quasar    intrinsic
luminosity, the points are all  located inside  an envelope, shown by the
dotted lines in Fig.~\ref{mag}. The  existence of this envelope suggests
that, for a given host magnitude, there is a maximum nuclear luminosity.
This was already observed in earlier studies, e.g.  by \citet{floyd04} or
\citet{jahnke04}, and is explained by the black hole mass - bulge mass
relation   \citep{mclure02}: for a  given bulge  mass (and hence bulge
luminosity), the black hole  mass is roughly  fixed and the luminosity
of the nucleus is more or less limited to the  Eddington   luminosity (see
Section~\ref{arate}).

Several objects in the  sample
had  been previously observed   by various   teams. \citet{kuhl02} and
\citet{jahnke04} observed some of the  objects with the ESO 3.5m
New Technology Telescope in the near-IR,  and  decomposed the data
into point sources plus 2D analytical  profiles  for  the  hosts (exponential 
disc  or de Vaucouleurs).  Two objects, HE\,1302-1017 and HE\,1029-1401,
were  also observed  with {\it HST} by  \citet{bahc97} and classified  as
elliptical,
both with small close companions.  HE\,0956-0720  was classified as  a
disc by  \citet{percival}.    Our recent {\it HST}/ACS  images  of  several
objects in the sample has also  allowed to detect obvious spiral arms
in four  hosts,  namely HE\,0306-3301, HE\,0354-5500,  HE1503+0228 and
HE\,2345-2906 (Letawe et al., in prep).  Table~\ref{morph} gives a summary of
the morphological
characterizations available, while the adopted morphology is given in Table
\ref{sampletab}.

\begin{table}
\centering 
\caption[]{Summary of the hosts morphological types and stellar populations. 
In each column an ``E'' indicates an elliptical galaxy,  ``D'' stands for  disc
and ``U'' for undefined. (1-) Object name, (2-Lit) morphology according to
\citet{bahc97} and \citet{percival}, (3-NIR) near-IR morphology
(\citealt{kuhl02}; \citealt{jahnke04}),
(4-HST)   {\it HST}/ACS  F606W morphology  from  our  own ongoing program,
(5-Dyn) 
morphology  as derived from  the velocity curve  of the host (``no D''
means  that a rotating  disc is  ruled  out), (6-) stellar population from
Section~\ref{stellpop}:
``Y'' for young spiral-like population,  ``O'' for old elliptical-like
population, and ``I'' for  intermediate age  population. HE\,0450-2958
is not classified because no host is detected. }
\begin{tabular}{lcccc|c}
\hline
Object name & Lit. & NIR&HST & Dyn.&Stell. pop.  \\
\hline
HE\,0132$-$4313  &-&-&-&D&I\\
HE\,0203$-$4221  &-&-&-&U&Y\\
HE\,0208$-$5318  &-&-&-&-&Y\\
HE\,0306$-$3301  &-&E?&D&U&Y\\
HE\,0354$-$5500  &-&-&D&U&Y\\
HE\,0441$-$2826  &-&E&-&-&Y\\
HE\,0450$-$2958  &-&-&-&-&-\\
HE\,0530$-$3755  &-&D?&-&-&-\\
HE\,0914$-$0031  &-&D&-&D&Y\\
HE\,0956$-$0720  &D&E&-&-&-\\
HE\,1009$-$0702  &-&D&-&D&Y\\
HE\,1015$-$1618  &-&D&-&-&-\\
HE\,1029$-$1401  &E&E&-&no D&O\\
HE\,1228+0131    &-&E&-&-&-\\ 
HE\,1302$-$1017  &E?&D?&-&U&Y\\
HE\,1434$-$1600  &-&E&E&no D&O\\
HE\,1442$-$1139  &-&D&-&-&I\\
HE\,1503+0228  &-&D&D&D&Y\\ 
HE\,2258$-$5524  &-&-&-&U&I\\
HE\,2345$-$2906  &-&-&D&D&Y\\
\hline
\end{tabular}
\label{morph}
\end{table}

\section{Overview of the spectral analysis}
\label{overv}

Because of the  extreme diversity in the  gas and stellar contents  of
quasar host galaxies, the amount of scientific  information that can be
extracted  from our images  and  spectra is  variable  from object  to
object.   Most of the  analyses proposed  in the present work
are performed  only on subsamples presenting  similar characteristics,
such as the presence of emission  or absorption lines, or the presence
of a measurable velocity field.  We summarize in Table~\ref{overvtab}
which types of measurements were possible for each object.

In  spectroscopy,  three  objects  have poor  quasar-host  separation,
either  because  the quasar  reaches  the   non-linear regime  of  the
detector (1 object, HE\,1228+0131), or because of poor PSF determination 
(2 objects, HE\,0956-0720 and HE\,1015-1618).
However, even with such  defects, we would be  able to detect galactic
emission  lines if they were sufficiently  prominent and extended. For
the    three object, we     do  not detect   any  underlying  emission
lines. These  hosts are therefore  very compact or  gas-poor galaxies.

Among the 17  remaining objects, 16  display  gas emission lines.  For
those, we determine  the ionisation source  of the gas. We classify  9
galaxies as  ``{H}{II}  galaxies'', i.e.   with  gas mainly ionized  by stars
(Section~\ref{ion}),  and estimate the  star formation rates for these
hosts as well as their mean metallicities. The methods we use in order
to determine  these parameters rely  on the gaseous properties of
the {H}{II} regions only  (Sections~\ref{ssfr} and
\ref{gmeta}).   

Twelve host  galaxies have spatially  extended emission lines, that we
use to measure their velocity fields as described in Section~\ref{rc}.
Among them, 5 are symmetrical about the central quasar and are good candidates
for  mass  modelling through  the rotation curves
(Section~\ref{dyn}). As a by-product, those five regular galaxies give
the opportunity to compare the redshift of the object as deduced from
the quasar emission lines and as obtained  using the emission lines of
the host (Section~\ref{redsh}).

\begin{table}
\centering 
\caption[]{Summary of the main spectral characteristics for the 20 host
galaxies.    (1-) Object name, (2-E)   presence of emission lines, (3-A)
 detection of absorption lines, (4 to 7)  type of analysis performed on the
spectrum: gas metallicity (met), star formation rate (sfr), extraction
of a  radial  velocity curve  (rvc), and mass   modelling (mass).  The
ionization source   of  the gas   (AGN  or young stars)  could be
detemined for all objects with emission lines, while Lick indices could
be measured in all objects with absorption lines.}
\begin{tabular}{lcccccc}
\hline
Object name & E & A &met&sfr&rvc & mass \\
\hline
HE\,0132$-$4313  & x&x&-&-&x&x\\
HE\,0203$-$4221  &x&x&-&-&x&-\\
HE\,0208$-$5318  & x&x&x&x&-&-\\
HE\,0306$-$3301  & x&x&x&x&x&-\\
HE\,0354$-$5500  & x&x&-&-&x&-\\
HE\,0441$-$2826  & x&x&x&x&-&-\\
HE\,0450$-$2958  & x&-&-&-&-&-\\
HE\,0530$-$3755  &x&-&-&-&-&-\\
HE\,0914$-$0031  & x&x&x&x&x&x\\
HE\,0956$-$0720  & -&-&-&-&-&-\\
HE\,1009$-$0702  &x&x&x&x&x&x\\
HE\,1015$-$1618  &-&-&-&-&-&-\\
HE\,1029$-$1401  & x&x&-&-&x&-\\
HE\,1228$+$0131  & -&-&-&-&-&-\\ 
HE\,1302$-$1017  &x&x&x&x&x&-\\
HE\,1434$-$1600  & x&x&-&-&x&-\\
HE\,1442$-$1139  &-&x&-&-&-&-\\
HE\,1503$+$0228  & x&x&x&x&x&x\\ 
HE\,2258$-$5524  &x&x&-&-&x&-\\
HE\,2345$-$2906  &x&x&x&x&x&x\\
\hline
\end{tabular}
\label{overvtab}
\end{table}

Fifteen galaxies out of the 17 that can be analysed exhibit measurable
absorption lines that we use to infer the stellar content of the hosts
(Section~\ref{stellpop}). We correlate  the Lick indices measured on
the absorption lines with the  Nucleus-to-Host ratios (N/H) determined
in our  VLT images, and  find  that the  absorption lines are detected
only in galaxies with N/H $\la$ 15.  This converts to a limit of about
N/H $\leq$ 45 in the  continuum of the  spectra, because of the larger
impact of  slit losses on  the host than on  the quasar.
 We are therefore not able to measure  the stellar population of
the galaxies   with the  highest  N/H, including   the peculiar object
HE\,0450-2958 for which no continuum is detected at all \citep{mag05}.

Finally, we analyse in Section~\ref{quasar}  the quasar spectra themselves
and measure  some  of   their  main physical   properties  such   as  
luminosity,  mass of the  central black hole and accretion
rate.

\section{Stellar and gas content}
\label{anispec}

We can use our host spectra  in order to  carry out a comparative study of  the
 gaseous and stellar contents in
active and quiescent galaxies.
 Prior to the analysis, let us note that for several objects,
the  galactic H$\alpha$ emission line  is not perfectly separated from
the  quasar emission  because of the combination of (1) the overcrowding of
atmospheric lines in this spectral region for objects with the highest
redshifts and, with a smaller impact, (2) the  very high  nucleus-to-host flux
ratio  in  this wavelength range (N/H up to  180).   As a  consequence, the
determination of the amount  of reddening in  the host galaxy, as well
as other measurements involving H$\alpha$ fluxes and Equivalent Widths
(EWs),  are not accurate  for  all objects in   the sample.  For  this
reason we  present  all our results without  any   correction for dust
extinction. This has little  influence on most of  the characteristics
of our host spectra, as we carefully select the estimators of the various
galactic properties analysed here to be as independent on extinction as
possible.

\subsection{Stellar content}
\label{stellpop}

\begin{figure}
\centering
\includegraphics[width=8.5cm]{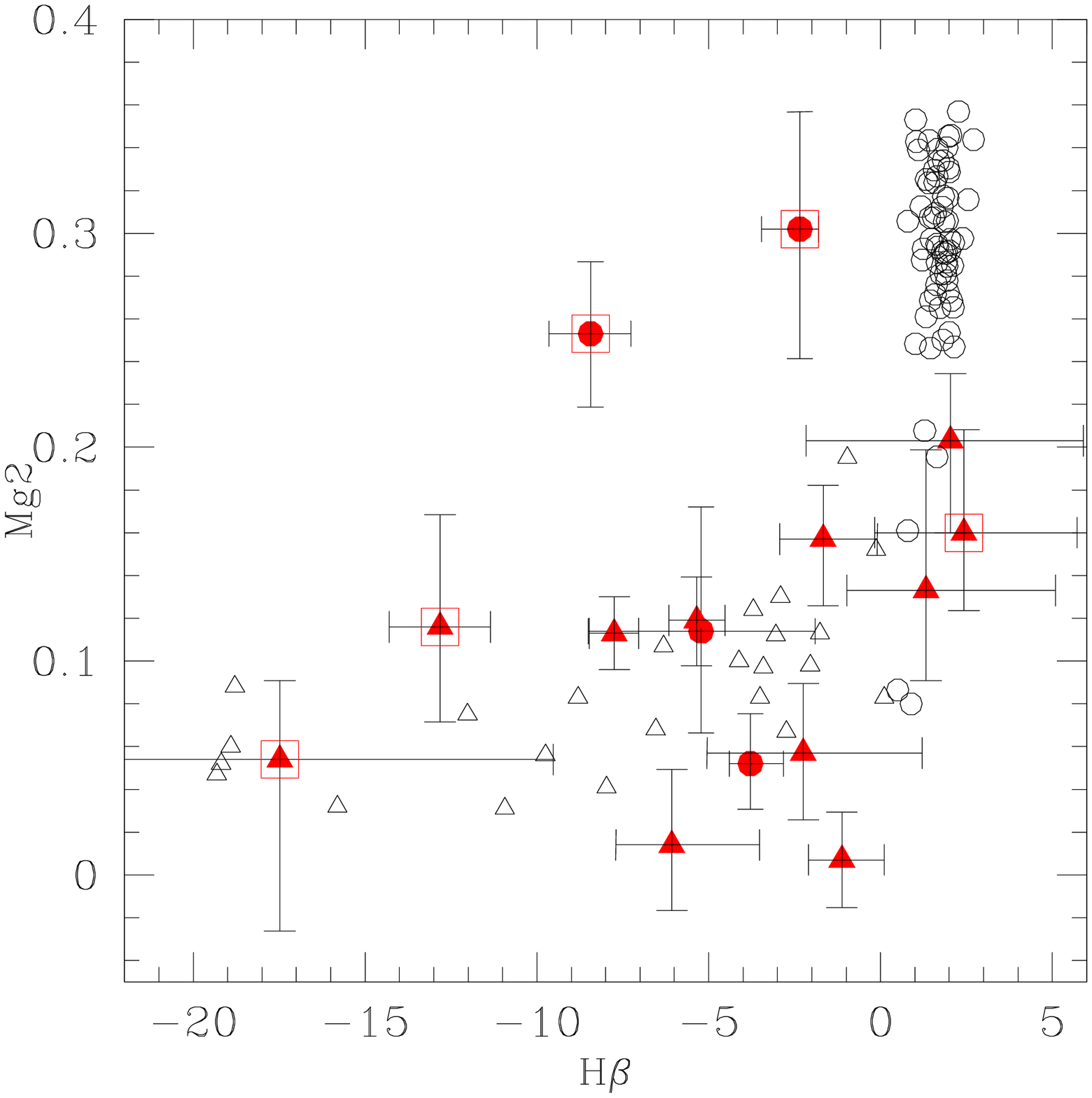}
\includegraphics[width=8.5cm]{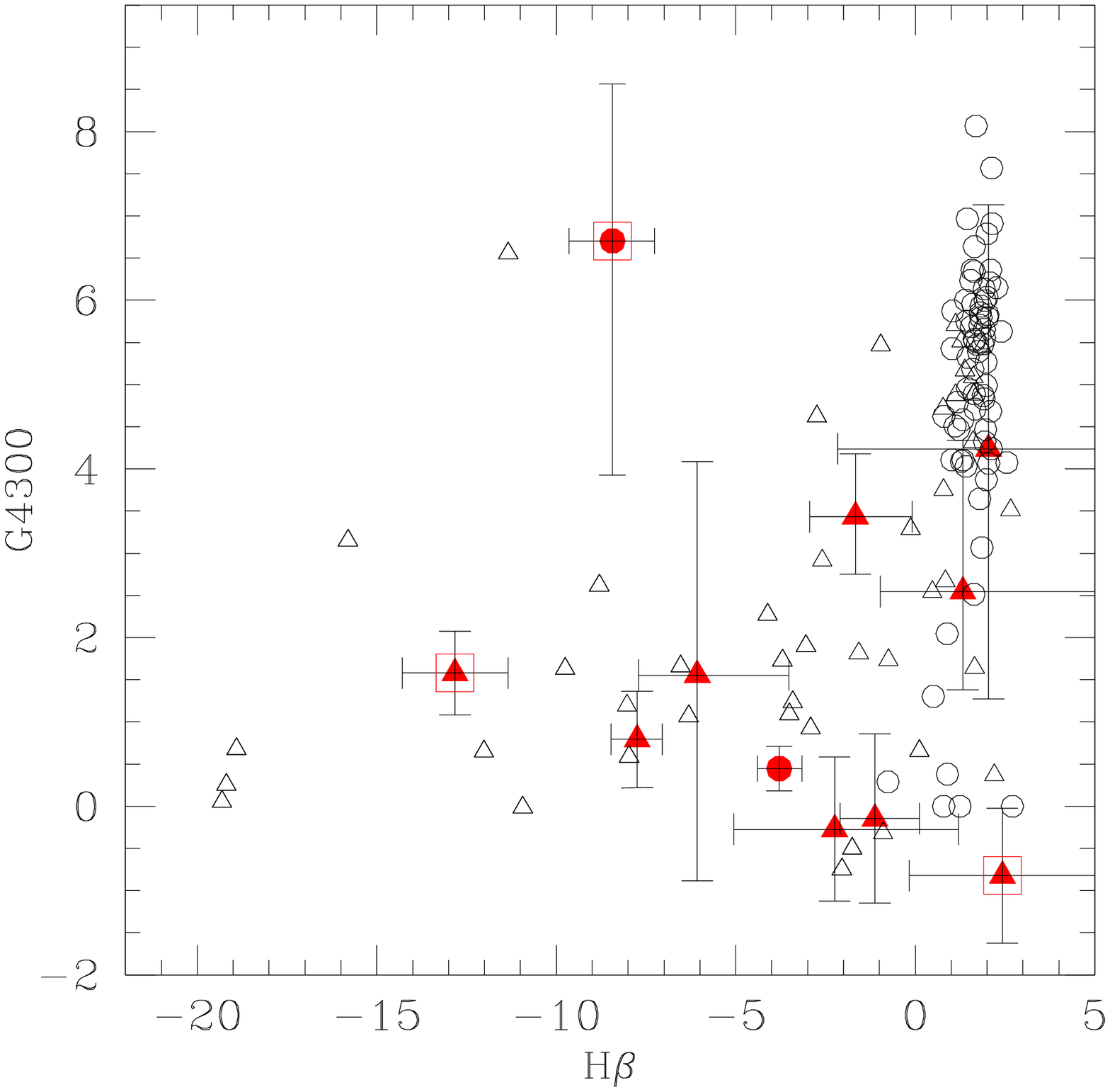}
\caption{Comparison of the Lick indices measured in our quasar hosts (filled
symbols) 
with two samples of  non-active  galaxies.  Open squares identify  the
galaxies  where the gas  is found to  be ionized by the central quasar
(Section~\ref{ion}).  The  indices for the  samples of Kennicutt (K92)
and Trager (T98) are plotted as open  symbols.  For all three samples,
circles indicate   ellipticals galaxies and  triangles  indicate other
types  of  galaxies. The error bars are the combination of the
uncertainties on the index measurements given the data dispersion and the
uncertainties from deconvolution (see Section~\ref{error}). Our quasar  hosts  compare well with the
late-type galaxy  sample, matching the  subsample of Sc  or later emission line
galaxies.}
\label{indexmg}
\end{figure}

 The main absorption features are detected in  the continuum of 15 out
of our 20 galaxies, allowing us to  measure several of the Lick indices
(\citealt{worthey}) and to compare them with the  ones obtained for two
samples  of nearby  quiescent  galaxies: Kennicutt (1992ab; hereafter K92) 
 for late-type
galaxies and \citet{trager98} for early-type galaxies.  The
\reduceme\,\footnote{http://www.ucm.es/info/Astrof/software/reduceme/\\
reduceme.html}   package  was  used   to  carry  out  the  measurements
\citep{cardiel}.    We  use   the published  indices  of
Trager's  sample, while  we make  our  own measurements on the Kennicutt atlas
of galactic spectra.  As only few absorption
features    are  detected in  our spectra,   estimates  of the stellar
velocity dispersions are not reliable.  We therefore do no correct our
measurements for the  velocity dispersion.  However, it should be pointed out
that such a correction would remain within the error bars, and would
not affect the global characterization of the sample.

The most prominent absorption lines  detected in  our spectra are  the
{Ca}{II} H  and K  lines (3969 \AA\,  and  3933 \AA), the  MgII  triplet
around 5174 \AA\, and the G-band of CH at 4303 \AA.  The latter is not
detected in all galaxies. Hydrogen absorption lines, when present, are
always blended with emission.  In  Fig.~\ref{indexmg} the Lick
indices Mg2 and G4300 are plotted versus the  H$\beta$  index (see
\citealt{worthey} for a detailed  definition of the indices).  A negative index
stands for emission. CaHK is
not presented because  it is not available in  T98.    
 The Mg2 index windows are  modified
with    respect  to the  original    definition,   in  order  to avoid
contamination from the  [{O}{III}]4959 \AA\, emission  line: we change
the    original     continuum     windows   [4895.125$-$4957.625]  and
[5301.125$-$5366.125]    \AA\,   to     [4895.125$-$4947.625]      and
[5301.125$-$5356.125] \AA.
 
In Fig.~\ref{indexmg}, the   elliptical galaxies from  Trager's sample
(open circles) all  show large Mg2 and  G4300 indices, indicative of the
high  metallicities associated with old stellar population.   They  also     
show H$\beta$   in    absorption,
representative of gas-poor galaxies. We restricted the Kennicutt
sample (open triangles)  to spiral galaxies  showing emission  lines, as
almost all  galaxies in our  sample  show prominent  emission
lines. The K92 spiral galaxies are located at lower metallicities than
the  T98 ellipticals  and  generally display  negative H$\beta$, accounting  
for globally young stellar  populations  and an important ionized  gas
content.

Our  QSO hosts  (filled symbols)  follow  well the distribution of
late-type galaxies, compatible    with   a subsample  of normal     Sc
population or later because  of their similar H$\beta$ emission lines.
The   hosts   of  HE\,0132-4313,  HE\,1442-1139,   and  HE\,2258-5524,
presenting faint or barely detectable H$\beta$  emission lines, lie at
the limit between  the K92 and T98 samples.   They likely host stellar
populations characteristic of  galaxy types between   E  and Sa.   Among the
galaxies   identified  as ellipticals   from  their morphology (filled
circles;   \citealt{bahc97}; \citealt{kuhl02}),   only   two  hosts,
HE\,1029-1401 and  HE\,1434-1600,  stand in  a range  of metallicities
corresponding  to ellipticals,  but out  of  the  distribution of the T98
comparison sample because of their important H$\beta$ emission.

The ISM of some of our QSO hosts is found to be
ionized   by  the central AGN   itself   (see Section~\ref{ion}).  We
represent these objects with an  open square around the filled symbol in Fig.
\ref{indexmg},
as their  location in the  diagrams may be  biased by  a    significant
contribution  of   H$\beta$   emission induced  by the   nuclear
ionization. It  is the  case for the   old  population hosts,  and
probably for   the two objects  with  extreme H$\beta$ emission which,
without  this contribution, would rather  be  located in the bulk of
the distribution.

Only two  objects  in our sample have   stellar populations characteristic of
genuine
elliptical galaxies (see  Table~\ref{morph}), and  3 have populations
intermediate
  between E  and Sa.    The other  10   objects display  stellar
populations    typical of  young   discs rather  than bulge-dominated
galaxies.

In the following, we classify  the objects
according  to their host morphology.   When  the classification of  an
object  is secure, we  refer to it as Elliptical  or  Disc.  Among the
unclassified hosts,  we indicate the  galaxies with young stellar
population  (YP) with a distinctive  symbol to distinguish them from the
objects for which
we have no secure information at all. In all  but one object for which we have
  both morphological and   stellar population characterization, the
young stellar population of the host is associated to disc morphology.

\subsection{Ionization source}
\label{ion}

\begin{figure}
\centering
\includegraphics[width=8.5cm]{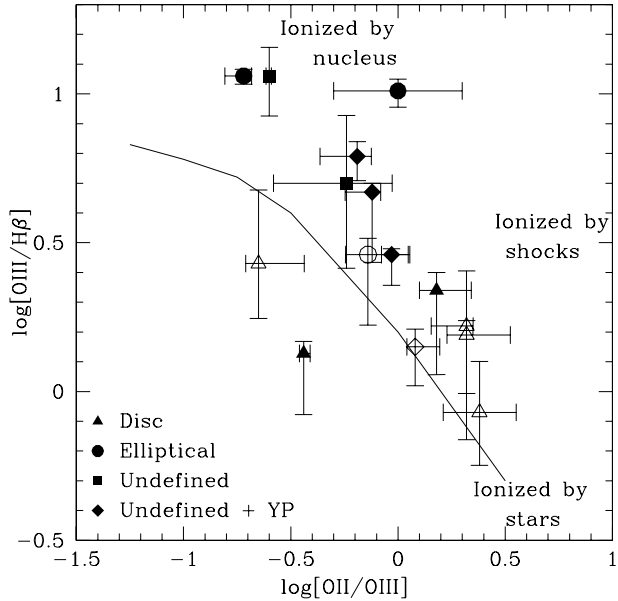}
\caption{Example of diagnostic diagram  constructed following \citet{baldwin},
for our sample of quasar hosts. Filled  symbols are for hosts that
display signs  of interactions, while  open symbols indicate isolated
hosts. Error bars are a quadratic combination of errors on the measurement
and on the deconvolution process (see Section~\ref{error}).  The  solid line 
follows the  locus of  typical {H}{II}
regions.  Most of the interacting hosts are in  the upper part of the diagram,
where the  central  AGN is   responsible  for the gas   ionization.  A
correction  for reddening  would shift  the points horizontally towards
the
right, i.e,  further  away  from the  HII region
locus.}
\label{diagdiagfig}
\end{figure}

\begin{figure}
\centering
\includegraphics[width=8.5cm]{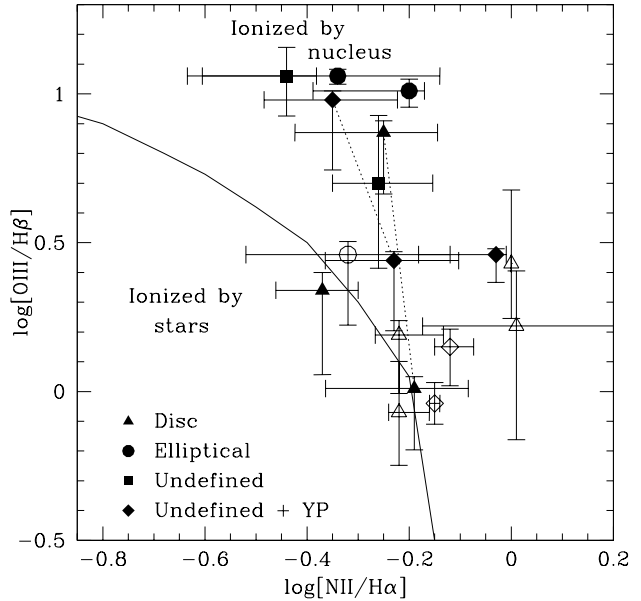}
\caption{Diagnostic diagram similar to Fig.~\ref{diagdiagfig} but for the
[N{II}/H$\alpha$] ratio. In this case, the  solid  line is the  separation
between
ionization of  the  gas by stars (lower left) and by the  central  quasar
(upper right). Filled symbols are for hosts which display signs  of
interactions, while open
symbols  indicate isolated hosts. Some  of the galaxies  in our sample
are not shown because  of unreliable  spectral separation between  NII
and H$\alpha$.  The symbols connected by dotted lines correspond to the same
hosts, in which different regions display different ionization levels (see
Table~\ref{diagdiagtab}).}
\label{ion_merg}
\end{figure}

Diagrams comparing the intensity  ratios between different  emission lines of
the host
galaxies allow to  infer the source  of  the gas ionization  in the
ISM.  Such diagrams (\citealt{baldwin}; \citealt{veilleux}) are often used
to discriminate  between  Seyferts and star-forming galaxies,  when
only     a global   quasar-plus-host   spectrum   of   the   object is
available. Here we construct these diagrams for the host spectra only,
after  the nuclear component  has been removed. We therefore determine
the main source of ionization of the gas accross the galaxy, excluding the
AGN  itself.  As far as possible, we choose  flux ratios of lines
with similar wavelengths, in order to minimize the  impact of  dust
extinction.    Our     diagrams    are    constructed  using     the
[{O}{III}]/{H}{$\beta$}, [{N}{II}]/{H}{$\alpha$} and
[{S}{II}]/{H}{$\alpha$} line  ratios, the latter two being sometimes less
accurate because   of the  difficulty to  separate    the quasar  and host
components  {H}{$\alpha$} region when it is overcrowded by atmospheric lines. 
Although it is more
sensitive to reddening,   the [{O}{II}]/[{O}{III}] ratio  is also used
because it can be easily measured in the majority of our quasar hosts.
Such diagrams  are presented in Figs.~\ref{diagdiagfig} and \ref{ion_merg},
and the measured values are listed in Table~\ref{diagdiagtab}.

Although reddening by dust cannot be corrected  for exactly, its impact
on  the position of the  points in Fig.~\ref{diagdiagfig} can  be estimated. 
For that purpose, we selected the subsample of  objects that have accurate
quasar-host separation in  the H$\alpha$ region  and we  estimated their
reddening from the Balmer decrement.  Using a Whitford reddening curve
(\citealt{mm}), we then corrected the [{O}{II}]/[{O}{III}] ratio.  These
reddening corrections range from 0.05 to 0.44.  Applying them would result in
shifting the points to the right by this amount in Fig.~\ref{diagdiagfig}.

We find two  major sources of  ionization: the hard radiation field of
the active nucleus itself (power-law ionization source) and ionization
by the  softer  photons produced by young   stars  in {H}{II} regions.
However, when a reddening correction is applied, most of the points in
Fig.~\ref{diagdiagfig} move  further away  from
the {H}{II} locus.  Does this mean that none of  our hosts is ionized by
pure stellar  emission?  Indeed, most   of the  low ionization  objects (low
[OIII]/H$\beta$ ratio) move, when   reddening is taken    into
account,  into  the    region which,   according   to   \citet{baldwin},
corresponds  to shock-heated gas.  The  objects further  away from the
HII region locus  after reddening correction (i.e. those with  highest
reddening) are  also galaxies undergoing a minor merger, in which a small
companion is being captured by the host.  Those  small encounters can possibly 
create  shocks, which produce an additional source of ionization, above the
young stars themselves.

 Table~\ref{diagdiagtab}
gives the intensity ratios  integrated over the whole galaxy  (without
nucleus) and the  corresponding classification of the main ionization source, either
hot stars (H), the AGN (A) or shocks (S).  The availability of 2D spectra
allows, for objects with the largest spatial extension and sufficient S/N in
the emission lines, to measure the intensity ratios in different parts of the
galaxy.  The three objects in the second part of Table~\ref{diagdiagtab} are
those in which the diagnostic varies significantly with distance from the
centre.  It is striking that, for all three objects, while the centre region
shows predominantly {H}{II} regions ratios (or intermediate ones), the
ionization of the outer parts is clearly dominated by the AGN radiation field. 
While it may not necessarily mean that the AGN ionization power is larger at
large distances, it suggests that the range of AGN ionization is larger than
the range of star formation (and hence of {H}{II} regions).  The latter
appears more concentrated towards the centre, while the quasar radiation field
reaches larger distances.  Note that these three hosts are involved in
gravitational interactions (see Section 8).

All  but one of the elliptical host galaxies harbour  gas and tend  to present
high ionization levels, i.e., have ISM ionized  by the quasar.  Even though
our sample contains only 4 elliptical galaxies, there might be a
connection between  this tendency and the  fact  that elliptical hosts
exhibit    bluer    colours    than  quiescent   elliptical     (e.g.,
\citealt{jahnke04}).  This blue colour excess is often explained  by the
presence of  a
young stellar  population,  but since the  ellipticals  in  our sample
show large  amounts    of  ionized  gas  and  globally    old stellar
population,   we  investigate if the strong    gas  emission lines can
account for the  observed  bluer colours.   We  therefore estimate the
restframe  $B-V$ colours of  our   elliptical hosts  directly  on  the
integrated spectra, in two ways: 1- using the 1D spectrum of the host,
and 2- using  a modified version  of the  spectrum where  the emission
lines  have been removed. We find  that the emission lines can account
for a  colour variation of  about 0.03 mag.  As
the average colour difference between active and quiescent ellipticals
is as large as $\Delta$(B-V)$=0.44$  \citep{jahnke04},  we can conclude  that 
the
presence of ionized gas  is not sufficient to  explain  the bluer  colours of
active galaxies compared to quiescent ones.

\begin{table}
\centering 
\caption[]{Logarithmic line ratios for the host galaxies with emission lines. 
The last  column gives our diagnostic on the main ionization  source: H stands 
for  normal {H}{II}  regions,  S for  shock heated gas,   A for ionization  by 
the active nucleus.   I stands   for {\it intermediate} between A and H, and
consists  of a mixture of different ionization sources.  Oxygen line ratios
indicated in  parenthesis are corrected for reddening.  The ratios  involving   
the H$\alpha$ line for   HE\,0441-2826   are only
indicative,   as the correction for atmospheric absorption is poor for
this object (see Fig.~\ref{spectro1}).  The second part of the table lists the
three objects for which the ionization source is found to vary significantly
with distance from the nucleus.  In those cases, the first line corresponds to
the integrated spectrum, the second to the central regions of the host and the
third to the external ones. }
\begin{tabular}{llcccc}
\hline
Object &   $\frac{{\rm O}{\rm  II}}{{\rm  O}{\rm  III}}$&  $\frac{{\rm
O}{\rm    III}}{{\rm   H}{\beta}}$&$\frac{{\rm   N}{\rm      II}}{{\rm
H}{\alpha}}$&$\frac{{\rm S}{\rm II}}{{\rm H}{\alpha}}$ &I.S.\\
\hline
HE\,0132$-$4313&\, -&-0.04&-0.15&-&H\\
HE\,0208$-$5318&\,0.08&0.15&-0.12&-&H\\
HE\,0306$-$3301&\,0.18 (0.62)&0.34&-0.37&-0.63&H/S\\
HE\,0441$-$2826&-0.14 &0.46 &(-0.32) &(-0.69) &H\\
HE\,0914$-$0031&-0.65&0.43&-&-&H\\
HE\,1009$-$0702&\,0.32 (0.44)&0.22&0.01&-0.28&H/S\\
HE\,1302$-$1017&-0.03&0.46&-0.03&-0.39&H/S\\
HE\,1503+0228&\,0.32 (0.43)&0.19&-0.22&-&H/S\\
HE\,2345$-$2906&\,0.38 (0.44)&-0.07&-0.22&-0.32&H\\
HE\,0530$-$3755&-0.24&0.70&-0.26&-&I\\
HE\,0450$-$2958&-0.60&1.06&-0.44&-0.28&A\\
HE\,1029$-$1401&\,0.02 (0.39)&1.01&-0.20&-0.02&A\\
HE\,1434$-$1600&-0.72&1.06&-0.34&-0.65&A\\
\hline
HE\,0203$-$4221&-0.19&0.79&-&-&I\\
{\it centre}&-0.16 &0.72&-&-&{\it I}\\
\vspace{0.1cm}
{\it 2.5 to 12 kpc} &-0.26 &0.92&-&-&{\it A}\\
HE\,0354$-$5500&-0.44&0.13&-0.20&-0.45&H\\
{\it centre}& -0.41& 0.01&-0.19&-0.39&{\it H}\\
\vspace{0.1cm}
{\it 9 to 20 kpc}& -0.65& 0.87&-0.25&-&{\it A}\\
HE\,2258$-$5524&-0.12 (0.21)&0.67&-0.22&-0.55&I\\
{\it centre}&-0.07&0.44&-0.23&-0.50&{\it H}\\
{\it 3 to 8 kpc}&-0.24&0.98&-0.35&-0.75&{\it A}\\
\hline
\end{tabular}
\label{diagdiagtab}
\end{table}

\subsection{Star formation rate}
\label{ssfr}

Star formation rates can   be  estimated only for galaxies presenting
 line ratios characteristic of
{H}{II}  regions,  thus avoiding  extra  emission arising from nuclear
ionization. We therefore restrict this section to the objects labelled
{\it H}  in Table~\ref{diagdiagtab}, excluding HE\,0132-4313  in which
[{O}{II}] is not detected. From the discussion of Section~\ref{stellpop}, we
note
that all galaxies in this subsample present a young stellar population. Five of
them are spirals, one is elliptical and the last two have undefined
morphologies.

\begin{figure}
\centering
\includegraphics[width=8.5cm]{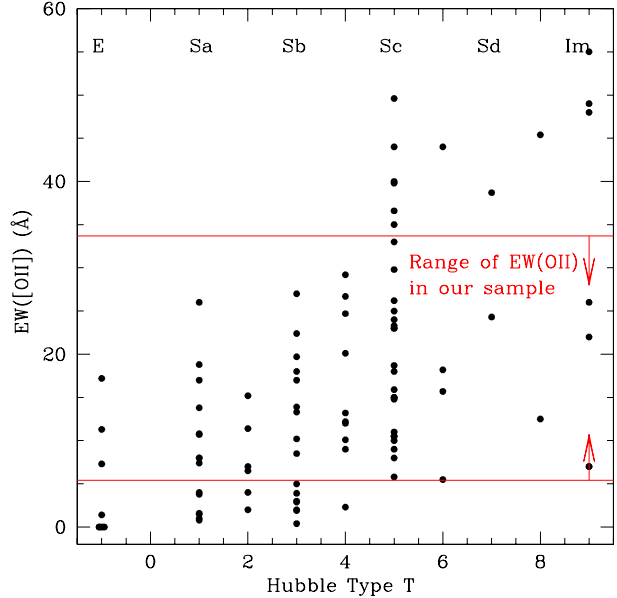}
\caption{EW({O}{II}) as a tracer of star formation rate, in the samples of
 galaxies  from \citet{ken92a}  and  from \citet{jansen} restricted to
the brightest part    of the luminosity  function.  The  range  of EWs
spanned by our {H}{II} galaxies  is shown by the horizontal lines,
suggesting  that there  are  no quasar hosts  with  low star formation
rate, nor extreme starbursts.  The majority of quiescent elliptical galaxies
presents no [OII]
emission line at all while most of our elliptical hosts do (not included in
this HII region-like subsample).}
\label{sfr}
\end{figure}

Several estimators  of the star  formation rate (SFR) are available in
the  optical range, one of  the most widely  used being the integrated
flux in the H$\alpha$  emission  line.  However,  because of the  poor
quasar-host separation  at  this wavelength  for  several objects,  we
adopt another     indicator,  [{O}{II}] 3727   \AA,  as   suggested by
\citet{kenOII}  and  \citet{kewley}. In order to get rid of the 
effect  of dust extinction,  we rely on EWs rather than fluxes and, thus,
directly compare  our  EW([{O}{II}]) with measurements
 in  non-active  galaxies,   sorted  by   Hubble   type. As a reference sample,
we use the atlas of
galaxies from \citet{ken92a} augmented by the sample of nearby galaxies of
\citet{jansen}. We  restrict ourselves  to the  brightest  part of the 
luminosity function with M$_B < -$18, as  EW([{O}{II}]) is slightly
correlated with magnitude.

Figure~\ref{sfr} shows the  measurements,  with a slight evolution  of
 EW([{O}{II}]) with Hubble type, later-type galaxies displaying
larger  EWs.  We  overplot on   the reference  sample  the range of
values spanned by  our quasar hosts. 
Figure~\ref{sfr} shows that while none of our hosts displays a low  SFR, none
of them has very large one, as could be found in starburst galaxies.  The 
EW([{O}{II}])  of our sample are compatible with  a
population  of regular disc-dominated  galaxies,  as  the   stellar
population analysis also suggests.
 
 Note that, for all but one host, the extreme values of over- or undersubtraction of the quasar
spectrum, derived from the method explained in Section~\ref{error}, do not
move the values outside of the EW range shown in Figure~\ref{sfr},
 thus leading to the same conclusions. Only HE\,1302-1017, at the
limit of oversubtraction of the nuclear component, would present an EW
compatible with extreme starburst.

\subsection{Gas metallicity}
\label{gmeta}
The oxygen abundance in the ISM, as an indicator of the enrichment of gas by
stars in the galaxies, allows to further characterize  our subsample of QSO
hosts in which stars constitute the main gas ionization source. We first
describe the methods used to evaluate the gas metallicity, then we compare our
sample to non-active galaxies.
 
The standard measurement of the gas metallicity, log(O/H)+12, can be evaluated
by the parameters R23 (ratio of collisionally excited emission-line
intensities) and O32 (indicator of ionization parameter):
\begin{eqnarray*}
R23& =& \frac{I([{\rm O}{\rm II}]3727\,{\rm \AA}) + I([{\rm O}{\rm III}]4959\,
{\rm \AA} +5007\,{\rm \AA})}{I(\rm{H}\beta)} \\
O32& =& \frac{I([{\rm O}{\rm III}]4959\,{\rm\AA} +5007\,{\rm\AA})}{I([{\rm O}
{\rm II}]3727\,{\rm \AA})}
\end{eqnarray*}
The R23-oxygen abundance relation is globally double-valued (see e.g.
\citealt{mcg91}). For a given value of O32, the parameter R23 corresponds to
two metallicities, one for metal-rich gas (on the so-called {\it upper
branch}) and one for metal-poor gas ({\it lower branch}). Gas with  log(O/H)
+12 $>8.5$ is considered metal-rich, while with log(O/H)+12 $<8.3$ it is
metal-poor. In the metal-rich regime, R23 decreases progressively with
increasing metallicity, since higher metal abundance implies a more efficient
cooling and thus a lower degree of collisional excitation. Below the
turnaround region around log(O/H)+12 $=8.4$, the oxygen abundance becomes the
dominant factor and the R23 parameter decreases with decreasing metallicity.
This R23 degeneracy is broken by the analysis of the ratios  NII/OII or 
NII/H$\alpha$, which both grow linearly with metallicity, distinguishing between 
high and low  metallicity regimes.  We use here the calibrations for upper and
lower branches from \citet{mcg91}, as given in Eqs. 7 to 10 of
\citet{kobul99}.

 While these estimators have been widely used to infer oxygen abundances in
well resolved local HII regions, \citet{kobul99} have shown that global
emission-line spectra could also indicate the chemical properties of distant
star forming galaxies.
  A reliable estimate of the metallicity can also be reached with EWR23 and
EWO32, the same parameters as R23 and O32 but computed on EWs rather than on
intensities \citep{kobul}. As the line EWs are insensitive to interstellar
reddening effects,  the EW version is to be preferred in our case since
extinction cannot be properly corrected for the whole sample.   The EWR23
method is however affected by an additional uncertainty of around 0.1 dex in
comparison to other methods \citep{moustakas}. 

The oxygen abundances in our sample deduced from this EW calibration are listed
in Table~\ref{logoh}. Most of these abundances correspond to the turnaround
region, with  NII/OII and NII/H$\alpha$ ratios pointing to the metal-rich
branch. HE 0441$-$2826, the most metal-poor object of the sample, is at the
extreme limit of validity of the upper branch. 

\begin{table}
\centering 
\caption[]{Gas metallicities expressed as log(O/H) + 12, computed from
\citet{mcg91} EWR23 correlation as given in \citet{kobul99}.  We corrected the
H$\beta$ EWs of the  last two objects for underlying stellar absorption. The
maximum bulge-to-host ratio is estimated as discussed in the text.}
\begin{tabular}{lccc}
\hline
Object name &  log(O/H)+12& 1$\sigma$ error &bulge/host \\ 
\hline
HE 0208$-$5318&8.41&0.11&0.8\\
HE 0306$-$3301&8.45&0.16&0.5\\
HE 0441$-$2826&8.28&0.12&-\\
HE 0914$-$0031&8.83&0.11&0.7\\
HE 1009$-$0702&8.34&0.14&0.2\\
HE 1302$-$1017&8.48&0.08&0.8\\
HE 1503+0228&8.53&0.05&0.3\\
HE 2345$-$2906&8.79&0.05&0.4\\
\hline
\end{tabular}
\label{logoh}
\end{table}
\begin{figure}
\centering
\includegraphics[width=8.5cm]{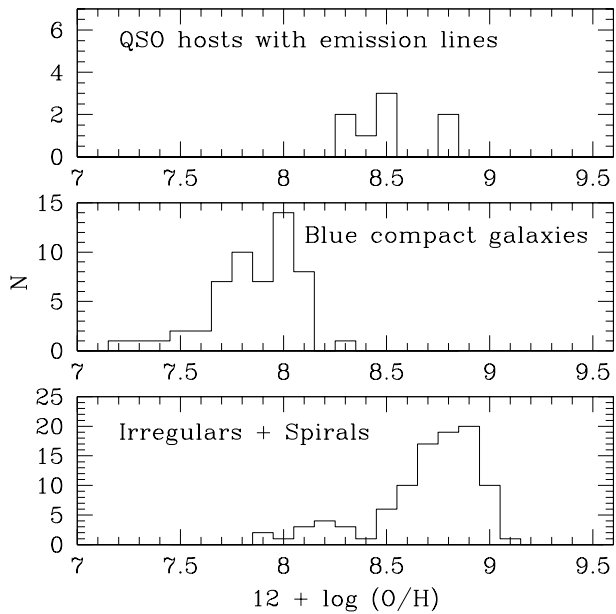}
\caption{Histogram of gas metallicities estimated on global spectra in our
sample (top panel), in blue compact galaxies (Isotov 1999, middel panel), 
irregulars and spirals (Kewley et al. 2005, bottom panel). While two galaxies
lie perfectly in the spiral distribution, the other hosts present intermediate
metallicities, in between blue compact and spiral galaxies.}
\label{metal}
\end{figure}
  We  compare in Fig.~\ref{metal} the oxygen abundances of QSO host galaxies
presenting gas ionized by stars to other samples of integrated spectra for
emission-line galaxies: a sample of blue compact galaxies, representing very
metal-deficient environments \citep{izotov}, and the sample of  nearby
irregulars and
spirals of \citet{jansen}, used in the previous section for SFR
characterization, with metallicities computed by \citet{kewley05}. This sample
contains all types of spirals, from S0 to Sd. The earlier-type spirals
(S0-Sab) tend to contain gas with higher metallicity, with  log(O/H)+12 around
9, while the small bump at lower metallicities (between 8 and 8.5) in this
distribution roughly corresponds to irregular galaxies.

The two most metal-rich galaxies of our sample, HE\,2345-2906 and HE\,0914-0031, sit in the middle of the spiral distribution. As for both we find
disc morphologies wity no trace of interaction and medium to low SFR, we
suggest that they have undergone a normal spiral galaxy evolution. In the case
of  HE\,0914-0031, the  cause of ignition of the quasar is thus still unknown.
Recent {\it HST}/ACS images of HE\,2345-2906, currently under analysis (Letawe et al., in prep.), reveal
a
prominent bar which could be an efficient driver of gas and dust to the
central region, thus triggering the QSO activity, as suggested by
\citet{crenshaw} for another type of AGNs, the Narrow Line Seyferts 1 (NLS1).

The rest of the sample has metallicities in between the blue compacts and the
more evolved spirals. The gas in these host galaxies is similar to the gas of
irregulars, thus suggesting a rather 
inefficient star formation. HE\,1302-1017, with the highest SFR of the sample,
an irregular radial velocity profile and  signs of small companions merging
in, might constitute a very young system where both nuclear activity and star 
formation have just been triggered by the mergers. For HE\,0306-3301, the minor
merger only induced nuclear activity without 
enhancing drastically the star formation. The four other galaxies show no trace
of gravitational interaction. They must also 
be young systems (compatible with their young stellar population, and with the
disc morphology found for most of them) but in which the origin of activity
remains, as for HE\,0914-0031, still unexplained.
 It is only at the limit of undersubtraction of the quasar light that the
metallicities would become compatible with normal spirals, the oversubtraction
still diminishing the host metallicities towards values typical of irregular
galaxies. 

\begin{figure}
\centering
\includegraphics[width=8.5cm]{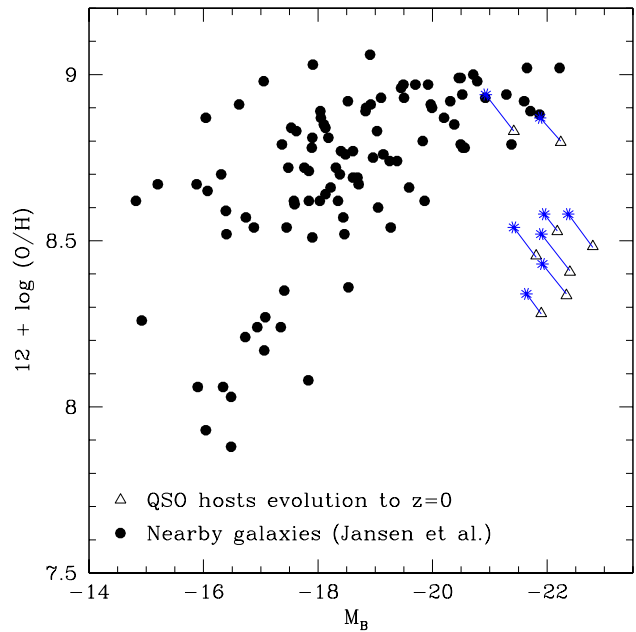}
\caption{ Metallicity-Luminosity diagram. The Jansen sample of nearby
galaxies is presented with solid circles, whereas our sample and its expected evolution is shown from the observed redshift (open triangle)
to $z=0$ (star-like symbols).}
\label{LZ}
\end{figure}

 The cosmological evolution of the comparison samples from $z\sim0.3$ to $z=0$ has
not been taken into account in the histogram. However, following PEGASE2
chemical evolution models  as presented in \citet{maier}, this cosmological
evolution on a standard metallicity-luminosity diagram does not exceed 0.1 dex
in metallicity and 0.5 dex in magnitude, as shown in Fig.~\ref{LZ}. The
aforementioned conclusions on galactic histories are thus not significantly
affected by a correction for evolution, as Fig.~\ref{LZ} shows that
evolutionary effects are not large enough to explain the low metallicities
observed at a given host luminosity.

Following the $M_{\rm BH}-M_{\rm bulge}$ scaling relation observed in active
and inactive galaxies \citep{mclure02}, the observed luminosities of the
quasars and thus their central masses ($>6\,.\,10^7 $M$_{\odot}$) imply a
substantial bulge (M$_V<-21.3 $) for all those host galaxies. The young
stellar and gaseous content inferred for them do not match a bulge population,
but rather a disc-like one. We estimate the bulge/host ratios from the  $V$
magnitudes of Table~\ref{sampletab}  and the bulge magnitudes from the $M_{\rm
BH}$ estimates of section~\ref{cm} (Table~\ref{logoh}). Given that these
ratios are upper limits because the host magnitudes are not corrected for
reddening and may be underestimated (section~\ref{analima}), they are fully
consistent with typical late-type spirals.

In the majority of the cases analysed in this section, the quasar activity
obviously appears in galaxies in which the star formation has had a low
efficiency so far. This observation could be linked to a selection effect, the
most metal-rich galaxies being more likely to contain more dust 
and thus suffer from extinction, falling below the magnitude criterion of our
sample selection.
But this also suggests that these spiral hosts are at an early stage of their
evolution and,
thus, that one of the mechanisms for triggering quasar activity may be linked
to the formation or very early evolution of (at least some) spiral galaxies.
As discussed in the next section, galactic collisions are obviously another
(and more powerful) triggering mechanism.

The origin of the triggering of QSO activity in the young spirals discussed in
the present section might  be confined to the central part ($<1$ kpc, i.e. the
pixel size of our data) of the galaxy.  In such a case, it would not be
detected from a study of the global host characteristics analysed here.

\begin{figure}
\centering
\includegraphics[width=9cm]{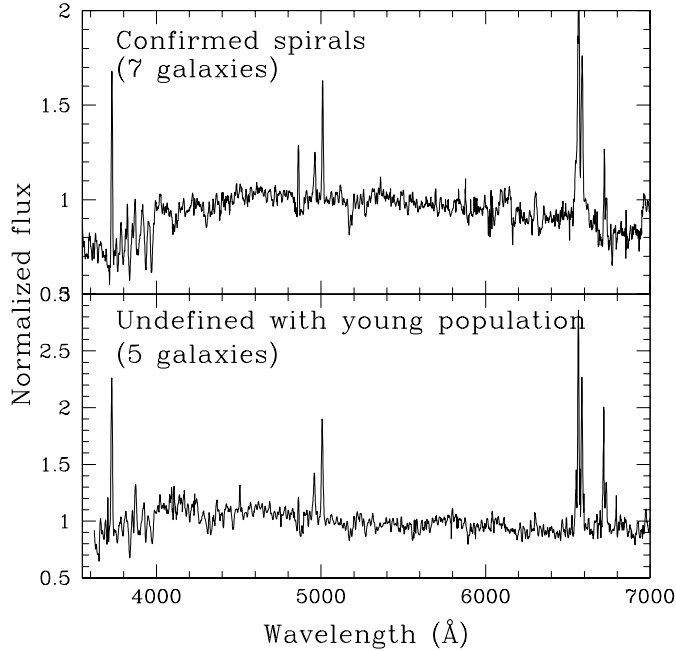}
\caption{{\it Top:} average spectrum for spiral hosts. {\it Bottom:} average
spectrum  for hosts with an   undefined  morphology but young  stellar
population.}
\label{moy1}
\end{figure}

\begin{figure}
\centering
\includegraphics[width=9cm]{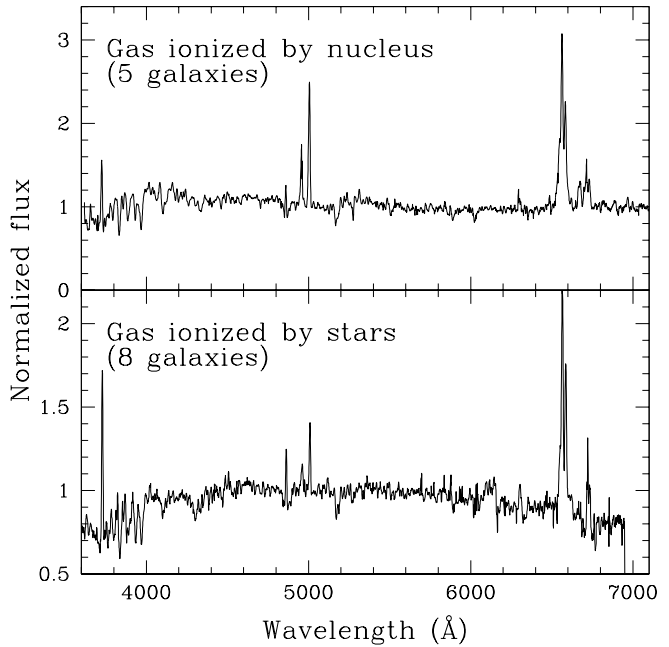}
\caption{{\it Top:} average spectrum for quasar hosts where the gas is 
ionized by the quasar. {\it Bottom:} average
spectrum  for hosts where the gas is ionized by stars.}
\label{moy2}
\end{figure}

\subsection{Average spectra}

For the  sake of illustration,  we computed average spectra for several
subsamples of  our 20 host  galaxies.  The  averages are  presented in
Fig.~\ref{moy1} for galaxies   classified as spirals  and for galaxies
with an undefined type and a young stellar population. Fig.~\ref{moy2}
shows average spectra for galaxies with an  ISM ionized by the central
quasar on one hand and  by stars on the other hand. The  four spectra with the
lowest S/N
have been removed from the averages,  as well as HE\,0450-2958 for which
no stellar continuum is  detected at all.   The averaging is performed
on the spectra in the rest-frame. They are all normalized to unit flux
at  5100 \AA\, and weighted according  to their S/N.  The two
average  spectra in Fig.~\ref{moy1}  present similar  global features,
indicating  that  an  important  fraction  of the    galaxies  with an
undefined morphological type could be disc-dominated. In Fig.~\ref{moy2}, the
average on galaxies  with gas ionized  by  stars has a blue  continuum with
prominent Balmer absorption lines
while  the  continuum for galaxies  where  the gas  is ionized  by the
nucleus is  flatter.   As expected in  HII  regions,  the blue  slope and
strong Balmer lines
indicate young stars, while ionization by the quasar appears in a mix
of galactic types with  a larger range of  slopes, hence producing on
average a flatter continuum.


\section{Dynamics and interaction}
\label{dyn}

\subsection{Radial velocity curves and mass modelling}
\label{rc}

We take  advantage  of  the very  different  spectral   widths  of the
emission lines in the  quasar  and in the   host galaxy  to
carry out an accurate quasar-host  separation in the   emission
lines.  This procedure, described in detail in \citet{courb02}, allows
 to extract detailed information on the velocity of the gas in the host galaxy.

\begin{figure*}
\centering
\includegraphics[width=17.5cm]{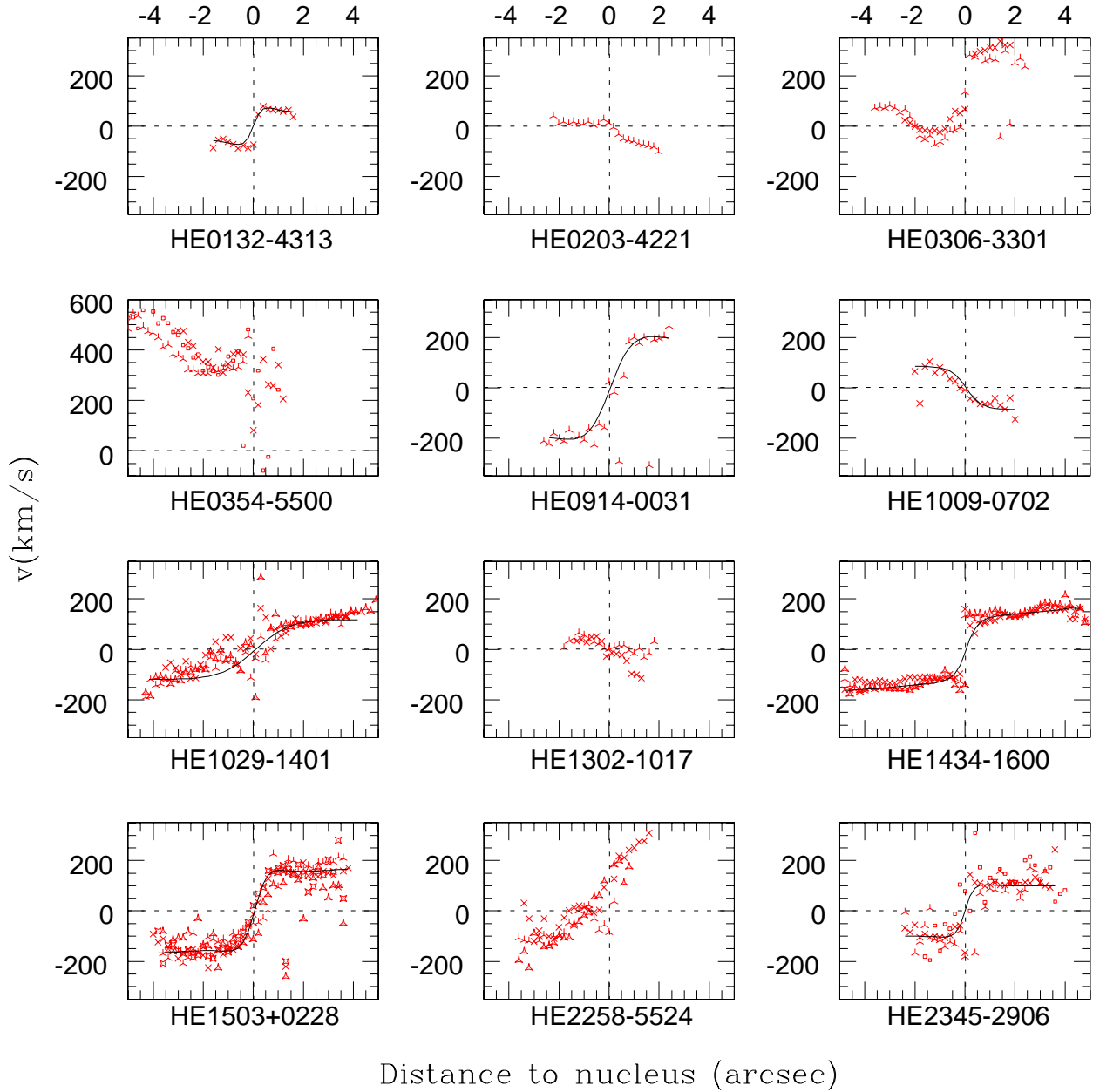}
\caption{Extracted radial velocity curves for all objects with sufficient 
S/N in the emission lines. The H$\alpha$ line is represented by 4-arms
crosses.   Since some grisms   have  overlapping spectral  ranges  the
H$\alpha$ line may appear twice. In such a case, the measurement of this line
in  the second grism is plotted
as 4-arms stars.  The [{O}{III}] line is represented by 3-arms crosses
(and stars when a second measurement is available) and  the  [{O}{II}] line is 
represented by  squares.  The
solid lines represent mass model fits when relevant (see text).}
\label{rotcurv}
\end{figure*} 

This spectral decomposition method    is applied to all   the  spectra
showing sufficiently high signal-to-noise (S/N  $>$ 5 per pixel) in the
external   parts  of the hosts.    Figure~\ref{rotcurv} presents the
extracted radial  velocity curves.  When several emission
lines  are extracted  for  the  same object, they   are represented by
different  symbols on the same  plot.  Figure~\ref{rotcurv} shows that
many radial velocity curves are distorted and thus cannot be fitted by
simple  mass models as  described in \citet{courb02}. These galaxies are
probably disturbed by ongoing  or past gravitational  interactions with
nearby companions.

Fitting a mass model to the velocity  curves requires knowledge of the
inclination and position angle (PA) of the  disc (if any is involved).
They are estimated using the acquisition images obtained before each spectrum,
using a  modified version of   the MCS  deconvolution algorithm  that
includes a 2D analytical model for the host galaxy.

The results of the mass model  fits on symmetrical velocity curves (see
\citealt{courb02})    are     represented  by      a  solid     line  in
Fig.~\ref{rotcurv}.   The corresponding mass  estimates  are  given in
Table~\ref{massgal}, along with the  inclination of the disc component
 on the line  of sight and the angle of the major axis relative to the slit.
The mass
model includes  a dark matter  halo,  a rotating stellar disc, and  a
fixed   central mass  (from  Section~\ref{cm}).  With our   spatial
resolution (0.2 arcsec per pixel, which translates into a linear scale
between 0.6  and  1.0 kpc  depending  on  redshift),  adding  a  bulge
component does not improve the fits and prior knowledge of the central
mass (e.g.,  estimates using the H$\beta$  line of the quasar)  do not
constrain the model significantly. The mass modelling does not allow to
weight the  relative contributions of  disc and bulge   in  the galaxies.
Table~\ref{massgal} shows that these hosts are rather massive spirals,
while not disproportionate relative to their inactive counterparts.

Although the  velocity curves of   HE\,1434-1600  and HE\,1029-1401  look
symmetrical, they cannot be fitted with a realistic set of parameters.
Close scrutiny of the images reveals, in both cases, elliptical galaxies
with neighbours merging   in or interacting, which  excludes the
assumption of  a simple   circular    motion.  The special  case    of
HE\,1434-1600 has been  presented in \citet{let04}.  HE\,1029-1401  is
seen nearly   face-on   (\citealt{kuhl02}),  and   interpreting radial
velocities as rotational velocities would  give unrealistic masses (up
to 10$^{17}$ M$_{\odot}$) given the inclination angle.

\begin{table}
\centering
\caption{Mass estimates for spiral galaxies with symmetrical rotation curve. 
The adopted model  consists of a rotating disc, a dark matter halo and 
a central mass.   All masses  are in  solar units.   $i$ is the inclination 
of the disc and  $\phi$ is the angle of the  major  axis of  the galaxy
relative to the slit. The determination of these angles is
the main source  of  uncertainty in our mass estimates.  The last two
galaxies have smaller  error  bars because  {\it HST} images are 
available to determine $i$ and $\phi$.}
\begin{tabular}{lccc}
\hline
Object &$\phi$&$i$ & $\log M$\\
& ($^\circ$)&($^\circ$)&{\scriptsize$(r<10$kpc)}\\[4pt]
\hline
 HE\,0132$-$4313&78&37&11.9$^{+0.2}_{-0.7}$\\[4pt]
 HE\,0914$-$0031&68&32&12.6$^{+0.5}_{-0.3}$ \\[4pt]
 HE\,1009$-$0702&66&29&11.7$^{+0.5}_{-0.3}$ \\[4pt]
 HE\,1503+0228&18&46&11.0$^{+0.1}_{-0.1}$\\[4pt]
 HE\,2345$-$2906&70&36&12.0$^{+0.1}_{-0.1}$\\[4pt]
\hline
\end{tabular}
\label{massgal}
\end{table}


\subsection{Interactions}
\label{interac}

Signs of interaction are found for nine out  of the twenty galaxies in
our sample, i.e.,  in 2 spirals, 2  ellipticals and  in 5 galaxies
with  undefined morphology.   The  hosts whose  images  (from this  or
previous studies) show  irregular  shapes, reminiscent of  interaction
are labelled ``interac'' in Fig.~\ref{img1}, while a classification of
the signs of   interactions is given  in  Table~\ref{sampletab}: close
companion outside the  galaxy  (1), tidal  tails (2)  or  merger, i.e.
companion inside the galaxy (3). A confirmation of those perturbations
is  provided  by  the  extracted  radial velocity curves  presented in
Fig.~\ref{rotcurv}.  All the  hosts  with asymmetric curves also  have
images  with irregular shapes.  We thus find that $\sim$ 50 per cent of our
host
galaxies display signs of interaction, which is compatible with several
previous studies (\citealt{smith}; \citealt{hutchings92}; \citealt{bahc97};
\citealt{sanchez}; \citealt{jahnke04}). This is also compatible with the
fraction of interacting but non-active galaxies according to \citet{dunlop03}.

The diagnostic diagram in Fig.~\ref{ion_merg} shows that {\em every galaxy
with long-range ionization by the nucleus is also interacting}. The mechanism
at work is probably  similar  to the  one  responsible for  the  Extended
Emission  Line  Regions (EELR)  usually   associated with  radio-loud
quasars   (\citealt{boroson85}; \citealt{stock87})   or Seyferts  with
radio emission \citep{evans}, as presented in \citet{stock02}.  During
the interaction,  the dust  surrounding  the quasar  is swept   out or
sufficiently perturbed, letting the  ionizing  nuclear beam reach  the
gas in the outer parts  of the galaxy  and ionizing it. Shocks induced
by the interaction are considered  as complementary sources of ionization.
If the radio jet alone was sufficient to produce EELR, this phenomenon
would be more common in Radio Loud  Quasars.  The present study widens
the  type of galaxies  where this phenomenon appears  to  the hosts of
radio quiet quasars (RQQ), as only 2 out of  the 7 cases of ionization
by the central quasar display radio emission, at a level under the
limit of classification as Radio   Loud. To our knowledge, almost   no
high  level ionization  in galactic gas   was previously found in  RQQ
(only one object  in \citealt{stock87}) probably  only because  of the
lack of a detailed spectral study of this kind of host galaxies.

The unused slits of MOS  spectrograph were placed  on galaxies in  the
field, looking for companions  at the same  redshift as the quasar.  While
this is far  from being an  exhaustive  analysis of the neighbourhood  of
each object, we find that 13 out of the 20 quasars have no detected neighbour,
4 quasars have one neighbour, 2 quasars have two and HE\,1434-1600 has
5 neighbouring galaxies with close redshifts \citep{let04}.


\section{Nuclear properties}
\label{quasar}

In this Section, we derive several properties of the quasars from their
spectra and   investigate    their relation with    host  type,
properties  or  environment.  As the amount of dust on the line of sight to the quasar cannot be estimate with precision for each quasar,  no correction for reddening is applied.

\subsection{Spectral characteristics}
\label{spch}

The  quasar  properties are listed below and summarized in Table~\ref{qso}:

\begin{itemize} 

\item The luminosities of emission lines: [{O}{II}] 3727 \AA\, and 
[{O}{III}] 5007 \AA, indicators of the amount of gas and ionization of
the Narrow Line  Region (NLR) as well as  H$\beta$ for the  Broad Line
Region (BLR);

\item The power law index $\alpha$   of the continuum 
(F($\lambda)\propto   \lambda^{\alpha}$), indicator  of  the
power of the  source. We evaluate  it from the   continuum slope in  the
$\log{F}$ versus $\log{\lambda}$ diagram between 4000 and 8000\AA;

\item The monochromatic luminosity at 5100 \AA, 
another indicator of the power of the nucleus, used for estimating the
bolometric luminosity;

\item The Balmer decrement H$\alpha$/H$\beta$. The Balmer decrement is dependent not only from extinction but also on physical density parameters in the BLR. It can however indicate the presence of dust.

\item The ratio [{O}{III}]/H$\beta$;

\item The FWHM of H$\beta$, useful for black hole mass determination 
(see Section~\ref{cm});

\item The presence of FeII lines, noticeable by emission line 
multiplets, such   as  for  instance  in   HE\,0132-4313  but  not  in
HE\,0203-4221 (see Fig.~\ref{spectro1}).

\end{itemize}

\begin{table*}
\centering
\caption{Nuclear spectral characteristics for the whole sample and derived 
values  of   the BH  mass and  accretion   rate.  Luminosities are  in
erg\,s$^{-1}$, $\alpha$  is  the  power  law  index of  the  continuum (F($\lambda)\propto \lambda^{\alpha}$), and the FWHM
of the broad H$\beta$ component is expressed in km\,s$^{-1}$.  Black hole
masses are  given in logarithm  of solar  masses and  accretion rates in
solar masses per year.  The atmospheric    absorption
overlapping H$\alpha$ in  HE\,0441-2826 spectrum was  corrected before
Balmer  decrement evaluation.}
\begin{tabular}{lccccccccc|cc}
\hline
Object &log L$_{[{\rm OII}]}$&log L$_{[{\rm OIII}]}$&log L$_{\rm H\beta}$&  $\alpha$&log$\lambda$L$_{51}$&FeII&  FWHM(H$\beta$)&$\frac{\rm{H}\alpha}{\rm{H}\beta}$&$\frac{[{\rm OIII}]}{{\rm H}\beta}$&log M$_{\rm BH}$&$\dot {\rm M}$\\
\hline
HE\,0132$-$4313  &41.24&40.93&42.60&-2.4&44.91&x&1056&3.2&0.02&7.21&0.85\\
HE\,0203$-$4221  &41.85&43.24&43.39&-2.1&45.33&&7153&7.4&0.63&9.15&2.16\\
HE\,0208$-$5318  &41.41&42.44&43.38&-2.5&45.07&&6075&3.0&0.11&8.83&1.17\\
HE\,0306$-$3301  &41.56&42.57&43.11&-2.3&44.98&x&2346&3.2&0.29&7.96&1.01\\
HE\,0354$-$5500  &41.66&42.60&43.23&-2.3&44.89&x&2264&3.4&0.09&7.86&0.81\\
HE\,0441$-$2826  &41.53&42.62&42.39&-2.2&44.99&&5771&5.3&1.69&8.87&1.12\\
HE\,0450$-$2958  &42.75&43.46&43.92&-2.3&45.58&x&1282&3.8&0.34&7.63&1.96\\
HE\,0530$-$3755  &41.80&43.13&43.11&-2.5&44.73&&1721&5.6&0.95&7.50&0.54\\
HE\,0914$-$0031  &40.89&42.34&42.95&-2.0&44.62&&3072&4.3&0.24&7.92&0.42\\
HE\,0956$-$0720  & -   &42.37&43.36&-2.2&45.03&&2636&3.7&0.1&8.07&1.06\\
HE\,1009$-$0702  &40.97&42.22&42.94&-1.6&44.71&x&1826&4.2&0.18&7.55&0.54\\
HE\,1015$-$1618  &41.08&42.52&43.31&-1.4&44.98&x&2212&3.1&0.17&7.89&0.98\\
HE\,1029$-$1401  &41.10&42.24&42.95&-1.3&44.69&&5887&4.1&0.19&8.60&0.61\\
HE\,1228+0131    & -   &42.30&42.88&-1.9&44.80&x&1403&3.8&0.26&7.41&0.73\\ 
HE\,1302$-$1017  &41.35&42.83&43.24&-2.0&45.33&&2945&4.7&0.36&8.38&2.16\\
HE\,1434$-$1600  &41.10&42.52&42.50&-1.4&44.19&&7068&7.4&0.95&8.37&0.17\\
HE\,1442$-$1139  &40.83&40.92&42.52&-1.7&44.30&x&2063&3.9&0.02&7.36&0.21\\
HE\,1503+0228    &40.14&41.64&42.35&-1.9&44.23&&2990&4.4&0.17&7.65&0.19\\ 
HE\,2258$-$5524  &40.93&41.54&42.69&-2.6&44.79&x&1352&3.4&0.07&7.31&0.59\\
HE\,2345$-$2906  &40.98&41.94&42.46&-2.0&44.15&&4138&4.0&0.29&7.87&0.15\\
 \hline
\end{tabular}
\label{qso}
\end{table*}

\begin{figure}
\centering
\includegraphics[width=8.5cm]{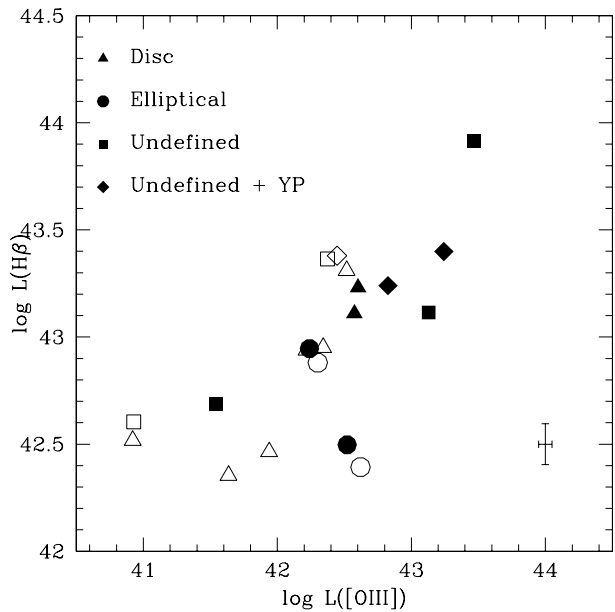}
\caption{Luminosity of the nuclear H$\beta$ versus luminosity of the nuclear 
[{O}{III}] 5007 \AA\, emission  line. The characteristics of  the host
galaxies  are as stated  in the legend. Filled symbols are for quasars hosted
by galaxies that display signs of interaction.  Typical 1-$\sigma$ error bars
are shown, including errors on  continuum subtraction for measurement
of luminosities. Luminosities are in erg\,s$^{-1}$.}
\label{lum}
\end{figure}

Some of these spectral characteristics are plotted in Fig.~\ref{lum},
which shows that the quasars involved in galactic interactions (filled
symbols) tend to have   more   emission from   the  gas   in  their  immediate
surroundings  (higher luminosities in   broad  and narrow  lines).  As
[OIII] is a good tracer of AGN power (\citealt{kauff}), we conclude that the
quasars in interacting systems  are  also generally more powerful.    This 
gives us direct
evidence that the gravitational  interactions constitute an  essential
process in bringing gas in the vicinity  of the black hole and feeding
it. Moreover, as we have seen from Fig.~\ref{ion_merg}, highly ionized
galactic gas is found in interacting systems, i.e.  the ones which
harbour the most powerful quasars. A high power of the quasar thus seems
essential for ionizing gas at large distance from the nucleus.

\paragraph*{Narrow line quasars} In examining the quasar spectral 
characteristics, we note that 5 quasars (HE\,0132-4313, HE\,0450-2958,
HE\,1009-0702,  HE\,1228+0131  and HE\,2258-5524)  are  fulfilling the
Narrow Line Seyfert 1 (NLS1) criteria, i.e.\   FWHM(H$\beta)< 2000 \kms$,
strong   {Fe}{II} multiplets and [{O}{III}]/H$\beta<3$ \citep{bian04},
even if their   luminosities are  characteristic  of  quasars and  not of
Seyferts.  We call them  ''Narrow  Line QSOs'' (NLQSOs), with  similar
properties as  NLS1s  but  at  higher  luminosities  (as suggested   by
\citealt{step03}).  The proportion  evaluated from  optically selected
samples  amounts to 10 \% only  \citep{grupe04},  while it reaches 
25 \% here, although   our sample is  also optically  selected. There  was no
selection bias on the spectral properties of the objects, but we have
only  6 quasars with  obvious  broad  components (FWHM(H$\beta)>  4000
\kms$),   while the rest  of  the sample  (9 objects) has  hydrogen lines of
intermediate widths. A  deeper investigation  of spectra in the whole HES
survey reveals  that 11 per cent of the quasars  
have FWHM(H$\beta) < 2000    \kms$, as  in other  optically   selected
samples. However, the FWHM measurements are  highly sensitive  to
spectral  resolution and  S/N (lower  in   HES than in this   present
study), and only  one of  the 5 NL objects of  the present  sample would be
classified as such from the original HES spectra. The actual proportion of
NLQSOs
is thus probably higher than 11 per cent in HES, and this argument might also
hold for  other  optical  surveys.  

The low-redshift QSO sample in  HES is purely colour-selected,
and there is no reason why the sample might be biased in favour of narrow-line
QSOs. We note that the distribution of H$\beta$ linewidths in the HES
is strongly peaked with a mode of about 2700 $\kms$ and a median
of 2940 $\kms$, which is not much more than the conventional
NLS1 criterion of 2000  $\kms$. Inspection of the spectra
of objects close to this limit confirms that it is not a physically
meaningful separation of distinct quasar sub-populations. For instance, we  could
add 4 more NL  candidates in our sample  if the limit was  situated at
2300 $\kms$ rather  than 2000 $\kms$ (all  quasars  but one presenting
prominent  iron lines have a H$\beta$ FWHM close  to this limit), reinforcing
our suggestion that the present NLQSO classification is largely arbitrary.
 The median
FWHM(H$\beta$) of our sample is 2500 $\kms$, which is fully
compatible with the full HES sample. We conclude that our sample is
a fair subset of the low-redshift quasar population.  It should however be pointed that the mean FWHM(H$\beta$) of the HES sample is substantially lower than found in the SDSS quasars ($\sim5300\kms$) of the same luminosity and mass range \citep{schneider}.  The reason for that difference is not known.

The  host of  those five
NLQSOs span  the whole range of magnitudes  of our sample, 2 out
of 5  are interacting,  3  have undefined  morphology (2 with
intermediate-age stellar population, 1  undefined), one is a spiral and one
an elliptical. The nucleus-to-host luminosity ratio estimated from images
ranges between   3  and  18, except   for the  peculiar  HE\,0450-2958
described in \citet{mag05} for which the upper limit on host magnitude
gives a minimum N/H of 85.  Mass modelling (Section~\ref{rc}), applied
to   two of the NLQSOs reveals   neither under- nor over-massive hosts
compared to broad  line QSOs.  There is thus   no clear tendency   for
NLQSOs to appear in a special type of host galaxy.


\subsection{Central mass}
\label{cm}

There exists a widely used  relation between the  black hole mass
($M_{\mathrm{BH}}$)  and characteristics of  the  Broad Line Region of
quasars \citep{kaspi00}:

\begin{equation}
M_{\mathrm{BH}}=R_{\mathrm{blr}}v_{\mathrm{blr}}^2 G^{-1}
\end{equation}

This  virial black hole  mass  estimate relies  upon the hypothesis that
dynamics in  the BLR  is dominated  by gravitation.  The reverberation
mapping method has  allowed to measure  directly the radius $R_{\mathrm{blr}}$ of 
the BLR in  several active  nuclei, and a  correlation has  been found between
$R_{\mathrm{blr}}$ and the AGN luminosity  at 5100 \AA.  The  velocity
dispersion $v_{\mathrm{blr}}$
is estimated from the FWHM of the broad  component of H${\beta}$. 

In a significant fraction of  our  quasars,  it is however  problematic  to
estimate accurately  the width of  the broad part of the hydrogen
line.  When the narrow component  cannot be  clearly detected and when
the broad  H$\beta$ itself tends to be  rather narrow, the separation
between both components  depends highly on assumptions that
are made on  flux ratio between [{O}{III}]  5007 \AA\, and the  narrow
H$\beta$ line, or on the adopted  fixed width for this narrow component
(see for  instance \citealt{grupe04}). Here we  choose to present mass
estimates  including global FWHM(H$\beta$) when  the separation is not
straightforward, assuming that in these cases  the contribution of the
narrow component  is negligible.   In    parallel, we  evaluate     the
uncertainty on these  values by making a second determination based on the
assumption that the  narrow H$\beta$ has the same width as [{O}{III}]
5007  \AA\, in order  to separate the former
from the   broad component. This    uncertainty on  narrow/broad  line
decomposition will mainly affect the quasars with the narrowest lines,
thus the lowest black hole masses, which might thus be underestimated.

\begin{figure}
\centering
\includegraphics[width=7.cm]{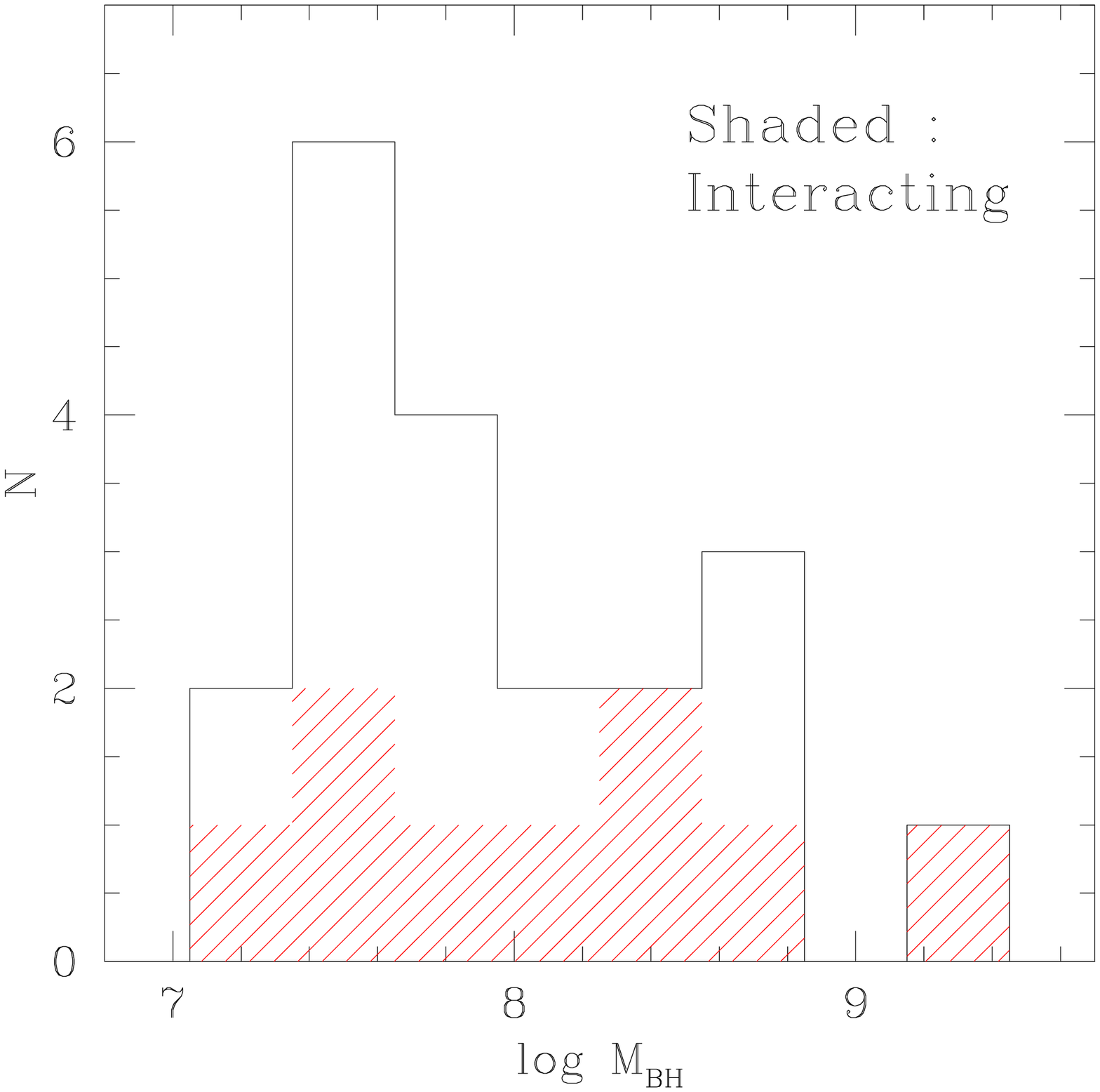}
\includegraphics[width=7.cm]{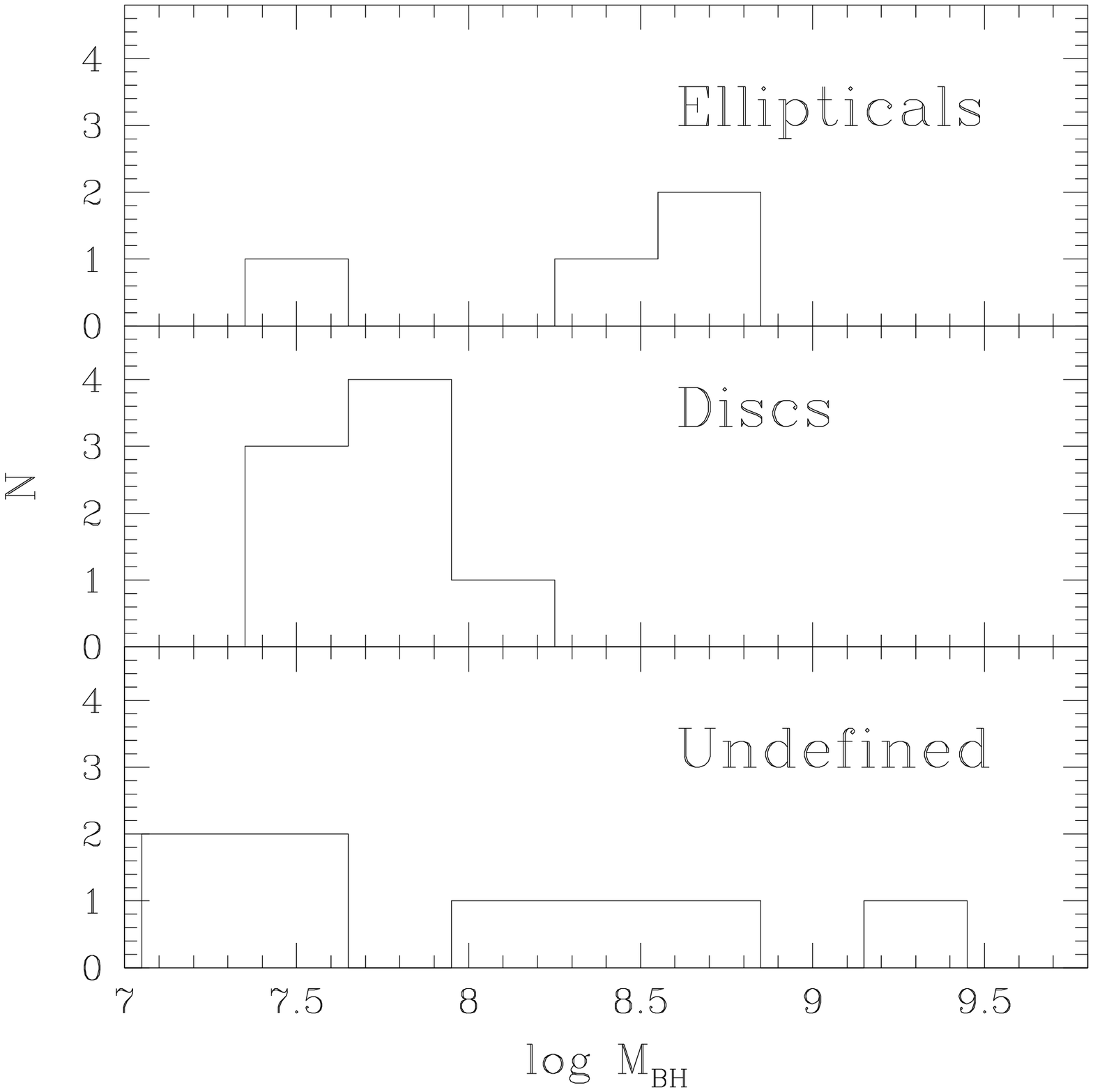}
\caption{Histogram of the black hole masses found with the virialized BLR
method. 
{\it Top}:  global  histogram where  the  black holes hosted in interacting
system are shaded. {\it Bottom}: histograms  of $M_{\mathrm{BH}}$ as a function 
of morphological classification. }
\label{histmbh}
\end{figure}

Figure~\ref{histmbh} shows histograms of  black hole masses using the
regression from  \citet{kaspi00}, after conversion of our luminosities
to   their     cosmology     ($H_0=75~\ho      $,   $\Omega_m=1$   and
$\Omega_{\Lambda}=0$). We find   that interacting systems are  hosting
slightly  more massive  black holes (on average, in solar masses, $\log{M_{\rm
BH}}=8.1 \pm 0.6$ 
in interacting hosts and $\log{M_{\rm BH}}=7.9 \pm 0.5$ for isolated ones).

This observation of higher mass black holes in currently interacting systems 
is  somewhat surprising. One might expect
that the central mass grows during the merger process and thus reaches
higher masses when  the system has had  time to relax. Those massive black
holes might however be  the result of previous mergers, being   in more 
crowded neighbourhoods, which would also  explain the
presence  of large amounts of  gas. However, the incomplete analysis of
the  neighbourhood  made with the slits available (Section~\ref{interac})
does not  allow  to give a  clear answer  to   this question.  Another
possible explanation is that there exist several scenarios to feed and
trigger the quasars, the most  efficient being a merger, creating  the
most massive black holes,  while lower mass  black holes would  follow
another evolution scheme (see Section~\ref{gmeta}).

Figure~\ref{histmbh}  also suggests that there  is a trend for spirals
to  host less massive black  holes. This is   consistent with the idea
that mergers bring fuel  to the nucleus. Those  disc galaxies have not
yet undergone major   interactions, as  mergers  would have   produced
larger spheroids \citep{stan}.

\subsection{Accretion rates}
\label{arate}

\begin{figure}
\centering
\includegraphics[width=8.5cm]{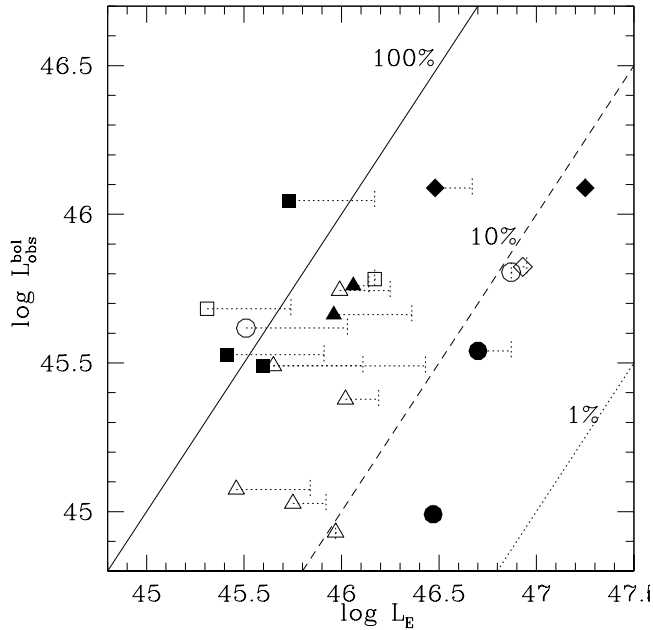}
\caption{Observed nuclear luminosity versus Eddington luminosity 
evaluated from the BH mass. The dotted, dashed and solid
lines are  at 0.01, 0.1 and 1  Eddington luminosity, respectively. The
one-sided error bars show where the objects would lie if we had separated the 
broad and narrow H$\beta$ as explained in the text. Symbols are as in Fig.
\ref{lum}: filled symbols for hosts with
interaction, morphology  of the  galaxies identified by  triangles for
discs, circles  for  ellipticals, squares when undefined  and diamonds
for undefined with young population.}
\label{acc}
\end{figure}

The nuclear luminosity measured on  quasar spectra can provide  direct
indications on accretion rates in the nucleus,  as $\dot M =L/\eta
c^2$, where $L$ is the bolometric luminosity of  the quasar and $\eta$
is the efficiency  of  the energy transfer  from  mass falling to  the
black  hole, that we assume   to be constant. The observed  bolometric
luminosity is  estimated  from  the optical  monochromatic luminosity:
$L_{\rm obs}^{\rm bol}\sim 10\lambda L_{5100}$ \citep{laor}. In Fig.~\ref{acc}
we  compare   $L_{\rm obs}^{\rm bol}$ to the   Eddington bolometric luminosity
$L_E$, i.e.  the maximum  luminosity of  a source that  is powered  by
spherical accretion. $L_E$  was estimated from  $M_{\rm BH}$ following
Eq.~\ref{edd},  stating   equilibrium between radiation pressure   and
gravitation:

\begin{equation}
L_E=1.26\; 10^{38}(M_{\rm BH}/{\rm M}_{\odot})\; {\textrm erg\, s}^{-1}
\label{edd}
\end{equation}

\begin{figure}
\centering
\includegraphics[width=8.5cm]{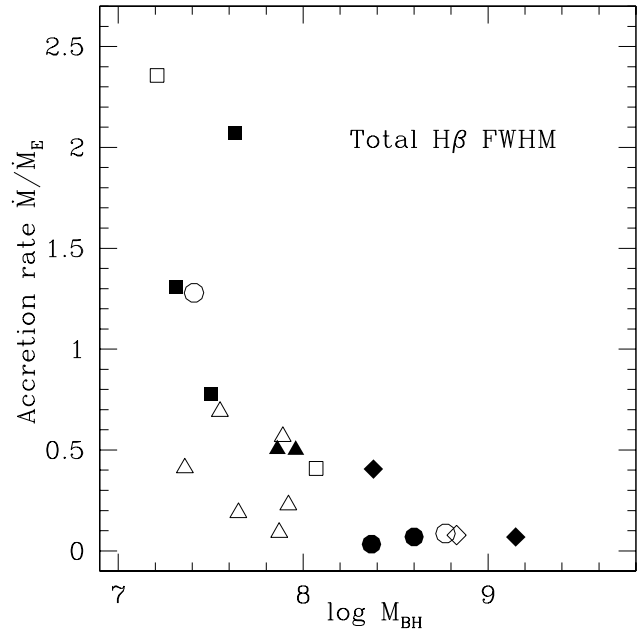}
\includegraphics[width=8.5cm]{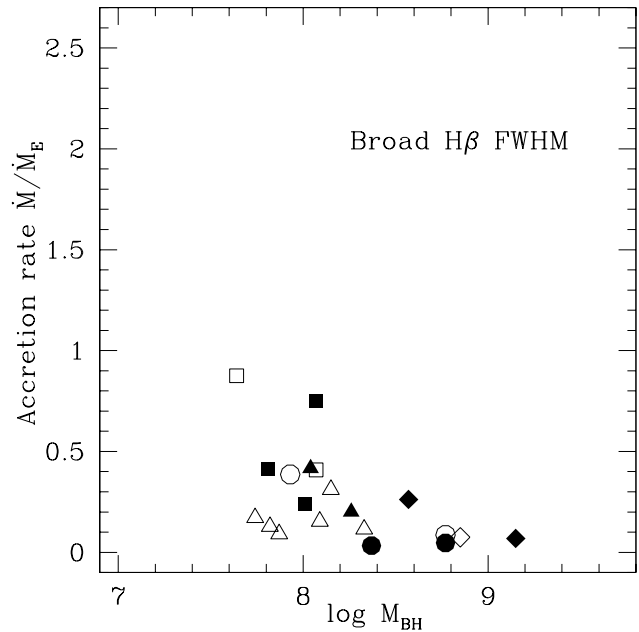}
\caption{Accretion rate in unit of $\dot M_E$ versus black hole mass in
 M$_{\odot}$.   Both  values     depend    on  the   estimation     of
FWHM(H$\beta$).  Symbols are as in Fig.~\ref{lum}. {\it Top:} estimates from
the total H$\beta$ FWHM. {\it Bottom: } estimates from the FWHM of  the broad
H$\beta$ only, after subtracting a narrow  component of the same
width as the [{O}{III}] line.}
\label{m_acc}
\end{figure}

From Figs.~\ref{acc} and~\ref{m_acc} it appears  that four quasars may
be accreting above the Eddington limit.  These ones are indeed among
the  NLQSOs discussed  in    Section~\ref{spch}. Those  objects may be
considered  as active nuclei   in  an  early  stage  of  activity.  The
accretion  rates are high while the  black holes  are not very massive
yet, and  those  rates probably  fall to  sub-Eddington ratios after a
short period of activity. This observation is however highly sensitive
to the definition of the broad H$\beta$ component. If we had chosen to
separate the broad  and narrow components of the  hydrogen
line as sketched in the previous subsection, no quasar would have been found 
with super Eddington  luminosity or accretion  rate, as  shown in Fig.
\ref{acc}
 by the dotted error bars and in Fig.~\ref{m_acc}, bottom. 
As can be seen  from  these figures, the objects most
sensitive to  this effect   are  the ones  with narrow  H$\beta$, thus
potentially super Eddington.  This points once  more to the ambiguity
of the classification  and    interpretation of the  Narrow   Line AGN
phenomenon.

The mean  accretion rate for quasars  in interacting  systems is 0.64
$\dot M_E$ while it is 0.58 $\dot M_E$ for non interacting ones
($\dot M_E$ being the Eddington accretion rate). If we
adopt   a  typical  value for the  efficiency  $\eta$  = 0.1,  these
accretion    rates     translate     to  $\dot      M=0.67    \pm 0.39
\,$M$_{\odot}\,$yr$^{-1}$ for non interacting and  $\dot M=1.11 \pm 0.77\,
$M$_{\odot}\,$yr$^{-1}$ for   interacting hosts.

 From  these  estimates  and from   Fig.~\ref{m_acc} which  shows  the
accretion  rate  plotted versus BH mass,  it  appears that on
average  the  accretion rate  is larger   when  there is gravitational
interaction, and that for a given central mass, the Eddington ratio is
higher in interacting systems. In contrast, isolated spiral galaxies have, on the average, lower accretion rates  onto lower mass BHs (as confirmed by Fig.~\ref{histmbh}).  This  is also  coherent  with the
BH-mass/bulge-mass relation (e.g.   \citealt{marconi}), spirals having smaller
bulges than ellipticals. 

Figure~\ref{m_acc} can be interpreted as showing the evolution of the quasar
during its lifetime. A young nucleus has high accretion rate and low-mass
black hole, that progressively grows while the accretion rate reduces. Two such sequences might be found in Fig.~\ref{m_acc}. First, at the bottom left, isolated spirals leading to moderate BHs ($\sim 10^8 $M$_\odot$). Secondly, at higher accretion rates, system undergoing strong gravitational interactions, leading to elliptical hosts with massive central BHs.


\section{Redshift determination}
\label{redsh}

 Analyses of  quasar emission  lines show,   as already observed  for
instance by  \citet{eracl}  and \citet{boroson05}, that  narrow  lines
such as [{O}{III}] are generally blueshifted relative to broad hydrogen lines.
The question then is to know which lines are  most suitable for measuring the
actual  redshift of  the  whole object.  The availability  of accurate
symmetrical rotation curves in  our sample provides a good opportunity
to answer this  question.  Taking advantage of  the separation  we are
able to make between nucleus and host emission lines, we have computed
accurate redshifts   for those galaxies  which show  the  most regular
rotation  curves    (HE\,0132-4313,  HE\,0914-0031,     HE\,1009-0702,
HE\,1503+0228 and HE\,2345-2906), averaging  shifts from points of the
extracted rotation  curves symmetrical around  the centre. After that,
we have searched in the quasar spectra which lines provides the best
estimator of the host galaxy redshift.  For all five  of them, it is the
wavelength corresponding to the tip of the H$\alpha$ line which best
matches the galactic redshift (within 0.2 \AA, or a fifth of a pixel),
followed with  nearly the  same accuracy by   the  H$\beta$ tip.  The
observed shifts are summarized in  Table~\ref{shift}. Assuming those 5
objects are representative of   undisturbed quasar host   galaxies and
that the quasar is motionless in the galaxy,  we conclude that the tips
of the quasar H$\alpha$ and  H$\beta$ lines provide the  best indicators  
of galactic redshift. Variations of  redshifts in  the  quasar emission  lines
can thus be  interpreted as bluer or  redder in comparison with the global
object  velocity  and  not only  in between  each other.   The redshifts
measured from the position of the H$\alpha$ tip for the whole sample
are presented in Table~\ref{sampletab}.

\begin{table*}
\caption{Observed shifts of the main quasar emission lines relative 
to the  galactic lines,  expressed in $\kms$.  Each  line position is
estimated  by a gaussian profile fitting (first line,  {\it  g}) and from the
wavelength corresponding to the tip (maximum intensity, second line, {\it
t}).}
\begin{tabular}{lccccccc}
\hline
&HE\,0132-4313&HE\,0914-0031&HE\,1009-0702&HE\,1503+0228& HE\,2345-2906& Mean
shift&$\sigma$\\
\hline
H$\alpha$(g) & +22&-104&-97&-117&-52&-70&41\\
H$\alpha$(t) &+7&-3&+7&+4&-30&-3&16\\[4pt]
H$\beta$(g)  &0&+5&-189&-133&-40&-71&84\\
H$\beta$(t)  &+5&+5&+24&+4&-35&+1&21\\[4pt]
OIII$_1$(g) &-&-27&-&-85&-109&-74&42\\
OIII$_1$(t) & -&-9&-&-91&-99&-66&50\\[4pt]
OIII$_2$(g) &-&-27&-&-79&-103&-70&39\\
OIII$_2$(t) &-&-5&-&-79&-98&-61&49\\
 \hline
\end{tabular}
\label{shift}
\end{table*} 


\section{Comments on individual objects}

\begin{itemize}

\item{HE\,0132-4313:  }Host galaxy with undefined morphology and  intermediate     
age   stellar    population.    No  sign     of
interaction. Quasar spectrum typical of Narrow Line Quasar (NLQSO).

\item{HE\,0203-4221:  }This object shows all signs of merger-induced and
enhanced activity: high ionization level found locally in the 
galaxy, young stellar population, 
evidence for interaction from significant distorsions in the image and
radial  velocity  curve.

\item{HE\,0208-5318: }Normal
disc-like    stellar  population, no  sign   of   interaction and gas
metallicity indicative  of  weakly enriched  medium. 

\item{HE\,0306-3301:  }Spiral galaxy with highly disturbed rotation curve, as a
likely sign of interaction. Young stellar population and
low metallicity.

\item{HE\,0354-5500:  }Merger in a spiral galaxy, inducing high level 
ionization and asymmetric rotation curve.

\item{HE\,0441-2826:  } Stellar content typical of disc. No information 
from acquisition image  (saturated quasar),  but radial  profile pointing  to
elliptical   morphology from a previous   NIR  study  \citep{kuhl02}. No
extended emission     line  suitable   for    radial   velocity  curve
extraction.  It is the  only  object with  a clear mismatch  between stellar
population (young) and morphology (elliptical).  The gas  metallicity  also  
points to  a
low enrichement by stars.  The  quasar spectrum itself is  peculiar,
with broad H$\beta$ blueshifted with respect to the narrow component ($\sim
2700
\kms$), probably revealing  an outflow.  Traces of shells or interaction would not be surprizing if
high  resolution imaging was  available.

\item{HE\,0450-2958:  }This object is a special case that has been
 addressed  by \citet{mag05}.   Upper    limits were placed  on    the
undetected stellar contribution to the spectrum of the host, and only a
cloud  of gas  ionized by  the  quasar was   detected. As  there is no
detection of any host, this  object is not   included in global  means or
detailed   galactic component   analysis.  Quasar  spectrum typical of
NLQSO.

\item{HE\,0530-3755:  }One of the faintest  hosts detected in the 
sample,  but    with a high Balmer decrement on quasar broad lines that  might suggest
obscuration by dust on the nucleus and for the host as well. Small  amount  of gas  partially
ionized by the  quasar; images suggesting  interaction with tidal tail
to  the West,  one  close companion  at  6 arcsec East, that  luckily
partially falls on the slit and  whose spectrum shows nearly identical
redshift  (faint galaxy,  low   continuum and  only  H$\alpha$ 
detected in   emission), plus   another putative companion  9 arcsec
South. The nucleus has an accretion rate above the Eddington
limit, but without matching all the NLQSO criteria.

\item{HE\,0914-0031:  }Normal spiral galaxy with high 
mass-to-light   ratio according  to  the  mass   modelling  of Section
\ref{rc}.

\item{HE\,0956-0720:  }Bad PSF, inducing poorly deconvolved spectrum. 
No   interpretation  possible  with such   a  low  S/N.  Morphological
classifications   from    previous   studies        are   contradictory
(\citealt{kuhl02}, \citealt{percival}).

\item{HE\,1009-0702:  }Normal late-type spiral galaxy according to morphology,
stellar and gasous content. Quasar spectrum typical 
of NLQSO.

\item{HE\,1015-1618:  }Poor separation from the quasar because of 
a bad PSF.

\item{HE\,1029-1401:  }The lowest redshift object of the sample, luminous
 elliptical with minor mergers, already considered in several previous studies.
Gas highly ionized by the nucleus. 

\item{HE\,1228+0131:  }Poor separation of the host from the quasar, 
because the  saturation level of the detector   was almost reached. No
interpretation possible. Quasar spectrum typical of NLQSO.

\item{HE\,1302-1017 (or PKS\,1302-102):  }
This object is the only radio loud quasar in the sample.
It cannot be properly classified concerning morphology. 
\citet{bahc97} classify the host as elliptical because of 
its smooth and  elliptical shape from {\it HST}  imaging, although they find
an   exponential profile to fit    their   data slightly better.    In
\citet{kuhl02}  it is  classified  as disc  from  2D  modelling on NIR
images, but  presents an irregular  radial  profile. It is  the most
luminous object of the sample. The stellar population of the host galaxy is
typical of Sc disc population. It  has  low gas  metallicity and the
highest SFR of the HII subsample. From \citet{bahc97},
it has two small companions merging in, perturbing the radial velocity
curve presented  in  this   paper and probably explaining the   irregular
radial
profile, but without high excitation of the large  amount of gas present
in this  host. If  it is  a morphologically  disc-dominated galaxy, the 
stellar  and
gaseous  analyses point to Sc-Sd where merger-induced star formation and
activity appear,  giving  a counter-example of a
luminous radio-loud quasar not hosted  by an elliptical. On the other hand, if 
it is an elliptical galaxy, it must anyway have undergone recent mergers,
which have largely  enhanced star formation and  brought a large  amount of
fresh gas, able to mask the old stellar populations from our analyses.

\item{HE\,1434-1600:  }Elliptical galaxy with gas ionized by the quasar at
large distance from the nucleus, 
discussed in \citet{let04}. 

\item{HE\,1442-1139:  }Probable S0-like galaxy, with no gas detected.

\item{HE\,1503+0228:  }Normal spiral host galaxy, studied as a test case 
in \citet{courb02}.

\item{HE\,2258-5524:  }Major merger resulting in highly ionized gas and 
asymmetric   rotation     curve.  Stellar  population  typical of  S0-Sa
galaxy. Quasar spectrum typical of NLQSO.

\item{HE\,2345-2906:  }Spiral host galaxy for one of the weakest quasars of the
sample 
(at the limit  of the quasar/Seyfert  separation). The narrow lines in this
the quasar are much more prominent relative  to the broad components than for 
all other objects of the sample. 

\end{itemize}

\section{Discussion}
\label{discu}
\subsection{Quasar hosts  characterization}
A first striking result is that most quasar host galaxies harbour large amounts
of  gas
 (ionized by stars, by the AGN or marginally by shocks), 
irrespective of morphological type. We can thus safely reinforce the
conclusions of \citet{scoville},
 based on molecular CO detection, that high luminosity quasars 
are generally {\it not} found in normal (i.e. gas-poor) ellipticals.  Previous 
mergers  or  collisions are probably   the
source of  gas for ellipticals. 

There is an obvious predominance of globally young stellar population 
(10/15 of the hosts with  known stellar content). As several hosts
 with a young stellar population have unknown morphologies, this  observation 
could  be linked to   the bluer
colours  of elliptical quasar hosts   in comparison to  inactive ellipticals
found     in  several       recent  multicolour    imaging     studies
(e.g. \citealt{kauff}; \citealt{jahnke04}; \citealt{sanchez}). We note
that the two    elliptical  hosts for which   our   stellar population
analysis shows  a   predominantly  old  population    (HE\,1434-1600  and
HE\,1029-1401) are also presenting this type of bluer colours, according to
\citet{jahnke04}.  One thus has to conclude that they contain a globally old
stellar population,  as deduced from Lick indices analyses, over which appears
an additional fraction of younger stars which account
 for the blue colour excess, without significantly modifying the spectral
absorption diagnostics of section~\ref{stellpop}.
As the bluer-than-expected ellipticals display a globally old stellar
population with an additional young component, and as most of the host
galaxies with a globally young stellar population are disc-dominated, we can
probably conclude that most of the hosts with undefined morphologies and young
stellar population have a significant disc component.\\
 From a statistical point of view,
 the proportion of discs found in this study (8 of
 the  19 radio quiet  quasars, or 7/14 if  we
 remove  the NLQSOs that are often excluded from quasar samples,   thus  around 
45 \%) is  
 compatible with   the   proportions  found  by  \citet{dunlop03}  and
 \citet{floyd04} (namely around 30 \% among  radio quiet quasars), given
 the limited number of objects in both samples. But as these last two studies
 have  used several constraints   on  target selection,  there  might
 indeed be  a  selection bias in  their   samples towards ellipticals,
 while no other constraint apart from luminosity was applied to ours. Given
   that our proportion is a lower limit, the radio quiet quasars probably
appear as
 often in spirals as in ellipticals.

The confirmed spirals show HII regions compatible with rather metal-poor Sc to
Sd types, regarding star formation and gas metallicity. Their black holes are less
massive than the ones found
in  ellipticals (as expected from the bulge mass - black hole mass scaling
relation, \citealt{mclure02}), and only a minority present signs of
interaction (2/8).

The elliptical host galaxies contain ionized gas (3/4), have massive black
holes and at least 50 \% of them  present signs of interaction. The only
{\it
normal} elliptical, with no gas detected, could
  not be well deconvolved, so that no detailed analysis of its stellar and gas
content has been possible.

It has been claimed  that  all quasars more luminous
 than  $M_{V}   = -23.5$   are
 hosted by massive ellipticals (\citealt{dunlop03};  \citealt{floyd04}).
Converting this limit to our cosmology
 gives   $M_{V}<  -23.25$.   As seen  from Table~\ref{sampletab}, all
 our quasars  but one are brighter  than this limit, and even
 if the most luminous AGNs lie in ellipticals, we  find almost half of them
inside
 disc-dominated galaxies in this  magnitude range. The mean nuclear
 $V$ magnitudes are $M_{V} = -24.5$ both for elliptical and for disc hosts,
even
 if we remove the potential NLQSOs identified as  discs from our sample
 (they were excluded from  Dunlop's sample). This observation that
 the   morphology of  luminous quasar host    galaxies are not   necessarily
 elliptical was also  put forward  by  \citet{sanchez}  in  the case of higher 
 redshift quasars ($0.5 < z < 1.1$).
 
\subsection{The quasar-host connection}
\subsubsection{Influence of galactic environment on the activity of the
nucleus}
As shown  in   Section~\ref{quasar},  interaction obviously    brings
material (gas, and probably dust) to the central part of the galaxy and enhances
the accretion rate and power of the quasar. Recent simulations of galaxy
mergers (\citealt{dimatteo}; \citealt{cattaneo}) lead to the same conclusions. 
The most massive black holes
are  hosted   in systems  presently  interacting,  while we might have
expected   them in systems  at the  end  of the merging process, as discussed
in Section~\ref{cm}.
 The quasar harboured by interacting systems
are also more powerful. 

On  the other hand,  half of our quasars  are found in non-interacting
  galaxies,  with a majority of  spiral  hosts (6 confirmed discs, 2
ellipticals and  3 undefined).  Most of them are young gas-rich systems.
The predominance of a disc component ensures that these system have not
undergone major
mergers, as the latter tend to produce spheroids \citep{stan}. 

  This suggests that mergers  are not  the only
mechanism   for  triggering    activity,    as already    proposed  by
\citet{jahnke04} and \citet{dunlop03}.  Alternative mechanisms may
be bar-driven accretion (found to be very efficient for AGN fueling
for instance in  NLS1s, \citealt{crenshaw}, or for HE\,2345-2958 in the present
study) or gravitational instabilities related to spiral arms formation. But
small scale
nuclear events, related to an early stage of formation of spiral galaxies,
might also trigger activity. They are
 undetectable with the spatial resolution of the present observations. Galactic
winds arising from central supernovae are plausible  drivers of
  fuel towards the central engine,
 given that the weak enrichment of gas in heavy elements is measured on global
spectra only.

There does not seem to be a single scenario
for the evolution of quasars and their  hosts, but different evolution schemes
lead to different characteristics of
 the central engine. If the system encounters
major merger events,  it  will lead  to a higher BH mass and a more powerful
nucleus, but the nucleus could be active as well with less dramatic and 
smaller scale events in young spiral galaxies,
 leading to smaller and less powerful quasars.
 
\subsubsection{Influence of the quasar radiation on the host galaxy}

In 7 galaxies out of 20, we find gas  ionized by 
the quasar far from the nucleus (in ellipticals,  spirals or  undefined hosts).
This phenomenon is independent from radio emission, as none of the objects
with such long-range ionization is radio loud.
The common characteristics of those galaxies are of course the presence of gas,
but also a powerful quasar,
and evidences for gravitational interaction, two properties that are linked.  
The   origin   of high
ionization may lie in  the fact that  gas and dust around the nucleus
are swept by the  interaction and let  the powerful nuclear ionizing beam reach
large distances throughout the galaxy.

The analysis of gas metallicity in quasar hosts with a HII region-like ISM
shows  abundances at the lower end of typical spiral metallicities.
The nuclear activity in these young systems could  slow down the star formation
efficiency in some way, hampering the formation of massive stars 
and slowing down the enrichment of the surrounding gas. 
Links between star formation and quasar activity \citep{jahnke04} are
preferentially found for elliptical hosts where activity
 seems to enhance star formation. However, in that case, neither quasar
activity nor star formation would be expected in an isolated host. Both are
made possible by the presence of gas, probably brought by mergers with younger
systems.  We agree with the conclusions of \citet{can01} who, from a study of
UltraLuminous IR Galaxies and QSOs, conclude that it is the merging process
that triggers both 
star formation and activity, without  finding evidence for any direct influence
from one of these onto the other. In contrast, it does not appear unrealistic
to suggest that the quasar activity might slow down the star formation in
young, isolated, gas-rich spiral galaxies.
\\

Further spectroscopic and high resolution imaging investigations  of larger
samples of quasars and
their hosts could allow to  confirm the trends  found here and allow a
sharper  tuning  of the interpretations  and hypotheses  on the quasar
ignition and evolution.


\begin{figure*}
\centering
\includegraphics[width=1.\textwidth,height=0.4\textwidth]{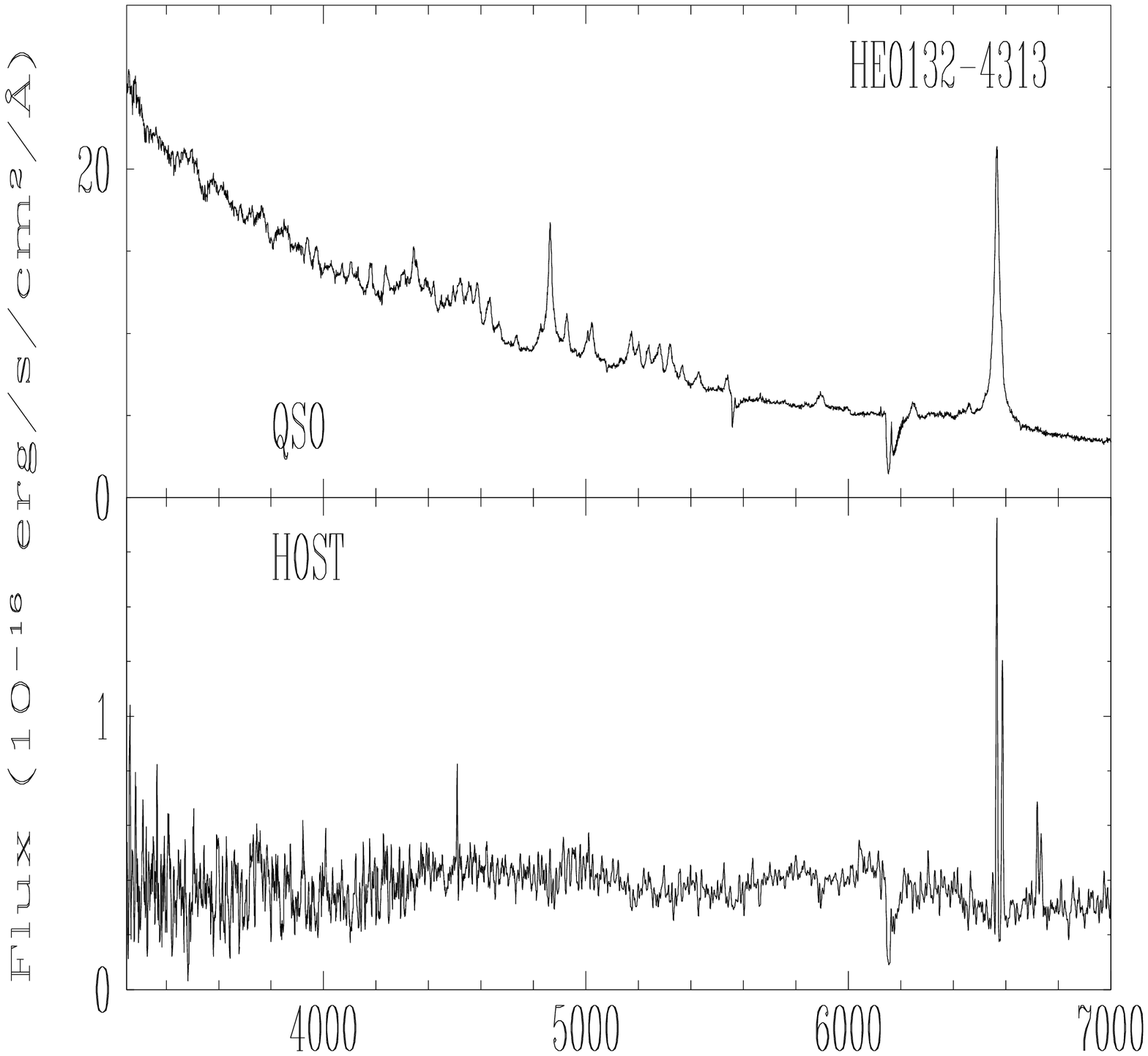}
\includegraphics[width=1.\textwidth,height=0.4\textwidth]{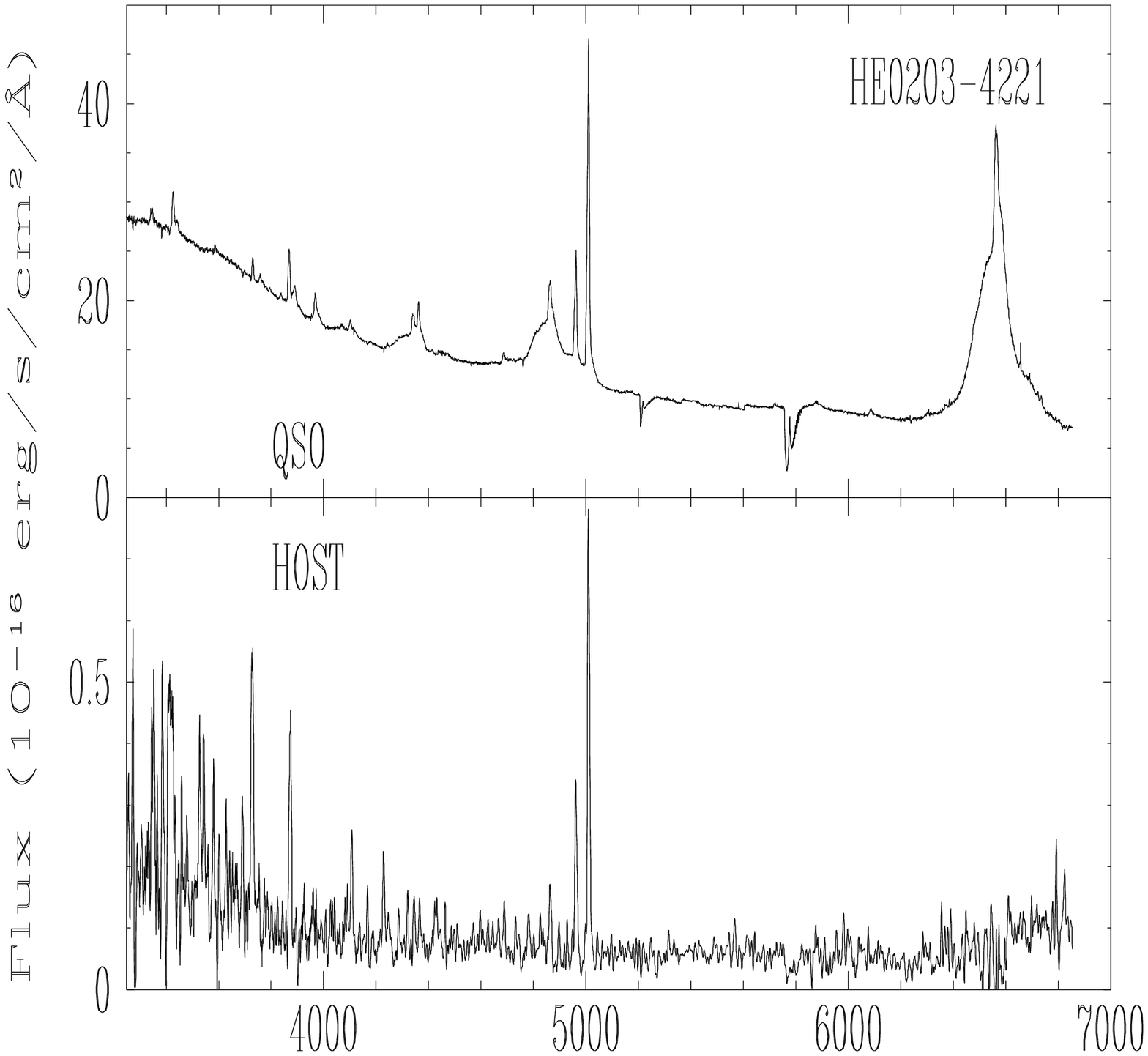}
\includegraphics[width=1.\textwidth,height=0.4\textwidth]{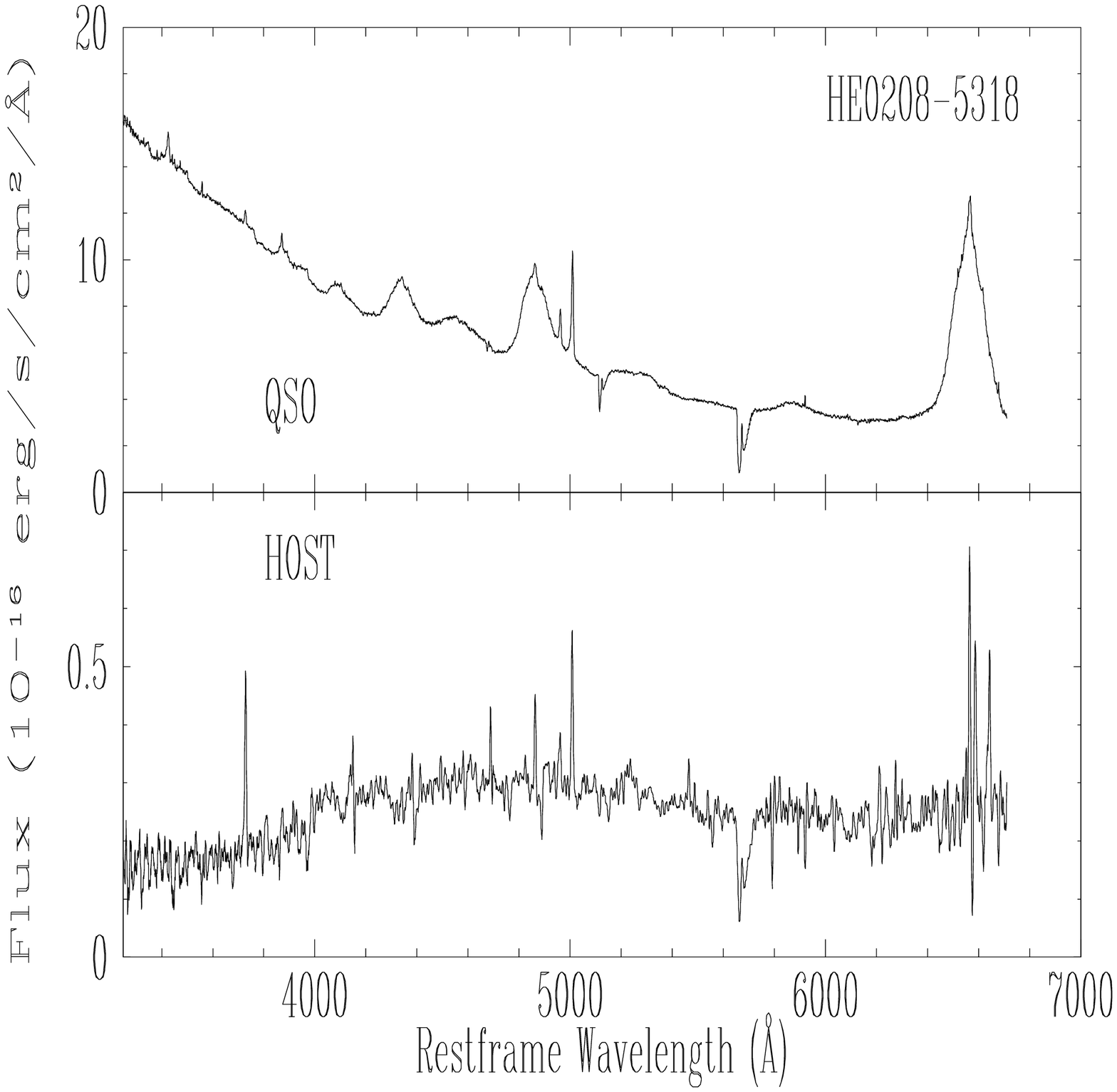}
\caption{For each object, integrated spectra of the quasar-only on top and 
galaxy-only at the bottom, after  deconvolution and separation, sorted
by increasing right ascension. The dotted galactic spectra give the limits of over- and under- subtraction of the nuclear component as discussed in Section \ref{error}.}
\label{spectro1}
\end{figure*}
\begin{figure*}
\centering
\includegraphics[width=1.\textwidth,height=0.4\textwidth]{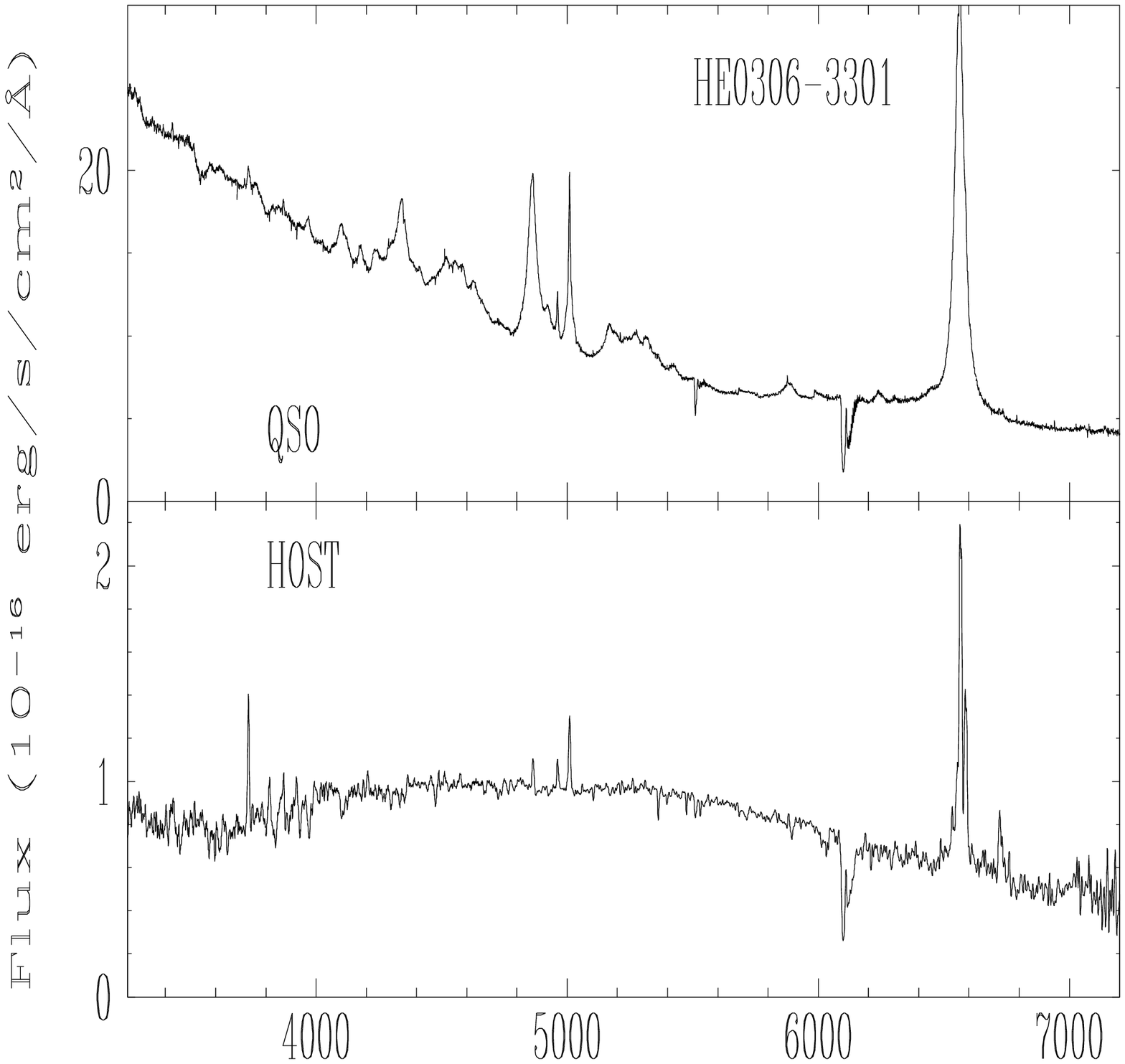}
\includegraphics[width=1.\textwidth,height=0.4\textwidth]{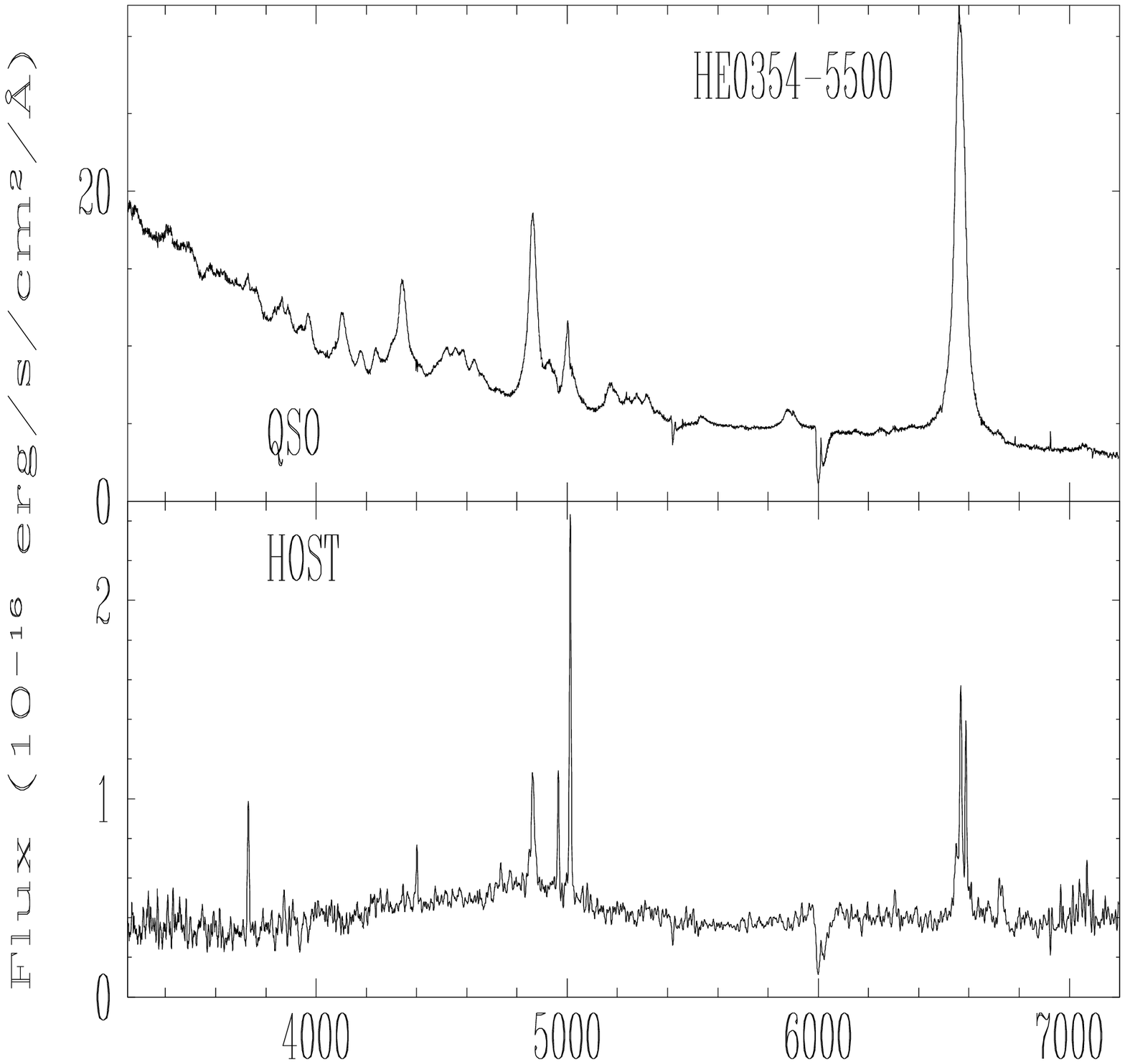}
\includegraphics[width=1.\textwidth,height=0.4\textwidth]{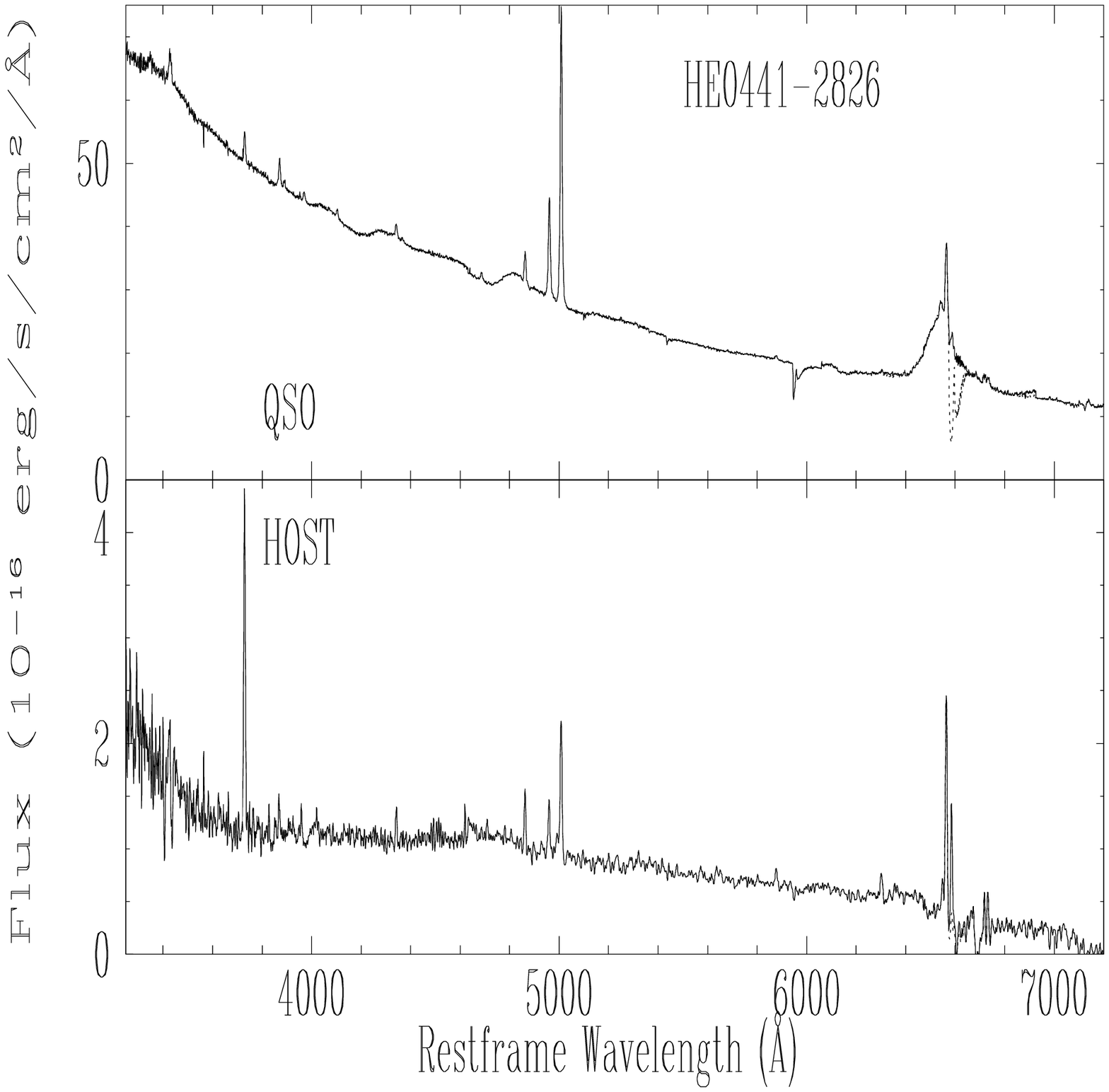}
\contcaption{\,\, There seems to be residual nuclear light in the 
continuum of the galaxy of HE\,0441-2826, because of a poorer PSF. The
blend  of H$\alpha$   with  atmospheric absorption  (dotted  line) was
partially corrected. }
\end{figure*}
\begin{figure*}
\centering
\includegraphics[width=1.\textwidth,height=0.4\textwidth]{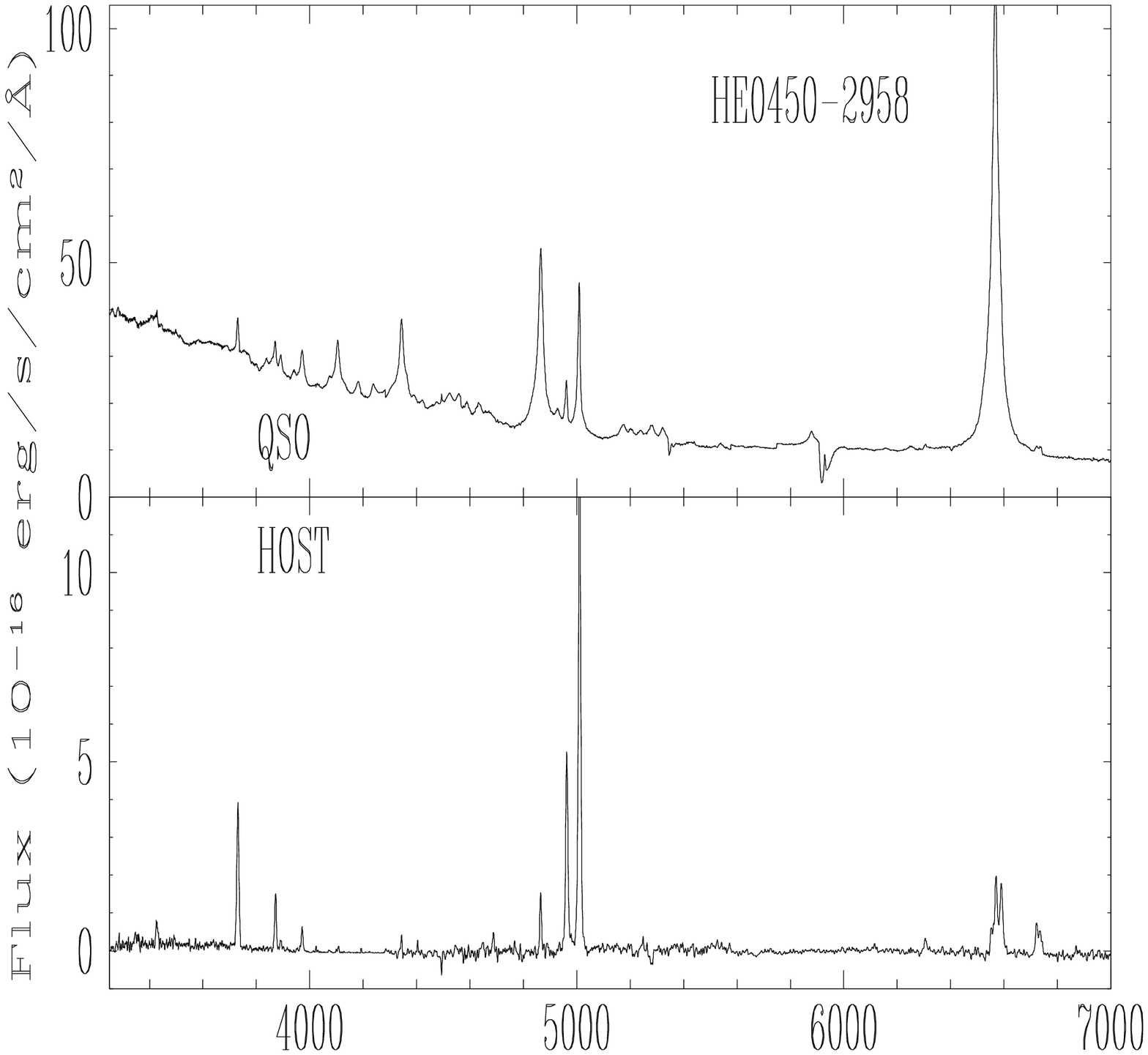}
\includegraphics[width=1.\textwidth,height=0.4\textwidth]{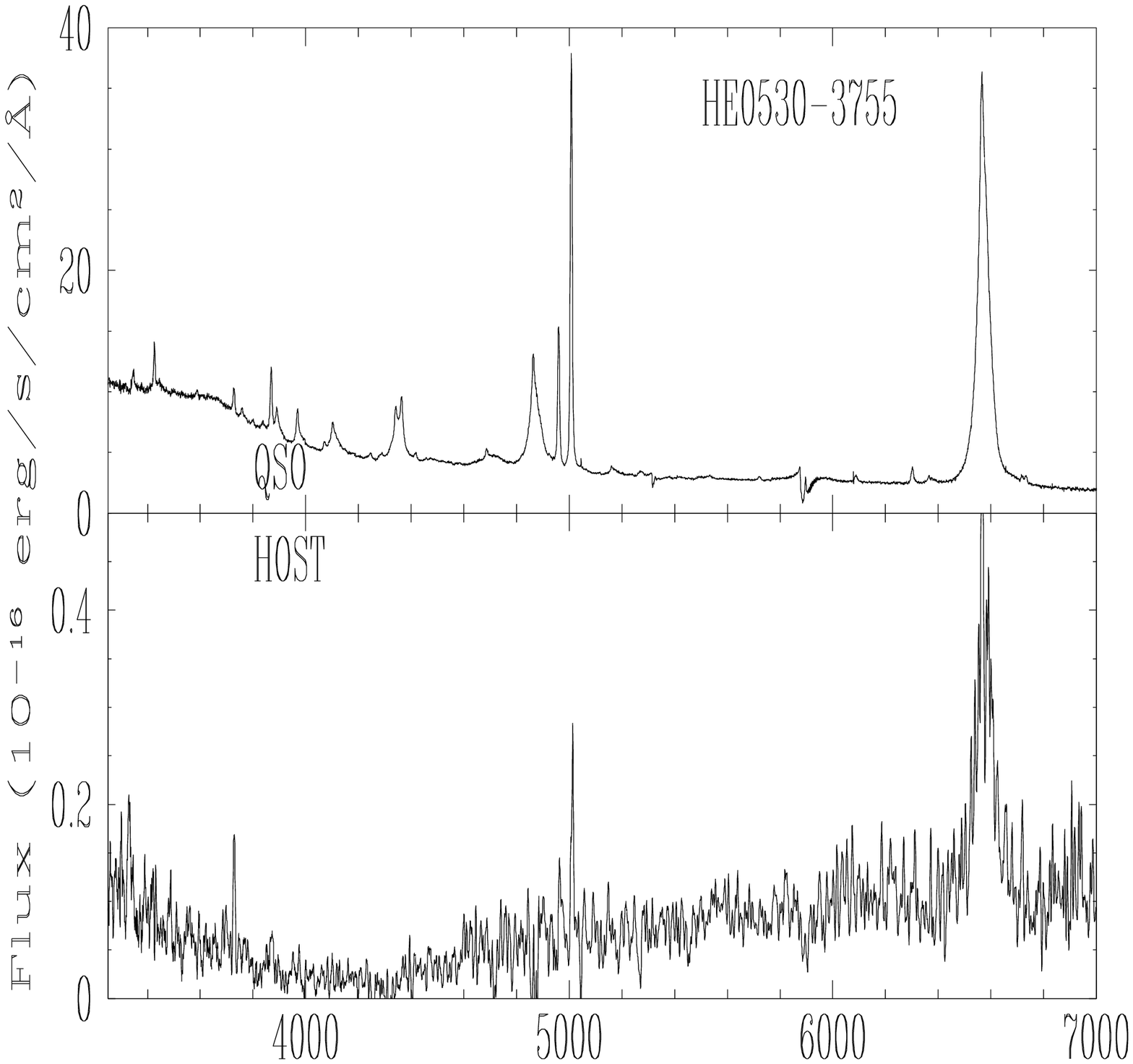}
\includegraphics[width=1.\textwidth,height=0.4\textwidth]{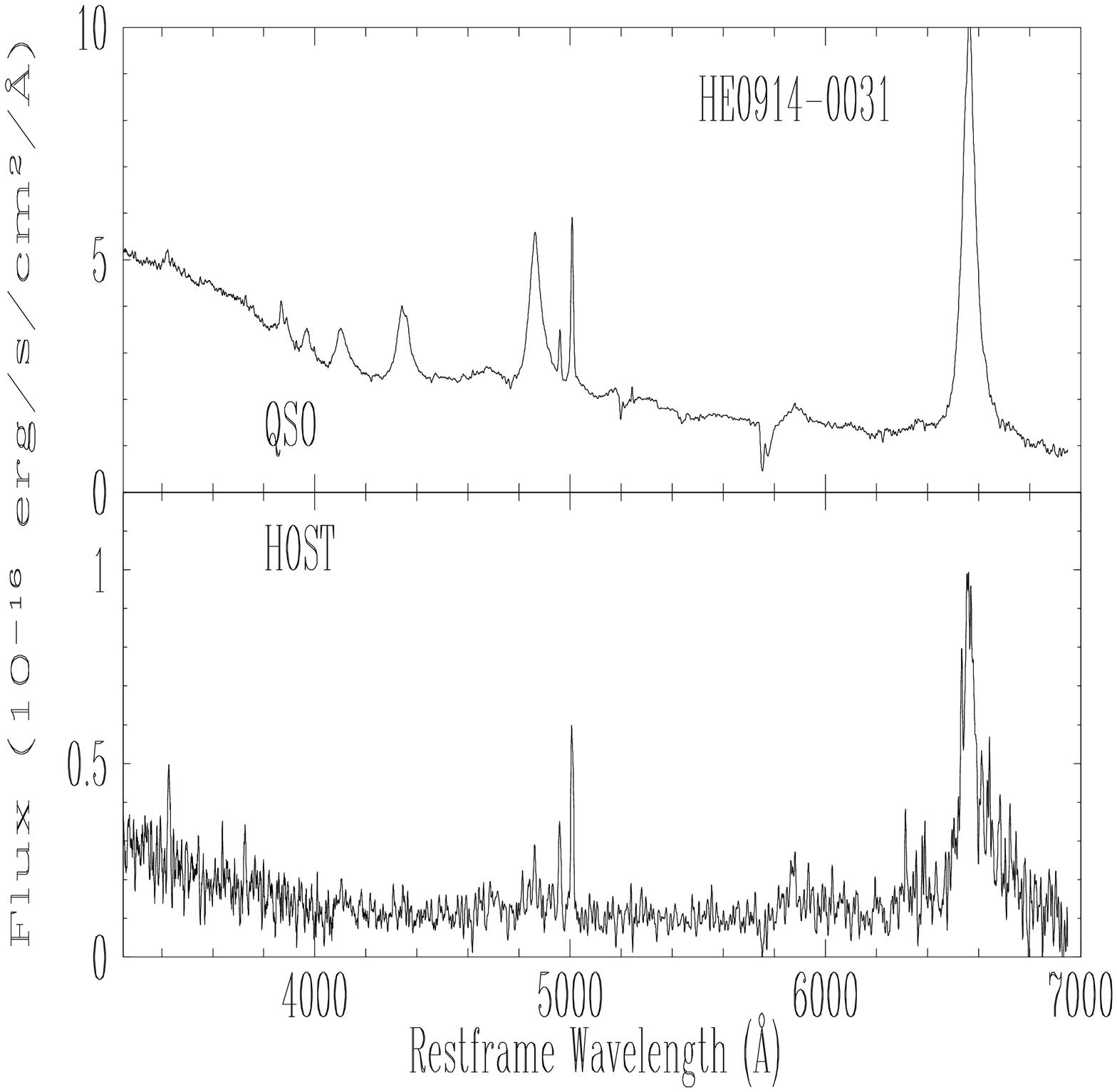}
\contcaption{Note the absence of continuum in the host of HE\,0450-2928.}
\end{figure*}
\begin{figure*}
\centering
\includegraphics[width=1.\textwidth,height=0.4\textwidth]{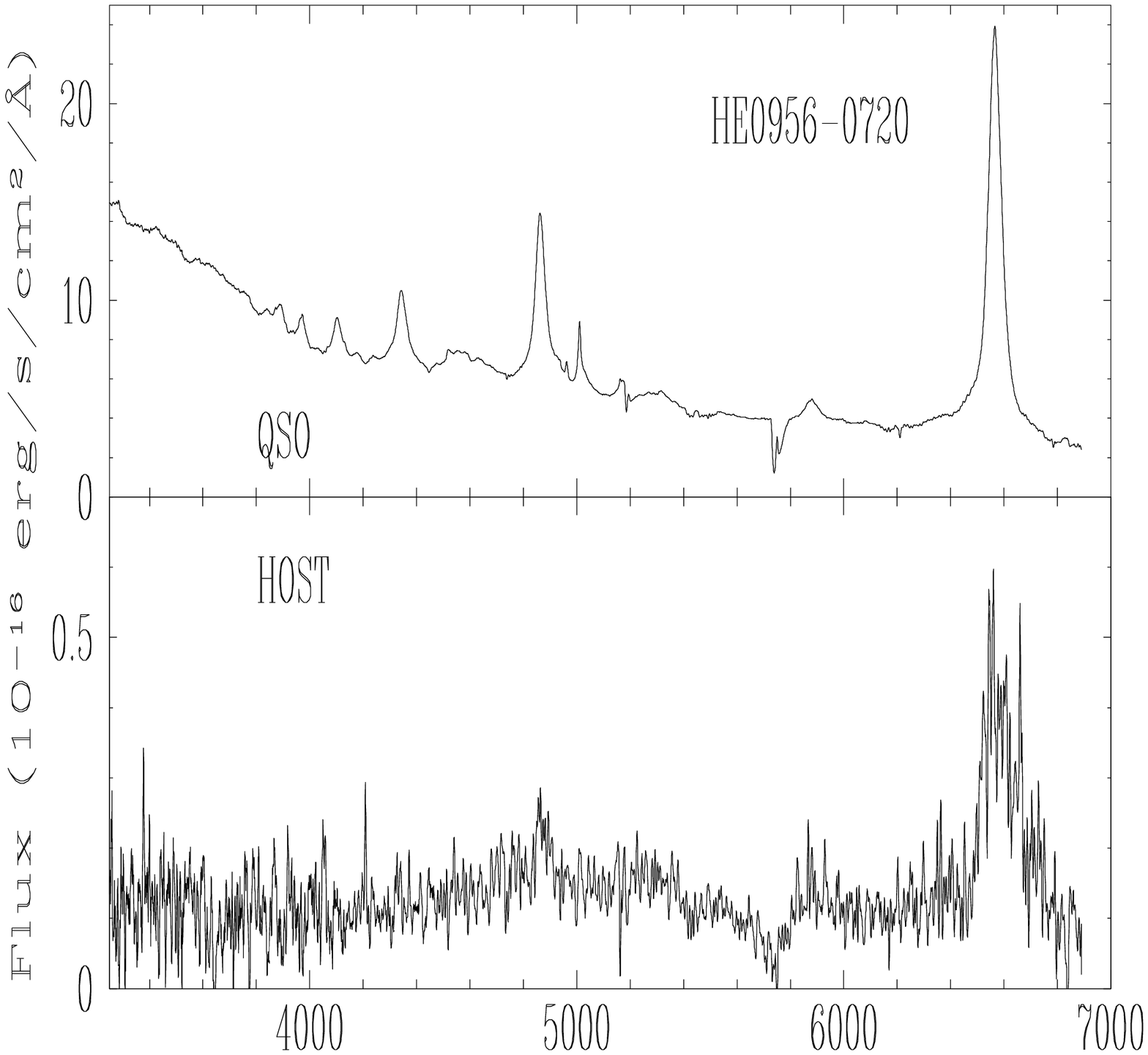}
\includegraphics[width=1.\textwidth,height=0.4\textwidth]{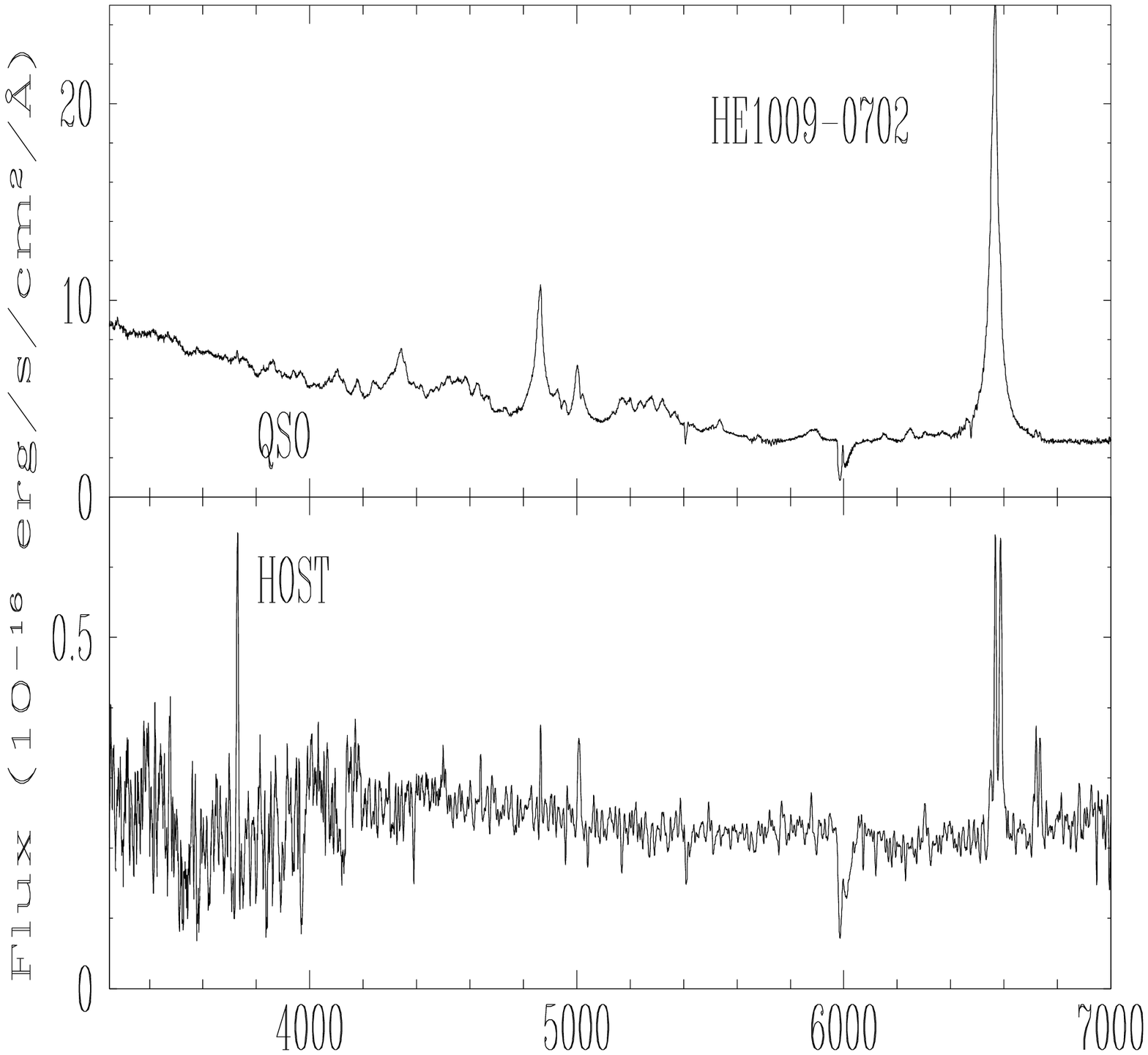}
\includegraphics[width=1.\textwidth,height=0.4\textwidth]{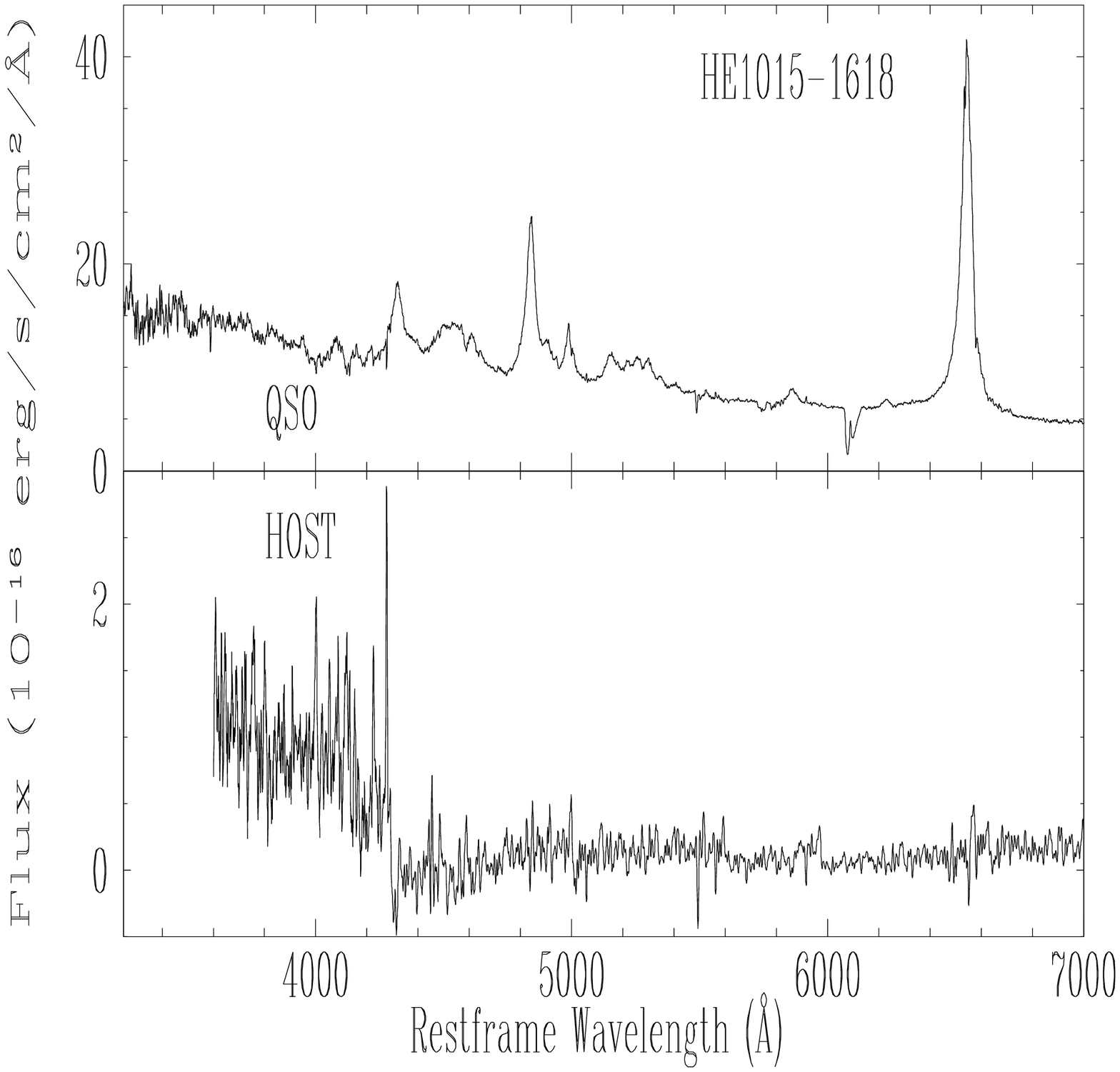}
\contcaption{\,\, HE\,0956-0720 and HE\,1015-1618 could not be properly 
deconvolved, because of high N/H and difficult PSF construction.}
\end{figure*}
\begin{figure*}
\centering
\includegraphics[width=1.\textwidth,height=0.4\textwidth]{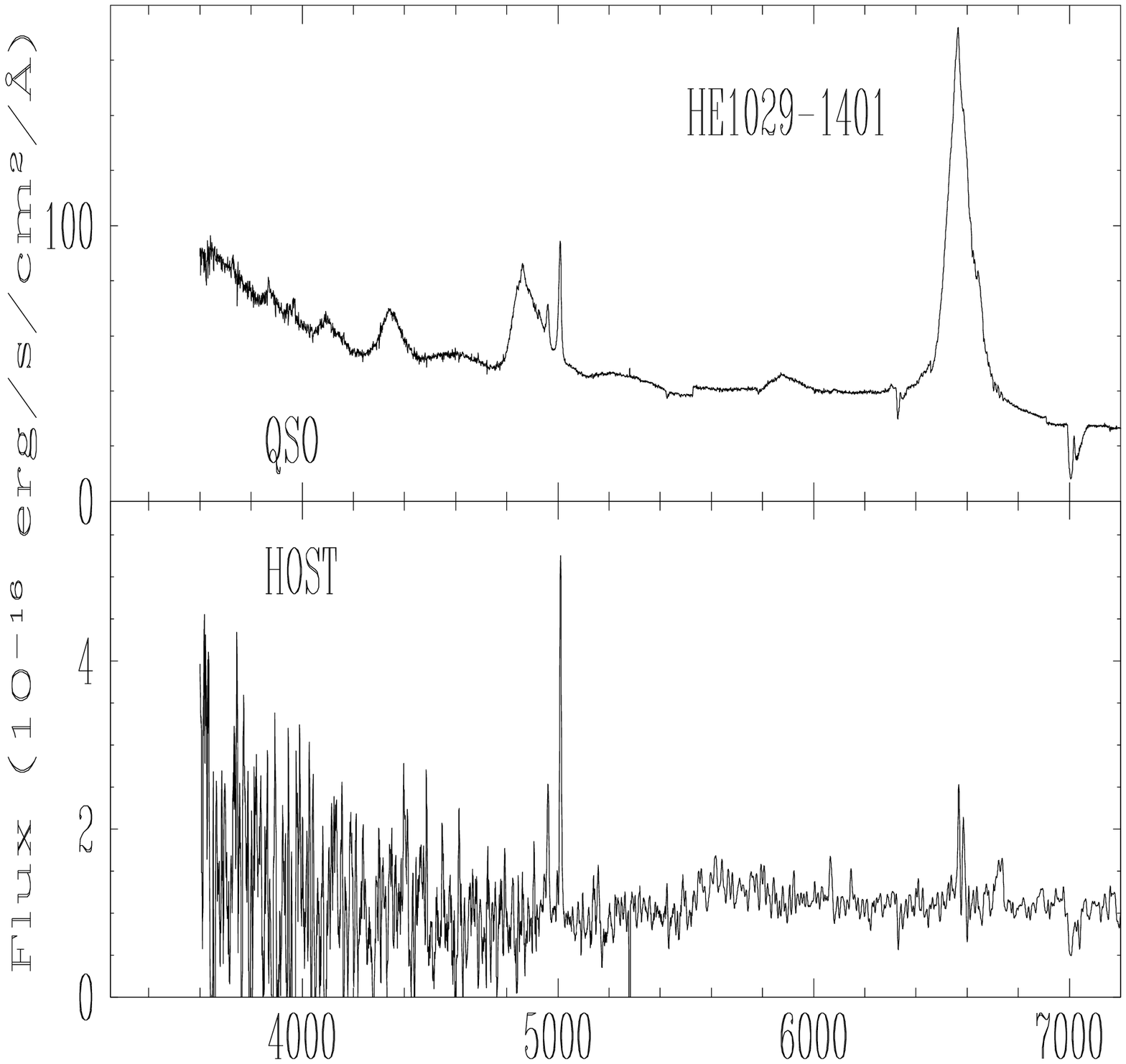}
\includegraphics[width=1.\textwidth,height=0.4\textwidth]{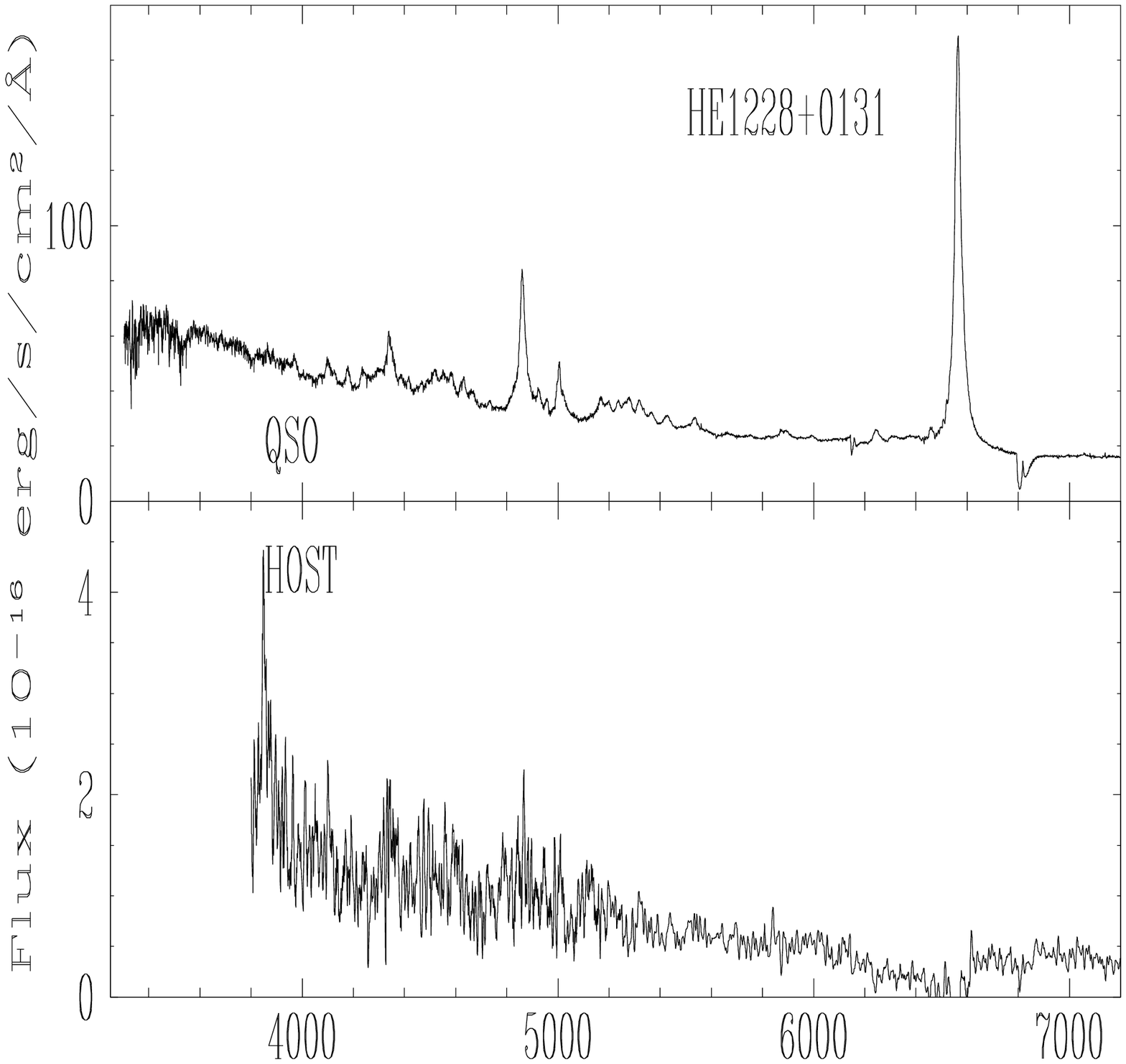}
\includegraphics[width=1.\textwidth,height=0.4\textwidth]{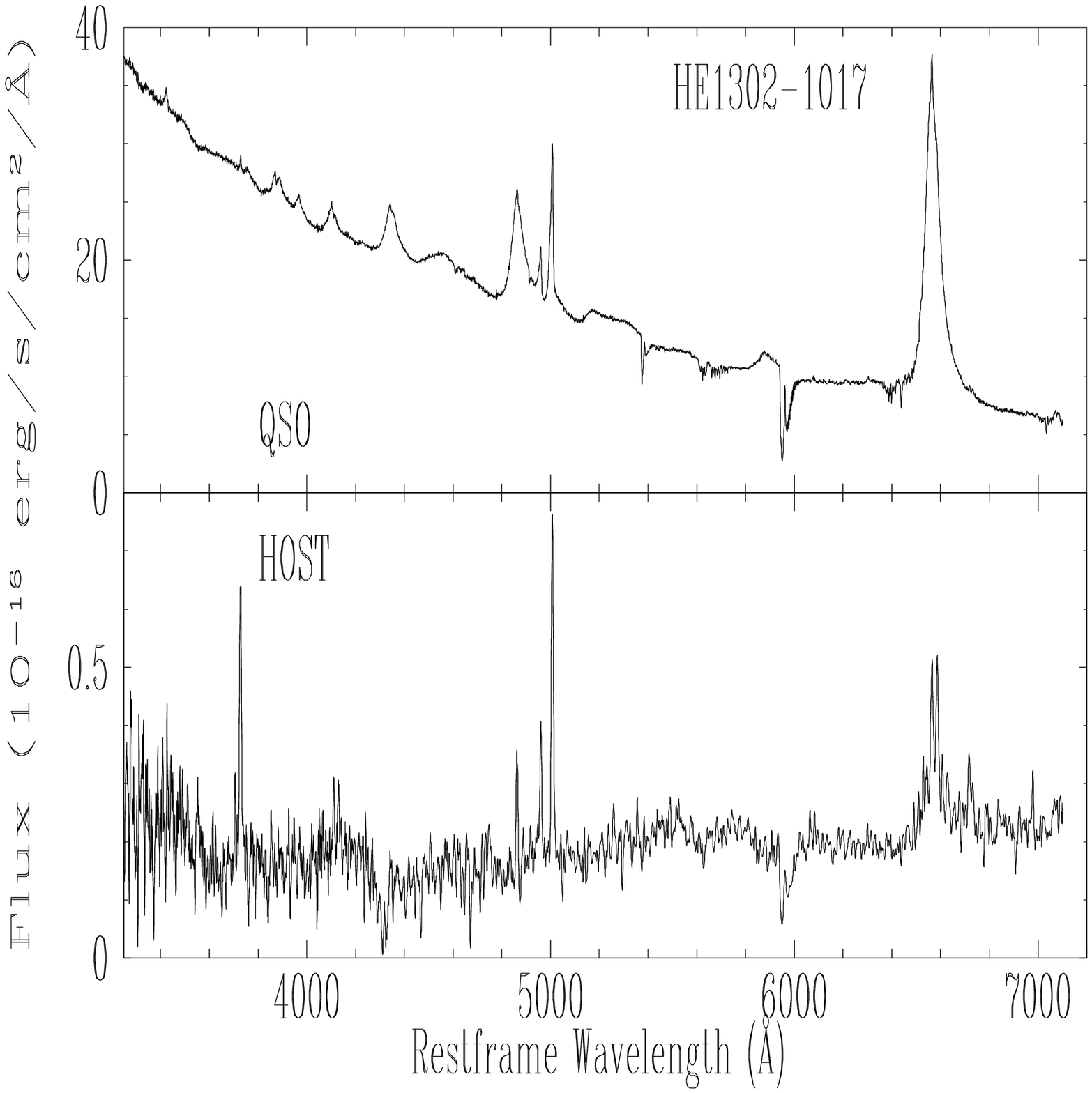}
\contcaption{\,\, Important noise in the bluer parts of the two first 
objects,   the PSF stars   being too faint   in the  blue for accurate
deconvolution in   this wavelength  range.  HE\,1228+0131 observations
were near saturation level on the detector, leading to non adequacy of
the PSF and thus poor deconvolution.}
\end{figure*}
\begin{figure*}
\centering
\includegraphics[width=1.\textwidth,height=0.4\textwidth]{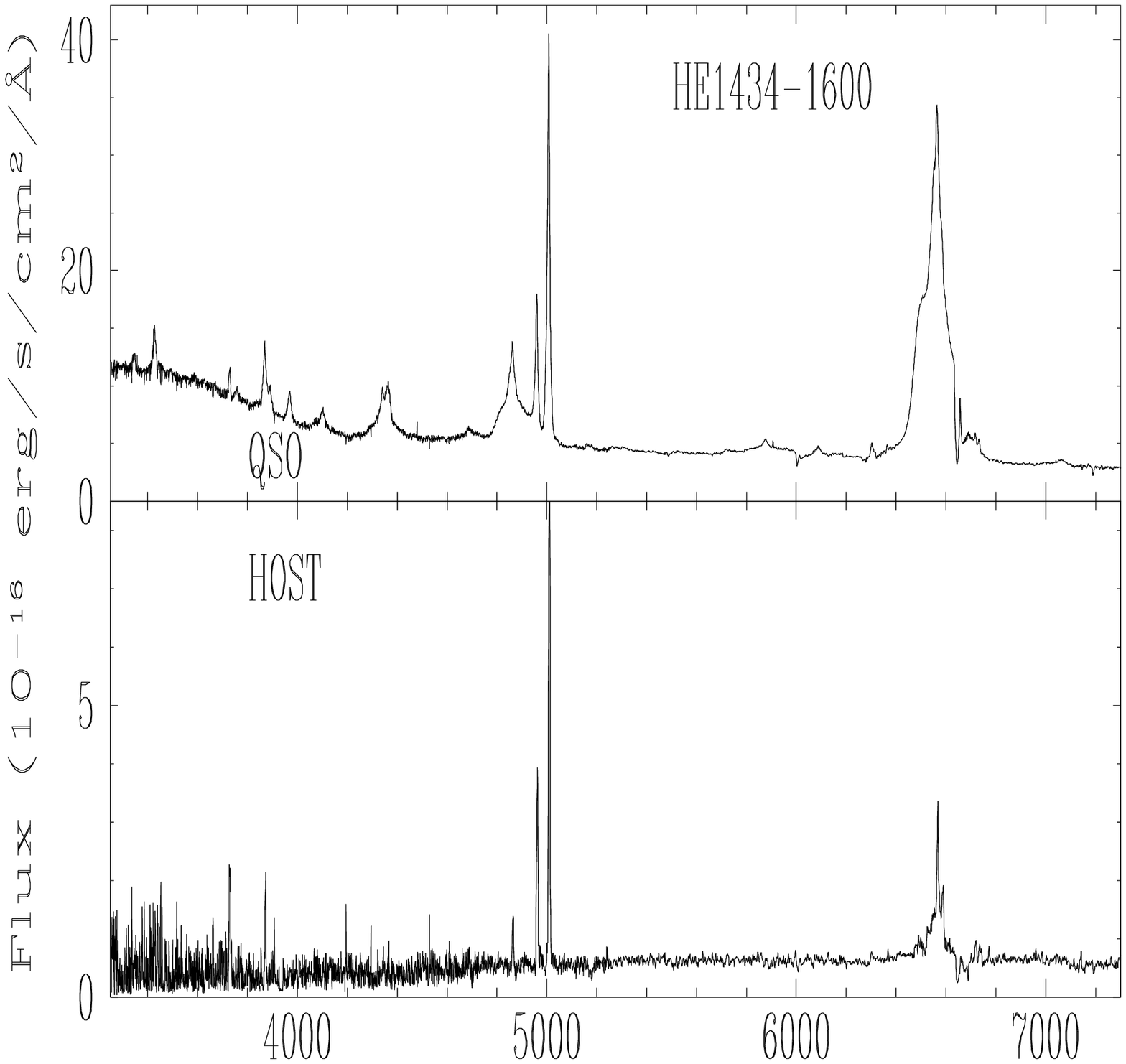}
\includegraphics[width=1.\textwidth,height=0.4\textwidth]{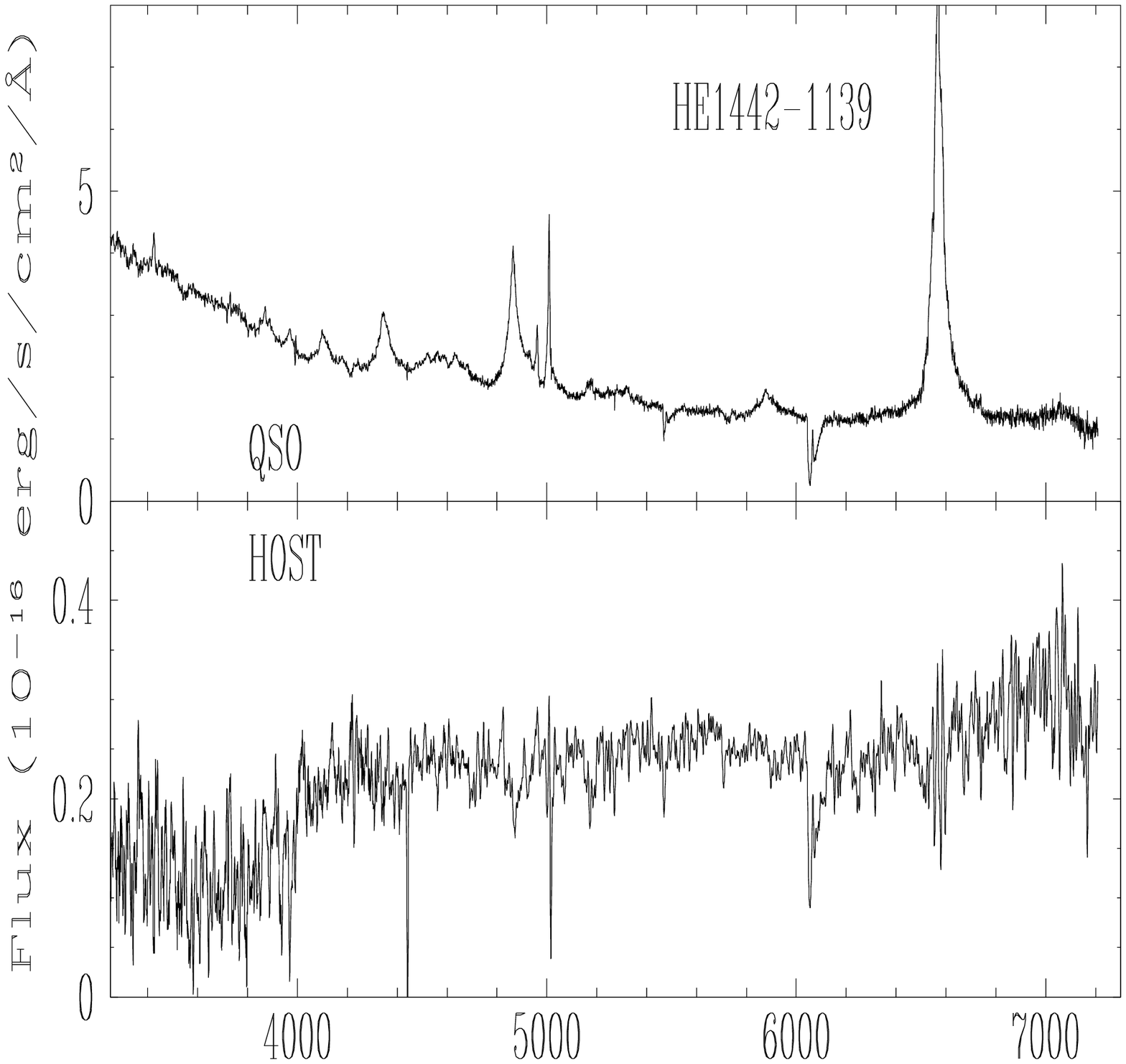}
\includegraphics[width=1.\textwidth,height=0.4\textwidth]{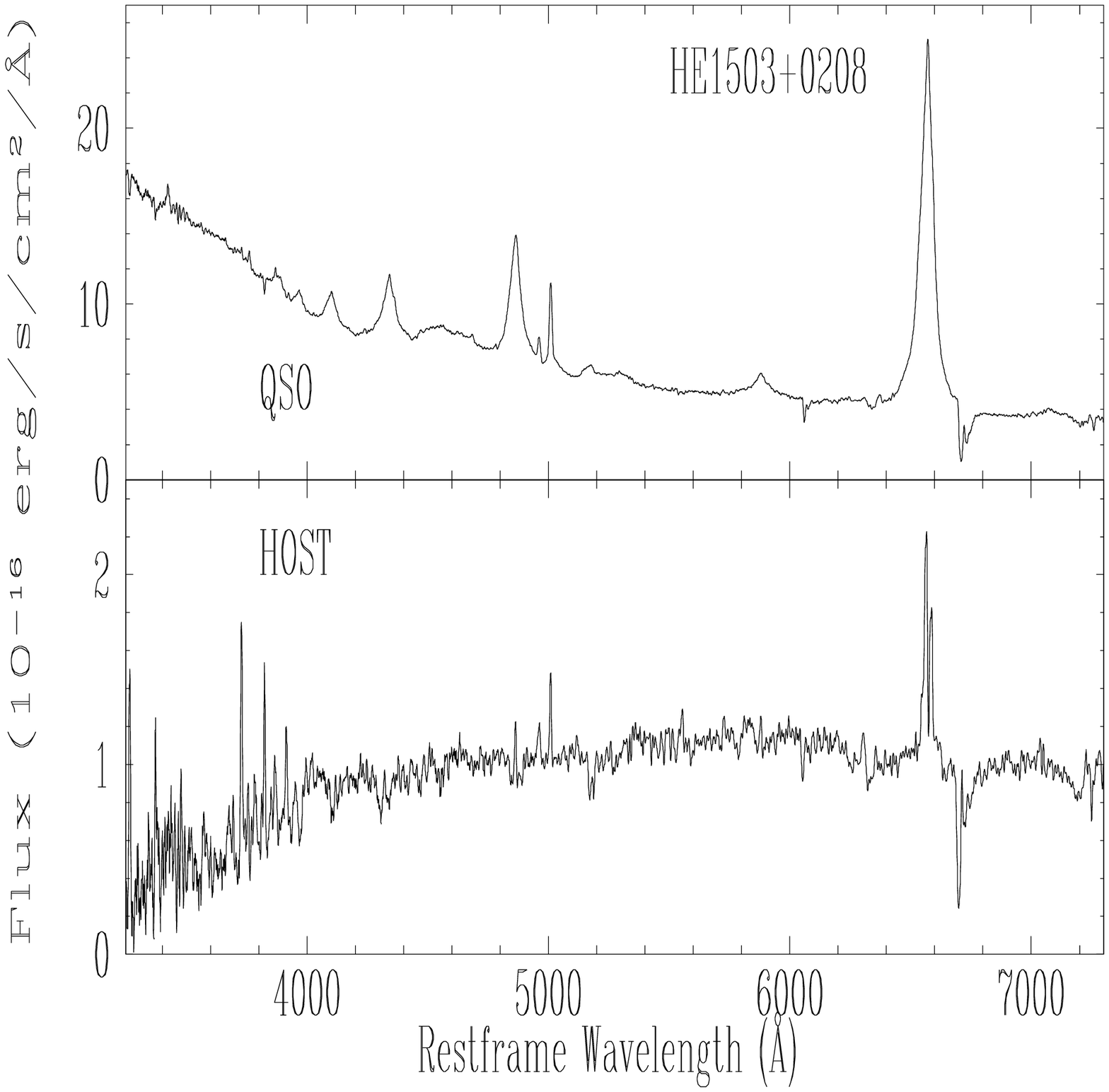}
\contcaption{}
\end{figure*}
\begin{figure*}
\centering
\includegraphics[width=1.\textwidth,height=0.4\textwidth]{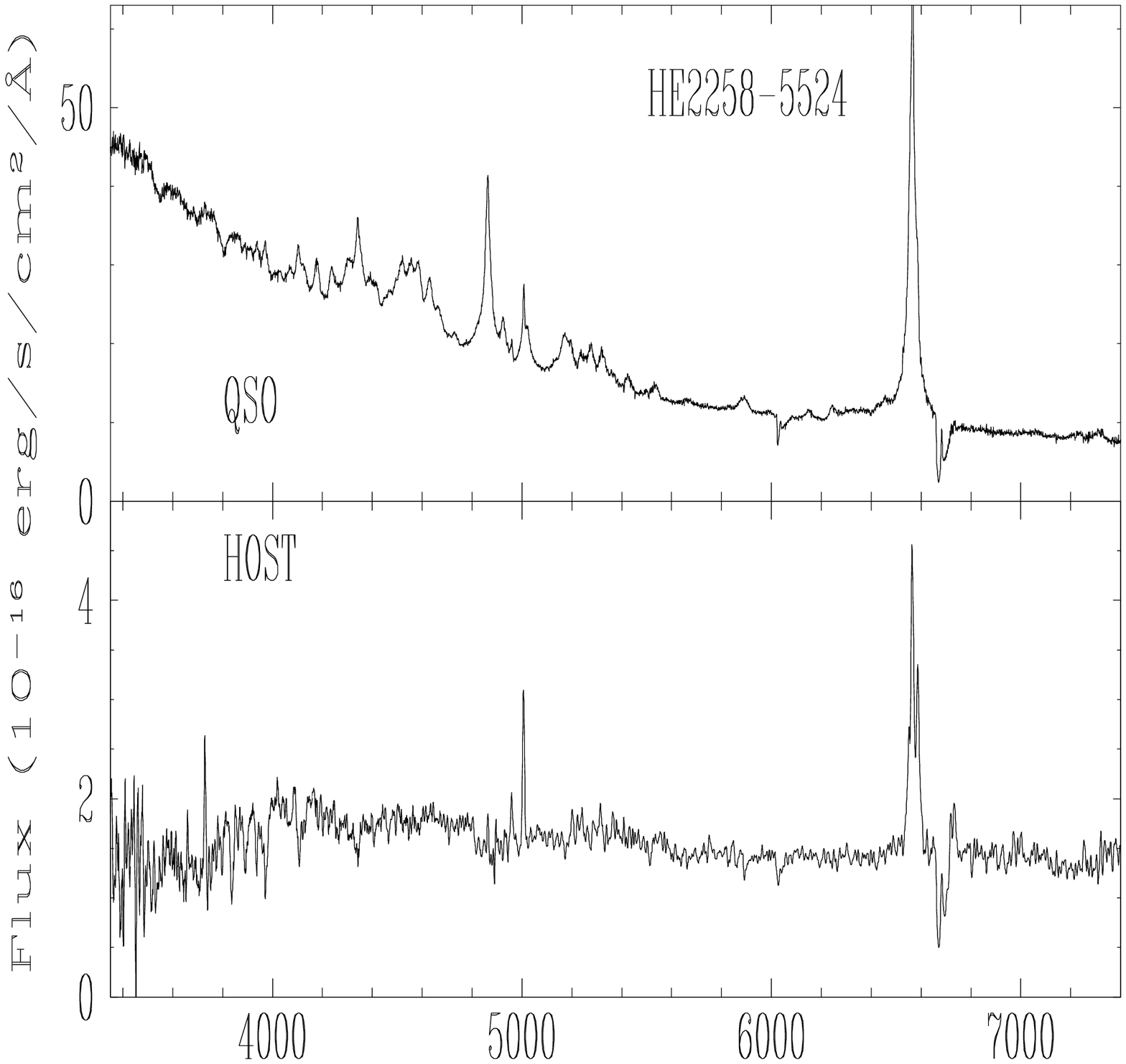}
\includegraphics[width=1.\textwidth,height=0.4\textwidth]{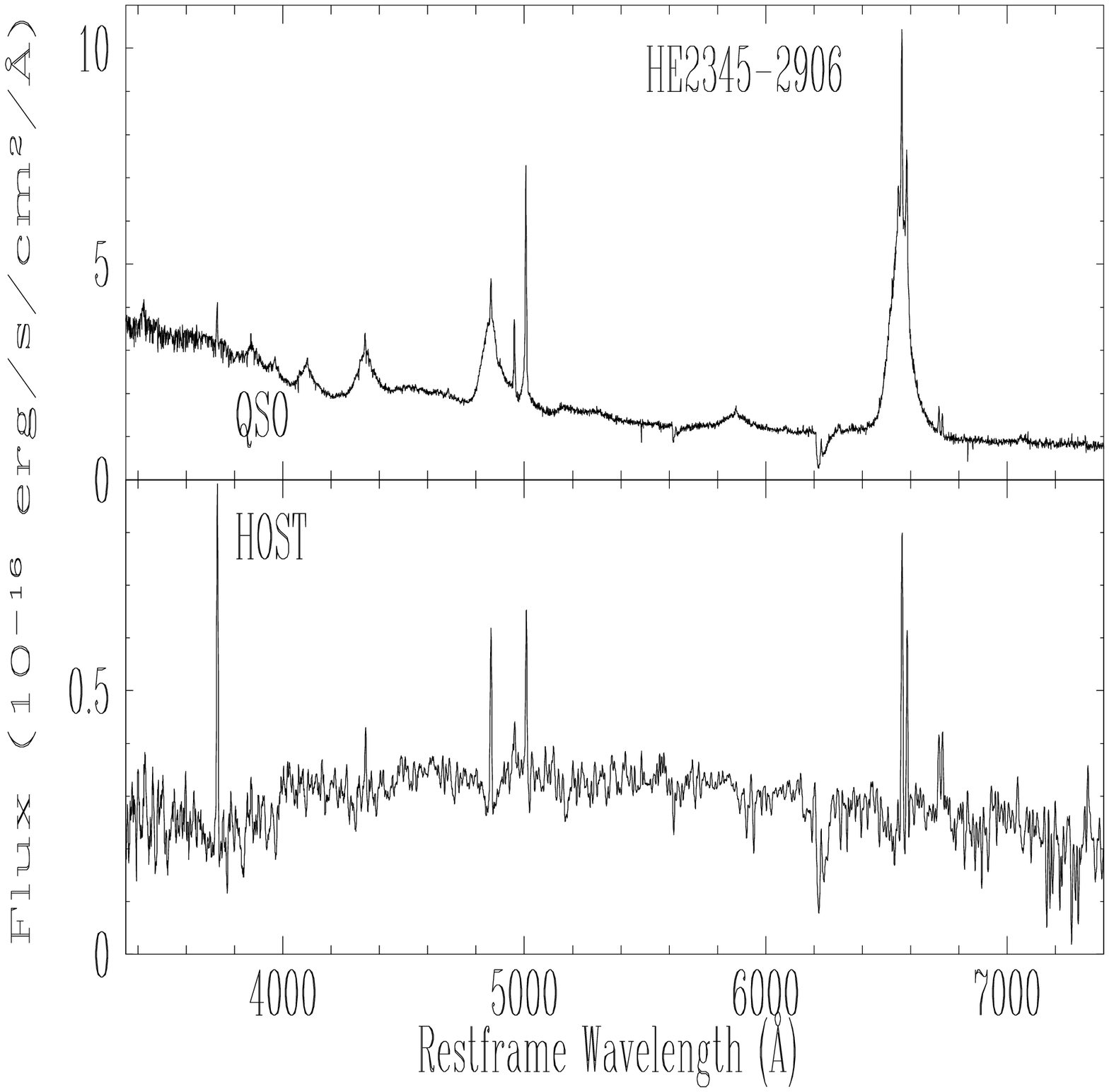}
\contcaption{}
\end{figure*}

\section*{acknowledgments}
G.~L. is a teaching assistant  supported by the University of Li\`ege,
(Belgium) .  The  P\^ole  d'Attraction Interuniversitaire, P5/36  and PRODEX
90195 (PPS Science Policy, Belgium, and ESA) contracts are thanked for
financial support.   F.~C., P.~J. and   G.~M. acknowledge support from
the Swiss   National Science Fundation.   K.~J.\  was supported by the
German DLR     under  project  number  50~OR~0404    and  DFG  project
SCHI~536/3-1.

\label{lastpage}
\end{document}